\documentclass[11pt,a4paper]{article}
\pdfoutput=1

\usepackage{jcappub}
\usepackage{slashed}
\usepackage{braket}
\usepackage{hyperref}
\usepackage{float}
\usepackage{bbold}
\usepackage{amsmath}
\usepackage{amssymb}
\usepackage{graphicx}
\usepackage{xcolor}
\usepackage{mathrsfs}
\usepackage{comment}
\usepackage{bm}
\usepackage[normalem]{ulem}
\usepackage{cancel}

\usepackage[scr=boondox]{mathalfa}

\newcommand{\gadgetcode}{\mbox{\tt Gadget-4}}
\newcommand{\gadgettwo}{\mbox{\tt Gadget-2}}

\newcommand{\classcode}{\mbox{\tt class}}
\newcommand{\cambcode}{\mbox{\tt camb}}

\newcommand{\concept}{\mbox{\tt CO{\textit N}\hspace{-1pt}CEPT}}

\title{Hybrid multi-fluid-particle simulations of the cosmic neutrino background}

\author[a,1]{Joe Zhiyu Chen,\note{Present address: 50 Martin Place, Sydney NSW 2000, Australia}}
\author[b]{Markus R. Mosbech,}
\author[c,a]{Amol Upadhye,} 
\author[a]{Yvonne Y. Y. Wong}

\affiliation[a]{Sydney Consortium for Particle Physics and Cosmology, School of Physics, The University of New South Wales, 
Sydney NSW 2052, Australia} 
\affiliation[b]{Sydney Consortium for Particle Physics and Cosmology, School of Physics, The University of Sydney, 
Sydney NSW 2006, Australia\\
ARC Centre of Excellence for Dark Matter Particle Physics}
\affiliation[c]{Astrophysics Research Institute, Liverpool John Moores University,
 146 Brownlow Hill, Liverpool L3 5RF, United Kingdom}

\emailAdd{jchen19940@gmail.com} 
\emailAdd{markus.mosbech@sydney.edu.au}
\emailAdd{a.r.upadhye@ljmu.ac.uk}
\emailAdd{yvonne.y.wong@unsw.edu.au}

\abstract{
Simulation of the cosmic clustering of massive neutrinos is a daunting task, due both to their large velocity dispersion and to their weak clustering power becoming swamped by Poisson shot noise.  We present a new approach, the multi-fluid hybrid-neutrino simulation, which partitions the neutrino population into multiple flows, each of which is characterised by its initial momentum and treated as a separate fluid. These fluid flows respond initially linearly to nonlinear perturbations in the cold matter, but slowest flows are later converted to a particle realisation should their clustering power exceed some threshold.
After outlining the multi-fluid description of neutrinos, we study the conversion of the individual flows into particles, in order to quantify transient errors, as well as to determine a set of criteria for particle conversion.  Assembling our results into a total neutrino power spectrum, we demonstrate that our multi-fluid hybrid-neutrino simulation is convergent to $<3\%$ if conversion happens at $z=19$ and agrees with more expensive simulations in the literature for neutrino fractions as high as $\Omega_\nu h^2 = 0.005$.  Moreover, our hybrid-neutrino approach retains fine-grained information about the neutrinos' momentum distribution.  However, the momentum resolution is currently limited by free-streaming transients excited by missing information in the neutrino particle initialisation procedure, which restricts the particle conversion to $z \gtrsim 19$ if percent-level resolution is desired.
}

\begin{document}

\begin{flushright}
	{\large \tt CPPC-2022-10}
\end{flushright}	
	
\maketitle

\section{Introduction}

The standard big bang predicts the existence of a relic neutrino background. At some 100 neutrinos per cubic centimetre per flavour today, these relic neutrinos are, next to the cosmic microwave background photons, the most abundant known particles in the universe. With the discovery of neutrino oscillations we now also know that at least some of these neutrinos must have masses and contribute to the present-day non-relativistic matter content an amount given by the reduced density  $\Omega_\nu \simeq \sum m_{\nu} / (93 \,h^2 \,  {\rm eV})$, where $\sum m_{\nu}$ sums over all thermalised mass eigenstates. Taking together the minimum and maximum mass values established from oscillation~\cite{deSalas:2020pgw,Esteban:2020cvm,Capozzi:2017ipn} and $\beta$-decay endpoint~\cite{Aker:2019uuj} experiments, the present-day reduced neutrino energy density is predicted to lie in the range $0.001 \lesssim \Omega_\nu \lesssim 0.03$.

The evolution of this neutrino background and particularly its spatial inhomogeneity under the influence of gravity is of interest to modern cosmology  because of the phenomenon of free-streaming---the escape of relic neutrinos from gravitational potentials due to their large thermal motion.  Free-streaming distinguishes relic neutrinos from the canonical cold dark matter (CDM) and imprints on the large-scale spatial distribution of matter in a distinctive, $m_\nu$-dependent way that makes it possible to infer neutrino masses from observations~\cite{Lesgourgues:2006nd,Wong:2011ip}.
The computational framework for quantifying these effects in the linear regime of gravitational evolution  is well established~\cite{Ma:1995ey}, on which basis 
measurements of the cosmic microwave background (CMB) anisotropies and the baryon acoustic oscillations (BAO) currently constrain the neutrino mass sum to $\sum m_\nu \lesssim 0.5$~eV in the $\Lambda$CDM class of cosmologies~\cite{Planck:2018vyg}.

In the nonlinear regime of evolution, many studies have also appeared in recent years, putting forward a variety of techniques both perturbative and non-perturbative to capture neutrino free-streaming effects on large-scale structure formation~\cite{Brandbyge:2008js,AliHaimoud:2012vj,Archidiacono:2015ota,Dakin:2017idt,Mccarthy:2017yqf,Bird:2018all,Chen:2020bdf,Chen:2020kxi,Inman:2020oda,Brandbyge:2008rv,Viel:2010bn,Villaescusa-Navarro:2013pva,Castorina:2015bma,Banerjee:2016zaa,Banerjee:2018bxy,Brandbyge:2018tvk,Bayer:2020tko,Elbers:2020lbn,Partmann:2020qzb,Elbers:2022xid}.
For neutrino mass sums not exceeding $\sum m_\nu \sim 1$~eV, several such schemes that combine $N$-body cold particles and a perturbative approach for neutrinos~\cite{Brandbyge:2008js,AliHaimoud:2012vj,Partmann:2020qzb,Chen:2020bdf,Chen:2020kxi} are even able to predict the power spectra of the cold matter (CDM and baryons) and the total matter to sub-0.1\% and sub-1\% accuracy, respectively, without incurring significant run-time or implementation overhead.

Less attention, however, has been directed to getting the {\it neutrino inhomogeneities themselves} in the nonlinear regime of evolution ``right''.  Indeed, those sub-\%-accurate methods alluded to above typically employ a grid-based {\it linearised} approach to track neutrino perturbations that also underestimates their nonlinear growth by, as we shall show, factors of a few to an order of magnitude, depending on the nature of the linear theory.%
\footnote{Amongst existing perturbative treatments of neutrino inhomogeneities in nonlinear structure formation, we distinguish between two types of linear theory: linear neutrino perturbations and linear response. The former approach~\cite{Brandbyge:2008js,Partmann:2020qzb} uses purely linear neutrino perturbations outputted from a linear Boltzmann code such as \cambcode{}~\cite{Lewis:1999bs} or \classcode{}~\cite{Blas:2011rf,Lesgourgues:2011rh}. The latter approach also linearises the equations of motion for the neutrino inhomogeneities, but allows the inhomogeneities to respond to a gravitational potential sourced by nonlinear perturbations in the cold matter distribution~\cite{Ringwald:2004np,AliHaimoud:2012vj,Chen:2020bdf,Chen:2020kxi}.}
The lack of attention to this aspect of nonlinear evolution is undoubtedly a consequence of the inability of conventional astronomical probes to directly measure the relic neutrino content of the universe.  Nonetheless, we emphasise that novel astronomical observations have been proposed that may reveal more about nonlinear neutrino evolution than do conventional $N$-point statistics. These include a dipole distortion in galaxy cross-correlations induced by relative velocities~\cite{Zhu:2013tma,Inman:2016prk,Zhu:2019kzb,Zhou:2021sgl}, a long-range correlation in galactic rotation directions~\cite{Yu:2018llx}, wakes~\cite{Zhu:2014qma}, non-Gaussianity~\cite{Liu:2020mzl}, suppressed mass accretion by dark matter halos~\cite{Wong:2021ats}, 
and an environment-dependence of the halo mass function~\cite{Yu:2016yfe}.
  From the particle physics perspective, $\beta$-decay endpoint spectrum measurements~\cite{PTOLEMY:2019hkd} and high-energy astroparticle probes~\cite{Weiler:1982qy,Ringwald:2005gx,Brdar:2022kpu} too are sensitive to relic neutrino clustering in the Milky Way and elsewhere in the universe through direct particle interactions. Thus, reliable predictions of the relic neutrino inhomogeneities are certainly desirable.

In this work, we explore the question of how and what it takes to compute the nonlinear evolution of neutrino inhomogeneities accurately (to be quantified), while using computational resources in an optimal way.
A complete $N$-body particle representation of the 3+3 neutrino phase space is generally taken to be the gold standard for tracking nonlinear evolution in the neutrino sector.  It is however resource-intensive, both in terms of time-stepping and memory requirements, if noise is to be kept at an acceptable level.  On the other hand, grid-based approaches---where equations of motion in the Eulerian frame are solved on a spatial grid---can circumvent these issues, and, as discussed above, a number of existing computational schemes already employ some form of grid-based linear theory in their modelling of neutrino inhomogeneities.
The companion paper to this work, Paper 2~\cite{Chen:2022cgw}, will take this grid-based modelling of neutrino inhomogeneities to higher orders in perturbation theory.  In a similar vein,  reference~\cite{Banerjee:2016zaa} and the \concept{} papers~\cite{Dakin:2015uka,Dakin:2017idt,Dakin:2021ivb} also employ a grid-based nonlinear fluid description, together with a closure condition, to achieve the same goal.

Here, rather than pursuing a complete representation of nonlinear neutrino inhomogeneities on the grid, we explore an alternative, ``hybrid-neutrino'' simulation approach that combines a grid-based linear response theory together with a partial particle realisation in both time and momentum space.  At the heart of the hybrid principle is that, for neutrino masses compatible with current cosmological observations, only a small fraction of the thermally-distributed relic neutrino population will have sufficiently low momenta to warrant a non-perturbative $N$-body treatment and, even then, only for a fraction of the full simulation time. The majority of neutrinos remains fast-moving and hence amenable to some form of linearised perturbative computation, either for the entire duration of the simulation or at least down to some suitably low redshift.  This means that a judicious partitioning of the neutrino population will allow us to concentrate scarce computational resources only to those slow-moving neutrinos that genuinely require an $N$-body realisation, without sacrificing accuracy on the fast-moving streams.

To our knowledge two existing, independent studies in the literature utilise, to various extents, the hybrid principle~\cite{Brandbyge:2009ce,Bird:2018all}; we shall elaborate on their exact implementations in section~\ref{sec:hybrid}.  Suffice it to say here, however, that our implementation of the hybrid principle  incorporates elements from both of these works and improves upon them in various ways.   In particular, the present work can be considered an extension of the 
the multi-fluid linear response (MFLR) method presented by some of us in~\cite{Chen:2020bdf}, itself based upon the multi-fluid perturbation theory of~\cite{Dupuy:2013jaa,Dupuy:2014vea,Dupuy:2015ega}; the MFLR method will serve as our grid-based linear response theory down to  a set of conversion redshifts at which portions of the neutrino population will be converted into $N$-body particles.
The main objectives of this work, therefore, are
\begin{itemize}
\item  To demonstrate concordance between MFLR and the $N$-body approach for the fastest neutrinos; 
\item Where concordance cannot be achieved, to determine the criteria for which neutrinos to convert from a MFLR description to $N$-body particles; and 
\item To investigate the issue of transients arising from improper particle neutrino initialisation, from which to determine the optimal redshift(s) for conversion to particles. 
\end{itemize}
As we shall show, the last point concerning transients---especially transients arising from the lack of representation of the neutrino anisotropic stress at initialisation---is in fact the current limiting factor on how late the conversion to particles can begin.

The paper is organised as follows.  We first elaborate on the hybrid-neutrino schemes of references~\cite{Brandbyge:2009ce,Bird:2018all} in section~\ref{sec:hybrid}, followed in section~\ref{sec:mflr} by the MFLR implementation of reference~\cite{Chen:2020bdf}, upon which we shall based our hybrid-neutrino scheme.  The execution of conversion from MFLR to particles is described in section~\ref{sec:conversion}.  We carry out an extensive set of convergence tests and discuss the resulting conversion criteria in sections~\ref{sec:criteria} and~\ref{sec:NLCriteria}. Conversion of multiple fluid flows at the same or at multiple redshifts is explored in section~\ref{sec:multiple_conversions}, 
where we also present our nonlinear neutrino power spectrum predictions and the corresponding changes to the cold matter and total matter power.
In section~\ref{sec:concept} we compare our predictions against the equivalent outcomes of the \concept{} approach, which, as mentioned above, employs a different, grid-based fluid description to model neutrino nonlinearities.  Section~\ref{sec:conclusions} contains our conclusions.  A more technical discussion of our particle initialisation procedure at conversion is given in appendix~\ref{sec:appendix}.

%%%%%%%%%%%%%%%%
%%%%%%%%%%%%%
%%%%%%%%%%%%%%

\section{Hybrid-neutrino simulations}
\label{sec:hybrid}

The terminology ``hybrid-neutrino'' refers to the use of a combination of $N$-body particles and a grid-based perturbation theory to model the evolution of the neutrino population under gravity, in contrast with a conventional, full-on $N$-body particle representation.  Computational resource constraints tend to limit the resolution achievable in the latter kind of simulations.  The hybrid scheme, on the other hand, takes advantage of the fact that high-momentum neutrinos do not cluster gravitationally significantly and yet contribute the most in noise and run-time. Its strategy, therefore, is to separate out the fast, mildly clustering neutrinos and model them using some form of perturbation theory, so that computational resources can be concentrated on achieving higher resolution in the slow neutrino subset that is more likely to cluster nonlinearly and hence requires a non-perturbative solution. In this section, we discuss the conceptual workflow of a hybrid-neutrino scheme and briefly review two existing implementations in the literature~\cite{Brandbyge:2009ce, Bird:2018all}. 

The central guiding principle of the hybrid-neutrino scheme can be summarised qualitatively as {\it reserve $N$-body particle modelling of massive neutrinos for when and whom it is absolutely necessary}. ``When'' pertains to the redshifting thermal velocities of the neutrinos as we move forward in time. At early times/high redshifts when these thermal velocities are large, the method of linear response suffices to capture the gravitational clustering of neutrinos, and there is little to gain from a $N$-body particle representation (but a lot to lose in terms of run-time and noise).
As the neutrino momentum redshifts with time and gravitational clustering progresses to the nonlinear regime, the need to switch to a non-perturbative $N$-body representation also becomes imperative.  However, the exact timing of the switch from perturbation theory to $N$-body representation is not only a matter of ``when'', but also a matter of ``whom'' requires it.  Indeed, just as there will be low-momentum neutrinos in the Fermi-Dirac-distributed relic neutrino population that will reach the conversion time
fairly early, there will also always be neutrinos of such high momenta that they may never slow down enough to warrant $N$-body modelling.

In a practical implementation of the hybrid-neutrino scheme, two points need to be considered:
\begin{itemize}
    \item Which is the most convenient neutrino perturbation theory to use at early times, given the understanding that a fraction of the neutrino population will undergo conversion to particles at later times, and 
    \item How to decide and execute this conversion to $N$-body particles. 
\end{itemize}
Our own implementation of the hybrid-neutrino scheme in this work\footnote{Our implementation of the hybrid-neutrino scheme in \gadgetcode{} is publicly available at \url{https://github.com/joechenUNSW/gadget4-hybrid_public}.}, as well as the two existing independent implementations in Brandbyge and Hannestad~\cite{Brandbyge:2009ce} and Bird et al.~\cite{Bird:2018all} differ from one another on both points.  We describe briefly the implementations of~\cite{Brandbyge:2009ce,Bird:2018all} below.  The scheme explored in this work will be detailed in sections~\ref{sec:mflr} and~\ref{sec:conversion}.

%%%%%%
%%%%%%%
%%%%%%%%

\subsection{Hybrid 1: Brandbyge and Hannestad}
\label{sec:Brandbyge}

The original undertaking of the hybrid-neutrino scheme was presented in reference~\cite{Brandbyge:2009ce}. Here, the base $N$-body code used to track the cold matter and neutrino particles is \gadgettwo{}~\cite{Springel:2005mi}, while the pre-conversion purely-linear grid-based neutrino perturbations are computed with the linear Boltzmann code~\cambcode{}~\cite{Lewis:1999bs}, using 15 Eulerian comoving momentum $q$ bins covering $q/T \in [1, 15]$ in integer steps, where $T=1.95$~K is the present-day neutrino temperature. Conversion from linear perturbation theory to neutrino particles takes place at a redshift~$z_c$ for all neutrinos of momenta below a cut-off $q_{\text{cut}}$, where $z_c$ corresponds to the time at which the thermal velocity of the $q/T=1$ bin falls below a factor $f_{\text{flow}}$ times the average gravitational flow velocity of the cold matter in the simulation box. 
The exact values of both $q_{\text{cut}}/T$ and $f_{\text{flow}}$ need to be determined from convergence tests.  For the benchmark case of $\sum m_\nu \simeq 1.2$~eV, reference~\cite{Brandbyge:2009ce} found the optimal parameters to correspond to a conversion redshift in the range $5 \lesssim z_c \lesssim 10$, in comparison with the simulation initial redshift $z_{\rm sim} = 49$.

An important point raised in~\cite{Brandbyge:2009ce} is that there is a non-trivial correspondence between the Eulerian frame of the linear perturbation theory~\cite{Ma:1995ey} and the Lagrangian frame of an $N$-body simulation which, if unaccounted, could lead to momentum non-conservation after a subset of neutrinos has been converted to $N$-body particles. 
In particular, in the Boltzmann equation for the neutrino phase space density contrast~\cite{Ma:1995ey},%
\footnote{The Boltzmann equation is given here in the conformal Newtonian gauge, whose line element is
${\rm d} s^2=-a^2(\tau) [- (1+2 \psi){\rm d} \tau^2 + (1-2 \phi) {\rm d} x^i {\rm d} x_i]$.}
\begin{equation}
\label{eq:boltzmann}
\frac{\partial \Psi}{\partial \tau}  + {\rm i} \frac{q}{\epsilon} (\vec{k} \cdot \hat{n}) \Psi + \frac{{\rm d} \ln \bar{f}}{ {\rm d} \ln q} \left[\dot{\phi}  - {\rm i} \frac{\epsilon}{q} (\vec{k} \cdot \hat{n} )\psi\right]=0,
\end{equation}
correct accounting demands that the background density $\bar{f}(q)$ be modified  to reflect the sharp edge at $q_{\text{cut}}$ post conversion. 
This modification was not however implemented in the study of reference~\cite{Brandbyge:2009ce}.

%%%%%%%%%%%
%%%%%%%%%%%%

\subsection{Hybrid 2: Bird et al.}
\label{sec:Bird}

The hybrid scheme received attention again in reference~\cite{Bird:2018all}, which puts forward three significant changes over the original implementation of~\cite{Brandbyge:2009ce}. Firstly, the grid-based modelling of the neutrino inhomogeneities is upgraded from a purely linear description to the linear response method of~\cite{AliHaimoud:2012vj}.  In essence a solution of equation~\eqref{eq:boltzmann} in the non-relativistic limit but with the gravitational potentials sourced by the nonlinear cold matter in the $N$-body simulation box, the neutrino inhomogeneities evolved in this matter can now pick up some degree of nonlinearity, in comparison with the purely linear description used in~\cite{Brandbyge:2009ce}.

Secondly, the neutrino population is segregated from the start into a slow and a fast portion, each tracked with its own linear response, thereby circumventing the issue of momentum non-conservation arising from partial conversion partway through the simulation as noted in~\cite{Brandbyge:2009ce} and discussed above in section~\ref{sec:Brandbyge}. Two parameters, $z_{\nu}$ and $v_{\text{crit}}$,
mark the conversion timing and the population partition between the slow and fast groups respectively.  Like the $q_{\rm cut}/T$ and $f_{\rm flow}$ parameters of~\cite{Brandbyge:2009ce}, these parameters are determined via user-defined tolerance criteria: in this case, the authors of~\cite{Bird:2018all} estimated the dimensionless power spectrum of the neutrino density fluctuations in each velocity bin and demanded that it remain below unity as a criterion for linear response validity.

The final change concerns how the conversion from linear response to $N$-body particles is executed. Rather than spawning neutrino particles at $z_{\nu}$, the implementation of reference~\cite{Bird:2018all} places neutrino particles in the simulation volume as tracers {\it already from the very beginning of the simulation} ($z_{\rm sim}=99$ in this case). These tracers are initialised as a homogeneous distribution with  random thermal velocities but no bulk velocities, and are evolved alongside the cold matter $N$-body particles. They do not however contribute to the gravitational potential at $z>z_{\nu}$, nor are their trajectories accounted for in the determination of the long-range, Particle-Mesh (PM) force time-steps in \gadgettwo{}. Their purpose is to evolve to an attractor solution by the conversion redshift~$z_{\nu}$, so that when the slow-group linear response is switched off and the tracers become visible to the gravitational potential at $z_\nu$, the now-visible neutrino particles can evolve transient-free.  Thus defined and executed, reference~\cite{Bird:2018all} finds that conversion from a linear-response-based  to a particle-based computation of the gravitational potential can take place at a redshift as low as $z_\nu=1$.

%%%%%%%%%%%%%
%%%%%%%%%%%%%%%

\section{Multi-fluid hybrid-neutrino scheme: Preliminaries}
\label{sec:mflr}

Our hybrid-neutrino implementation improves upon previous schemes~\cite{Brandbyge:2009ce,Bird:2018all} in several aspects.  Firstly, we adopt the multi-fluid neutrino perturbation theory of~\cite{Dupuy:2013jaa, Dupuy:2014vea, Dupuy:2015ega}, which partitions the neutrino population into multiple flows characterised by their initial comoving momenta. Inherently a Lagrangian formulation in momentum space, this choice of perturbation theory easily circumvents complications associated with partial conversion from the Eulerian to the Lagrangian frame~\cite{Brandbyge:2009ce}.  We have previously adapted the sub-horizon, non-relativistic limit of this theory into a linear response theory for massive neutrinos in cosmological $N$-body simulations of cold matter, dubbed ``multi-fluid linear response'' (MFLR)~\cite{Chen:2020bdf}. The present hybrid-neutrino scheme is a natural extension of our previous work.

Secondly, once we have decided which of the MFLR neutrino flows to convert into $N$-body particles at some suitable conversion redshift $z_c$, the MFLR solution for the flow at $z_c$ itself can be used to generate initial conditions for the $N$-body neutrino particles at conversion.  Such initialisation eliminates the need to evolve tracers together with the cold matter particles for the entire duration of the simulation (such as in~\cite{Bird:2018all}), and particle spawning can take place at a much lower redshift than $z=99$.

%%%%%%%%%%%%%%%%%%%%%%
\begin{figure}[t]
	\centering
	\includegraphics[width=100mm]
	{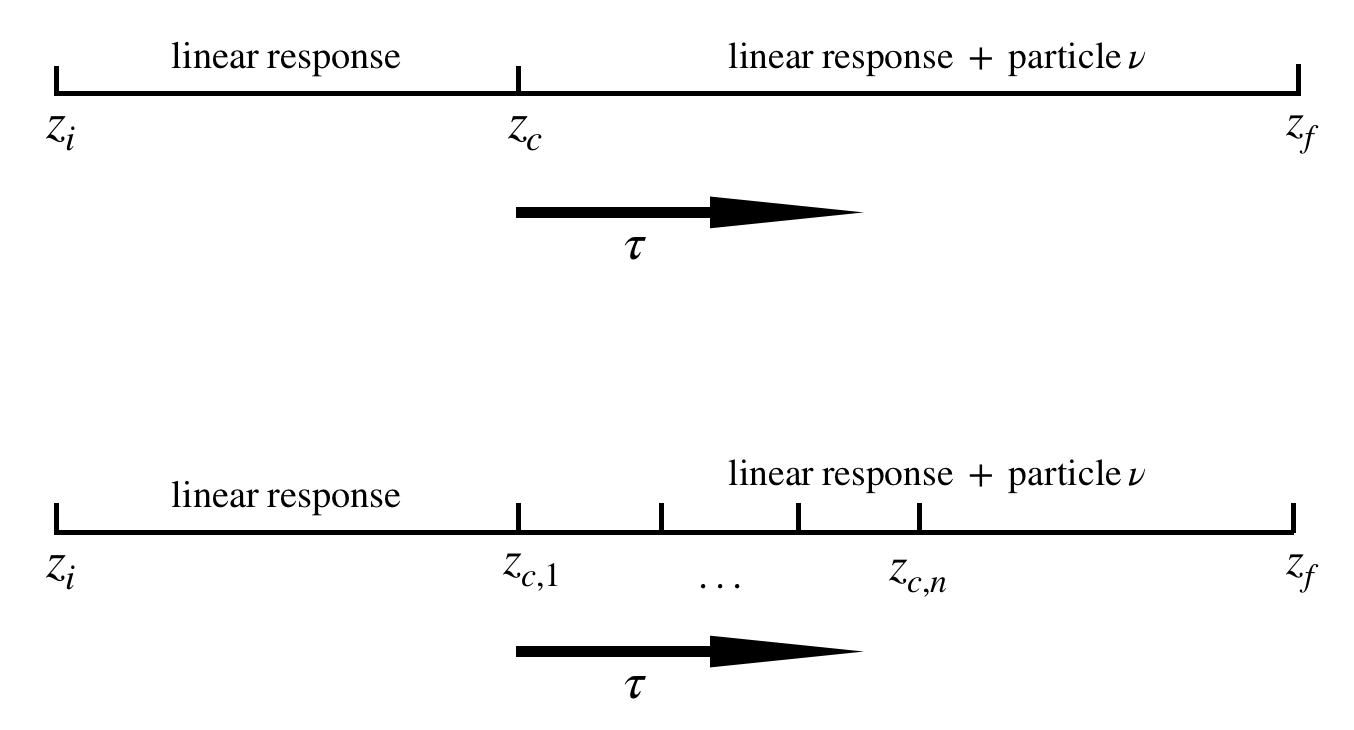}
	\caption{Schematic timeline of a hybrid-neutrino simulation.
The upper timeline shows the workflow in the case of a single instance of perturbation theory-to-particle conversion, where a subset of neutrinos is converted from a perturbative description to $N$-body particles at a redshift~$z_{c}$. The lower timeline shows a scheme 
with staggered conversion at multiple redshifts 
that is in principle possible using a multi-fluid decomposition of the neutrino population, such as explored in this work.~\label{fig:HybridWorkflowChart}}
\end{figure}
%%%%%%%%%%%%%%%%%%%%%

Lastly, the fine-grained nature of MFLR retains by default momentum information of the neutrino population, and offers the possibility of staggered MFLR-to-particle conversion at multiple redshifts---such as depicted in figure~\ref{fig:HybridWorkflowChart}---as the evolution of subsets of the neutrino population transitions progressively to the nonlinear regime. This possibility has not been explored previously and will be investigated in this work for the first time.

We describe first in this section the multi-fluid framework and the linear response adaptation of~\cite{Chen:2020bdf}, before extending the framework section~\ref{sec:conversion} to include conversion to neutrino particles in a hybrid-neutrino scheme.

%%%%%%%%%%%
%%%%%%%%%%%

\subsection{Multi-fluid perturbation theory in the sub-horizon, non-relativistic limit}
\label{sec:theory}

The multi-fluid perturbation theory was originally introduced in~\cite{Dupuy:2013jaa, Dupuy:2014vea, Dupuy:2015ega} as a fully relativistic theory. The sub-horizon, non-relativistic limit of the theory was discussed in~\cite{Chen:2020bdf} in the context of massive neutrino linear response in cold matter $N$-body simulations. Broadly speaking, the multi-fluid framework divides the total neutrino fluid into $N_{\tau}$ number of independent flows, where the flow $\alpha$  is characterised by its initial zeroth-order comoving momentum $\tau_\alpha$, also called the ``Lagrangian momentum'', and angular dependence encoded in $\mu \equiv \hat{k} \cdot \hat{\tau}_{\alpha}$.
These flows  evolve independently from one another, and develop perturbations in their energy density and velocity according to 
the fluid equations
\begin{equation}
\begin{aligned}
\frac{{\rm d} \delta_\alpha}{{\rm d} s} =& -{\rm i} \frac{k \mu \tau_\alpha}{m_\nu} \delta_\alpha - \frac{1}{m_\nu} \theta^P_\alpha, \\
\frac{{\rm d} \theta^P_\alpha}{{\rm d} s} =& -{\rm i} \frac{k \mu \tau_\alpha}{m_\nu} \theta^P_\alpha + a^2 m_\nu\,  k^2 \Phi
\label{eq:mflrFluidEqn}
\end{aligned}
\end{equation}
in the sub-horizon, non-relativistic limit.  
Here, $\delta_{\alpha}=\delta_\alpha(k, \mu, s)$ and $\theta_{\alpha}^P=\theta_\alpha(k, \mu, s)$ are, respectively, the Fourier $k$-space density contrast and momentum divergence, $s$ is the superconformal time related to the cosmic time via ${\rm d} t = a^2 {\rm d}s$, ${\cal H}(s)=a(s) H(s)$ is the conformal Hubble  rate, and we sometimes use the term ``Lagrangian velocity'' to denote $\vec{v}_{\alpha} \equiv \vec{\tau}_{\alpha} /(a m_{\nu})$ evaluated at $a=1$.%
\footnote{Reference~\cite{Chen:2020bdf} gives the fluid equations in terms of the time variable $s \equiv \ln (a / a_{\text{in}})$ for a reference $a_{\text{in}}$ and the  dimensionless velocity divergence $\theta_\alpha^{\rm dl}$ defined relative to our momentum divergence $\theta^P_\alpha$ via $\theta_\alpha^{\rm dl} \equiv -\theta_\alpha^P/(a m_\nu {\cal H})$.   The equations of motion are otherwise completely equivalent in the sub-horizon, non-relativistic limit.}

The decomposition in $\tau_\alpha$ is user-defined, as long as the totality of neutrino fluids respects the relativistic Fermi-Dirac distribution in the limit $N_{\tau} \to \infty$. The decomposition in $\mu$, however, is most conveniently handled by expanding $\delta_\alpha(k, \mu, s)$ and $\theta_\alpha(k, \mu, s)$ in terms of Legendre polynomials~${\cal P}_\ell(\mu)$, i.e., 
\begin{equation}
\begin{aligned}
\label{eq:legendre}
X(\mu) & =\, \sum_{\ell =0}^\infty (-{\rm i})^\ell {\cal P}_\ell(\mu) X_{\ell},\\
X_\ell &=\,  \frac{{\rm i}^\ell}{2} (2 \ell+1) \int_{-1}^1 {\rm d} \mu \, {\cal P}_\ell(\mu) X(\mu).
\end{aligned}
\end{equation}
The equations of motion for the $\ell$th Legendre moments then read~\cite{Chen:2020bdf}
\begin{equation}
	\begin{aligned}
		\frac{{\rm d} \delta_{\alpha, \ell}}{{\rm d} s} = & \,  \frac{k \tau_{\alpha}}{m_\nu} \left( \frac{\ell}{2\ell - 1} \delta_{\alpha, \ell - 1} - \frac{\ell + 1}{2\ell + 3} \delta_{\alpha, \ell + 1} \right)- \frac{1}{m_\nu}\theta_{\alpha, \ell}^P  \, , \\
		\frac{{\rm d}\theta^P_{\alpha, \ell}}{{\rm d} s} = & \,  \frac{k \tau_{\alpha}}{m_\nu} \left( \frac{\ell}{2\ell - 1} \theta^P_{\alpha, \ell - 1} - \frac{\ell + 1}{ 2 \ell + 3} \theta^P_{\alpha, \ell +1} \right) + \delta_{\ell 0}^{(\mathrm{K})}  a^2 m_\nu \, k^2  \Phi \, , 
	\end{aligned}
	\label{eq:mflrFluidHierarchy}
\end{equation}
where $\delta_{\ell 0}^{(\mathrm{K})}$ is the Kronecker delta function. Details of the truncation scheme for the hierarchy in practical implementations can be found in reference~\cite{Chen:2020bdf}.

Observe in equation~\eqref{eq:mflrFluidHierarchy} that the $\ell$th multipole moment couples only to the two moments immediately adjacent to it, i.e., $\ell-1$ and $\ell+1$, within {\it the same} fluid flow~$\alpha$.
Beyond this coupling, {\it different} neutrino flows are visible to each other only via the coupling of all fluid monopole ($\ell=0$) moments in the gravitational potential~$\Phi$, i.e.,
\begin{equation}
\label{eq:pot}
	k^2 \Phi(k, s) = - \frac{3}{2} \mathcal{H}^2(s) \left( \Omega_{\text{cb}}(s) \delta_{\text{cb}} (k, s) + \sum_{\alpha = 1}^{N_{\tau} } \Omega_{\alpha} (s) \delta_{\alpha, \ell = 0} (k, s) \right) \,,
\end{equation}
where $\Omega_{\rm cb}(s)$ and $\delta_{\rm cb}$ denote the time-dependent reduced cold matter density and density perturbation respectively, and $\Omega_{\alpha}(s)$ is the reduced energy density in the neutrino flow~$\alpha$ such that $\sum_{\alpha=1}^{N_\tau} \Omega_\alpha=\Omega_\nu$.%
\footnote{We label the neutrino flows $\alpha=1, \ldots, N_\tau$ in this work, in contrast with reference~\cite{Chen:2020bdf}, which uses the convention $\alpha=0, \ldots, N_\tau-1$.}
 Generalisation of the theory to multiple non-degenerate neutrino masses is straightforward: we need simply to replicate the equations of motion~\eqref{eq:mflrFluidHierarchy} for each unique neutrino mass value and ensure that the corresponding monopole density contrasts $\delta_{\alpha,\ell=0}$ contributes to the gravitational potential~\eqref{eq:pot} weighted by the appropriate $\Omega_\alpha$.

Thus, the advantage of this ``semi-Lagrangian'' multi-fluid perturbation theory as the framework upon which to build our hybrid-neutrino scheme is immediately clear: should a subset of neutrino flows be removed from the perturbation theory and converted to $N$-body particles, no modifications to the equations of motion are required for the remaining perturbative flows to continue their evolution correctly. Importantly, the perturbation theory naturally retains fine-grained momentum information,
allowing for high-resolution sampling in the low-momentum region and hence more precise control over the conversion to particles. 

%%%%%%%%%%%%%%%%
%%%%%%%%%%%%%%%

\subsection{Multi-fluid linear response (MFLR) \texorpdfstring{$N$}{N}-body simulations}
\label{sec:LRSim}

Having described the multi-fluid perturbation theory, let us now turn to its adaptation as a linear response theory for massive neutrinos in $N$-body simulations~\cite{Chen:2020bdf}, which will serve as a foundation for inserting hybrid-neutrinos. The power spectra obtained from this MFLR approach for the cold matter and neutrino flows will also serve as a baseline against which to test their counterparts from the hybrid-neutrino simulation. Details of the implementation of massive neutrino MFLR into the $N$-body code \gadgetcode{}~\cite{Springel:2020plp} can be found in reference~\cite{Chen:2020bdf}.

%%%%%%%%%%%%%%%%%%%%%%
\begin{table}[t]
	\begin{center}
		\footnotesize 
		\begin{tabular}{lcccc}
			\hline
			\hline
			Cosmological parameter & Symbol & nu00 & nu05 & nu03\\ 
			\hline
			Total matter density &  $\Omega_{m, 0} h^2$ &  0.14175 &  0.14175 & 0.14175\\ 
			Baryon energy density & $\Omega_{b, 0} h^2$ &   0.02242 &   0.02242 & 0.02242 \\
			Neutrino energy density & $\Omega_{\nu, 0} h^2$ &   0.0 & 0.005 & 0.003 \\
			Reduced Hubble parameter &  $h$ & 0.6766 & 0.6766 & 0.6766 \\
			Scalar spectral index & $n_s$ & 0.9665 & 0.9665 & 0.9665\\
			RMS linear matter density fluctuation on $8\,{\rm Mpc}/h$ & $\sigma_8$ & 0.8278 & 0.7139 & 0.7594 \\
			Optical depth to reionisation &  $\tau_{\rm re}$  & 0.094551 & 0.094551 & 0.094551 \\
			\hline
			\hline
			Multi-fluid parameter & Symbol & nu00 & nu05  & nu03\\
			\hline
			Total number of neutrino fluid flows & $N_{\tau}$ & 0 & 20 & 20\\
			Number of multipoles per fluid flow & $N_{\mu}$ & 0 & 20 & 20\\
			\hline
			\hline
		\end{tabular}
	\end{center}
	\caption{Cosmological and multi-fluid parameters for the cosmologies nu00, nu05, and nu03. The corresponding neutrino mass sums are $\sum m_\nu=0$, $0.465$, and $0.279$~eV, respectively, assuming three equal-mass neutrinos. The $\sigma_8$ values are {\it derived} from the quoted cosmological parameters, for a primordial scalar amplitude fixed at $10^{9} A_s = 2.135$ at the pivot scale $k_{\rm pivot}=0.05$/Mpc.\label{table:SimCosmoParam}}
\end{table}
%%%%%%%%%%%%%%%%%%
%%%%%%%%%%%%%%
\begin{table}[t]
	\begin{center}
		\footnotesize 
		\begin{tabular}{l|ccccc}
			\hline
			\hline
Flows & \multicolumn{5}{c}{Lagrangian momenta [meV]}\\
\hline
$\alpha=1, \ldots,5$ & 0.120364 & 0.184696 & 0.229146 & 0.266695 & 0.300878 \\
$\alpha=6, \ldots,10$ &0.333306 & 0.364920 & 0.396378 & 0.428214 & 0.460921 \\
$\alpha=11, \ldots,15$ &0.495012 & 0.531078 & 0.569851 & 0.612317 & 0.659898 \\
$\alpha=16, \ldots,20$ &0.714811 & 0.780892 & 0.865788 & 0.98895 & 1.238590 \\
			\hline \hline
		\end{tabular}
	\end{center}
	\caption{Lagrangian momenta $\tau_\alpha$ of the $20$ flows used in our nu05 and nu03 simulations.\label{tab:tauflow}}
\end{table}
%%%%%%%%%%%%%%

Consider two spatially flat, $\Lambda$CDM-type cosmologies, one containing massless neutrinos labelled ``nu00'', and one with the same total matter content, but has part of the cold matter replaced with three equal-mass neutrinos summing to $\sum m_\nu = 0.465$~eV, which we label ``nu05''.  In both cases, we assume a primordial curvature power spectrum described by a simple power law with a spectral index $n_s$ and a fluctuation amplitude $A_s$.  Save for the partial substitution of cold matter with massive neutrinos, all other parameter values are held fixed between nu00 and nu05. See table~\ref{table:SimCosmoParam} for details.

We run MFLR $N$-body simulations  to compute the nonlinear power spectra for the nu00 and nu05 cosmologies.  The $N$-body component, representing the cold matter, uses $N_{\text{cb}} = 512^3$ particles in a simulation box of side length $L = 256 \, \text{Mpc}/h$ and $N_{\text{PM}} = 1024^3$ PM grid cells, together with a tree force softening length $r_{\rm soft} = 50 \, \text{kpc} / h$.  Where required, the massive neutrino population is decomposed into $N_{\tau} = 20$ equal-number flows with Lagrangian momenta displayed in table~\ref{tab:tauflow}, and each flow is subdivided into $N_{\mu} = 20$ multipole moments.  In all cases, we use the $z=0$ linear cold matter power spectrum from~\classcode{}~\cite{Blas:2011rf,Lesgourgues:2011rh}, scaled back appropriately to the simulation starting redshift $z_{\rm sim}=99$, to initialise the cold particles via the Zel'dovich approximation  on a grid of the same resolution as the PM grid.  Initialisation of the neutrino flows takes place earlier at $z_{\rm in}=999$, and is achieved by putting the $\ell=0$ moments of the density and velocity divergence on an almost-attractor solution given the scaled-back cold matter density fluctuations at that time; higher multipole moments---set to zero at $z_{\rm in}$---have between $z_{\rm in}$ and $z_{\rm sim}$ to be regenerated from the monopoles.
We refer the reader to references~\cite{Chen:2020bdf,Chen:2020kxi} for details on the initialisation procedure.

%%%%%%%%%%%%%%%%%%%%%%%%%%%%%%%%%
\begin{figure}[t]
	\centering
	\includegraphics[trim=0mm 2mm 10mm 5mm,clip,width=120mm]
	{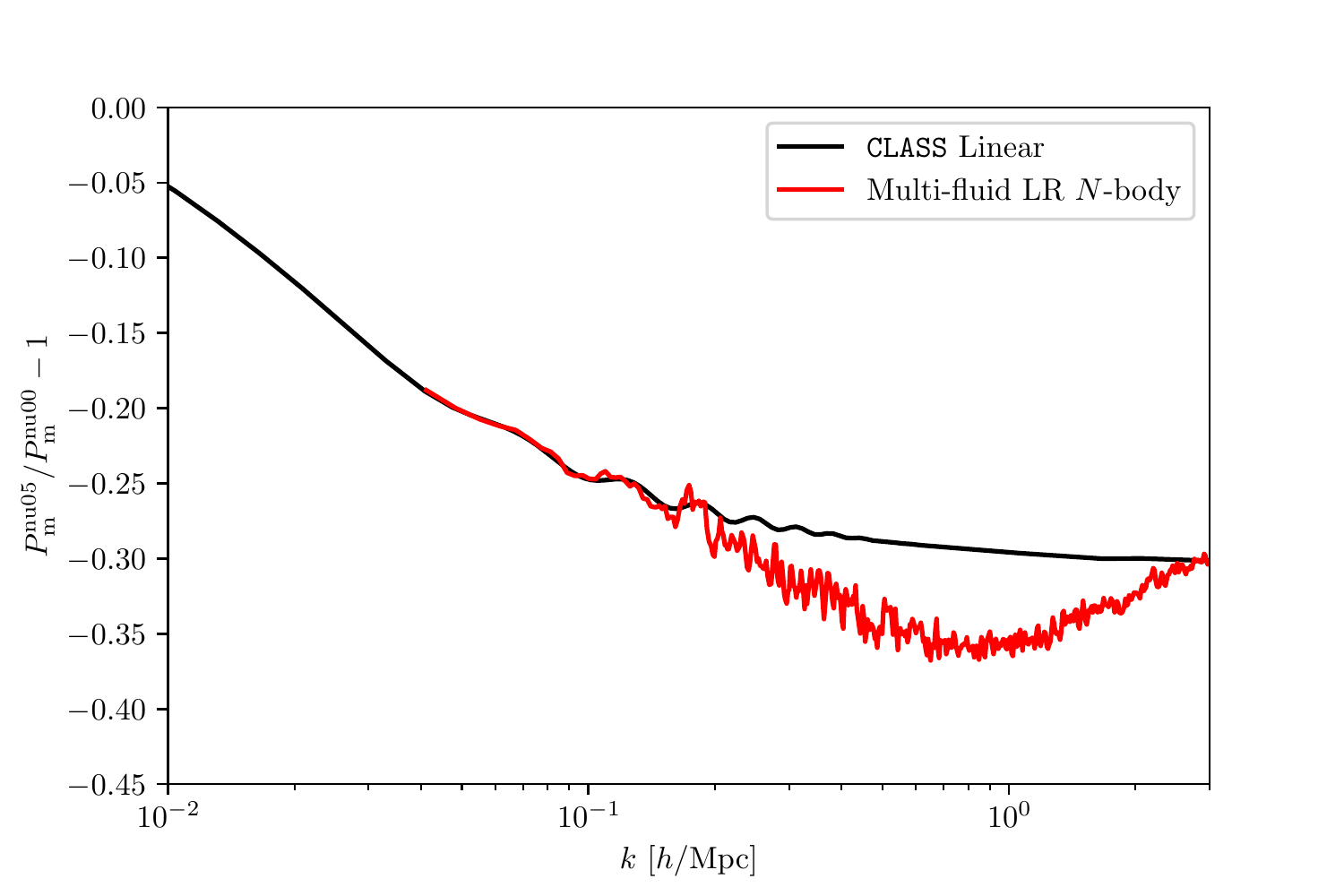}
	\caption{The $z=0$ total matter power spectrum of the massive neutrino cosmology nu05, relative to the massless case nu00. The red and black lines represent, respectively, the outcome of our MFLR $N$-body simulations and of linear perturbation theory computed with \classcode{}. 
	The two $N$-body relative power spectra exhibit a ``spoon'' shape on small scales not seen in linear theory.\label{fig:nu05LRSpoons}}
\end{figure}
%%%%%%%%%%%%%%%%%%%%%%%%%%%

Figure~\ref{fig:nu05LRSpoons} shows the resulting $z=0$ total matter  power spectrum of the nu05 massive neutrino cosmology $P_m^{\rm nu05}(k)$, estimated from the MFLR simulation via
\begin{equation}
P_m^{\rm MFLR} (k) = \frac{1}{\Omega_m}  \left|\Omega_{\rm cb}  \sqrt{\left\langle|\delta_{\rm cb}(\vec{k})|^2 \right\rangle}  + \frac{\Omega_\nu}{N_\tau}  \sum_{\alpha=1}^{N_\tau} \delta_{\alpha, \ell=0}(k)\right|^2   
\end{equation}
assuming the same phases for $\delta_{\alpha,\ell=0}(k)$ and $\delta_{\rm cb}(\vec{k})$, and expressed relative to its nu00 massless neutrino counterpart $P_m^{\rm nu00}(k)$ as per common practice.  For comparison we plot also the same power spectrum ratio computed from purely linear perturbation theory using \classcode{}. Clearly, the relative power spectra $P_m^{\rm nu05}(k)/P_m^{\rm nu00}(k)-1$ demonstrate good agreement between perturbation theory and simulations on large scales (i.e., at wave numbers $k \lesssim 0.1 \, h/\text{Mpc}$). On small scales, however, the non-linear, $N$-body relative power spectrum exhibits a ``spoon'' feature at $k \sim 0.8 \, h/\text{Mpc}$ not seen in its purely linear counterpart. 
This spoon shape has been consistently observed in other massive neutrino $N$-body studies~\cite{Brandbyge:2008js, Agarwal:2010mt, Adamek:2017uiq, Dakin:2017idt, Liu:2017now, Bayer:2020tko}, and can be explained in terms of the halo model of structure formation~\cite{Hannestad:2020rzl}.

%%%%%%%%%%%%%%%%%%%%%%%%%%%%
\begin{figure}[t]
	\centering
	\includegraphics[trim=0mm 2mm 15mm 5mm,clip,width=120mm]{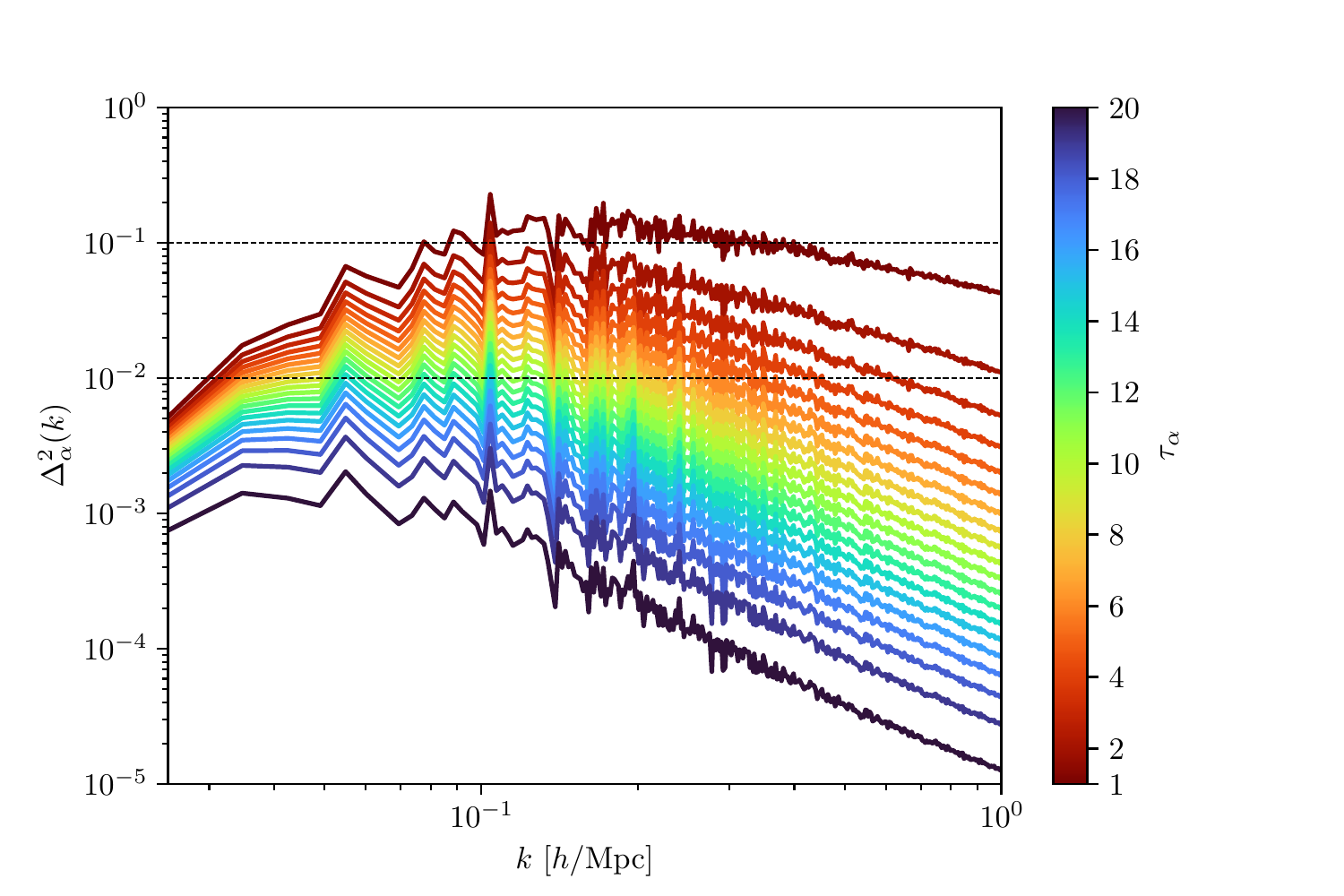}
	\caption{The $z=0$ dimensionless power spectra of the 20 neutrino flows from our MFLR $N$-body simulation of the nu05 cosmology. The smallest Lagrangian momentum is represented by the $\alpha=1$ flow in red, while the largest Lagrangian momentum is denoted $\alpha=20$ in violet.  Nonlinear evolution is likely important for flows with dimensionless power exceeding $0.1$, marked by the top horizontal dashed line.
	Even those flows crossing into the $0.01 \lesssim \Delta_\alpha^2 \lesssim 0.1$ region would likely display discernible nonlinear effects.  Then, by these criteria, some 50\% of the neutrino flows in this particular cosmology  warrant further investigation into potential nonlinear enhancements in the monopole power.\label{fig:nu05LRDnuRainbow}}
\end{figure}
%%%%%%%%%%%%%%%%%%%%%%%%%

Focussing on the nu05 cosmology, figure~\ref{fig:nu05LRDnuRainbow} shows the dimensionless power spectra $\Delta_\alpha^2(k) \equiv k^3 \langle |\delta_\alpha(\vec{k})|^2\rangle/(2 \pi^2)$ for all $N_\tau=20$ flows at $z=0$, which in the MFLR approach can be 
constructed from the monopole moments of the individual neutrino fluid flows as per
\begin{equation}
\label{eq:dimensionlessp}
\Delta_\alpha^{2\,{\rm (MFLR)}}(k) = \frac{k^3}{2 \pi^2} \left|\delta_{\alpha,\ell=0}(k)\right|^2. 
\end{equation}
 Unsurprisingly, flows of the smallest Lagrangian momentum $\tau_\alpha$ end up with the largest dimensionless power; at any given wave number $k$, we see that $\Delta_\alpha^2$ decreases as we increase~$\tau_\alpha$.  However, because MFLR uses linearised equations of motion to track neutrino inhomogeneities, the actual dimensionless powers displayed in figure~\ref{fig:nu05LRDnuRainbow} are likely only valid for small density fluctuations.  Reference~\cite{Chen:2020bdf} sets the MFLR validity criterion at $\Delta^2_{\alpha} \lesssim 0.1$, which we shall test explicitly against the hybrid-neutrino scheme in the coming sections.

Also of interest are those mildly nonlinear flows whose dimensionless powers fall between between 0.01 and 0.1. Here, genuine nonlinear effects are unlikely to dominate the evolution of these flows. Nonetheless, we might still expect some degree of nonlinearity to be discernible in the monopole power.  Again, the hybrid-neutrino scheme can be used to test the degree of nonlinearity in the evolution of these intermediate flows.  In fact, even in the case of nu05 characterised by a neutrino mass sum $\sum m_\nu = 0.465$~eV, we see in figure~\ref{fig:nu05LRDnuRainbow} that approximately 50\% of the neutrino fluid flows satisfy the criterion $\Delta^2_\alpha(k) \gtrsim 0.01$ at some wave number $k$.  One of our aims, therefore, is to formulate a more precise criterion to determine which momentum-subset of the neutrino population would require nonlinear modelling.

%%%%%%%%%%%%%%%%%
%%%%%%%%%%%%%%%%%%

\section{Conversion from MFLR to particle representation}
\label{sec:conversion}

Having introduced the MFLR $N$-body simulations, we are now in a position to convert some of the neutrino fluid flows into simulation particles. This section is devoted to describing how this conversion is executed.  Aside from the question of when this conversion should occur and which flow(s) should be converted, there are several practical considerations concerning how the information contained in the MFLR solution can be replicated as faithfully as possible in the particle representation, as well as limitations of the MFLR solution as currently implemented.  Indeed, the issue of conversion fidelity pertains also to older hybrid-neutrino schemes and to particle neutrino simulations in general.

In the following, we shall discuss first the issues of conversion fidelity and the limitations of MFLR, before detailing the actual MFLR-to-particle conversion procedure implemented in our hybrid-neutrino scheme.
The question of when and which flow(s) to convert will be addressed in section~\ref{sec:criteria}.

%%%%%%%%%%%%%%%%%%%
%%%%%%%%%%%%%%%%%%

\subsection{Preliminary considerations}
\label{sec:prelim}

At the interface between the MFLR solution and an $N$-body particle representation of the same, three issues come to the fore: perturbation phases, information contained in the MFLR multipole moments and extraction thereof, and mapping Eulerian perturbations to Lagrangian displacements.

\subsubsection{Perturbation phases}

While the MFLR solution can in principle contain all information about the neutrino population, in practical implementations some of this must invariably be sacrificed in favour of savings on computational resources.  In the particular implementation of reference~\cite{Chen:2020bdf}, we have averaged the gravitational potential $\Phi(\vec{k})$ at each wave number $k \equiv |\vec{k}|$ over all directions~$\hat{k}$, so that only one set of equations of motion per neutrino flow $\alpha$ per $k$ needs to be retained. An immediate consequence of this simplification is that the MFLR solution currently carries no phase information of the neutrino inhomogeneities at all. 

In a purely linear theory, the lack of phase information is clearly not a concern, as all neutrino perturbations $\delta_{\alpha,\ell}(\vec{k})$ and $\theta_{\alpha,\ell}(\vec{k})$ necessarily carry the same phase as the gravitational potential $\Phi(\vec{k})$, which itself does not change with time. Thus, if the conversion to neutrino particles was to take place at the same time as we initialise cold particles, i.e., at $z_{\rm sim} = 99$,
assigning the neutrino particles the same phases as for the cold particles likely suffices.  Late-time conversions, however, are necessarily affected by nonlinear evolution of the phases in the cold matter {\it along a neutrino's entire path}.  This is most easily seen in the formal solution to the fluid hierarchy~\eqref{eq:mflrFluidHierarchy},
\begin{equation}
\begin{aligned}
\label{eq:thetaPlaX}
 \delta_{\alpha,\ell} (\vec{k},s)  & = \, - (2 \ell +1) \, k^2  \int_{s_{\rm i}}^s {\rm d} s' \, a^2 (s') \, (s-s')\, \Phi(\vec{k},s') \,  j_\ell  \left[k \tau_\alpha (s-s')/m_\nu\right],\\
 \theta^P_{\alpha,\ell} (\vec{k},s)  & = \,  (2 \ell +1) \, m_\nu \, k^2  \int_{s_{\rm i}}^s {\rm d} s' \, a^2 (s') \,  \Phi(\vec{k},s') \,  j_\ell  \left[k \tau_\alpha (s-s')/m_\nu\right],
 \end{aligned}
\end{equation}
where we have reinstated the $\vec{k}$ dependence of $\Phi(\vec{k})$ and 
assumed $\delta_{\alpha,\ell}(\vec{k},s_{\rm i})=\theta_{\alpha,\ell}(\vec{k},s_{\rm i})=0$ at the initial time $s_{\rm i}$.  See appendix~\ref{sec:phases} for  the derivation of equation~\eqref{eq:thetaPlaX}.
The accumulation of phase changes along a neutrino's path means that not only will the phases of $\delta_\ell(\vec{k})$ and~$\theta^P_\ell(\vec{k})$ diverge in principle from the phase of $\Phi(\vec{k})$ and from one another at the same time, different multipole moments will also end up carrying different phases. 

Having said the above, however, further analysis of the solutions~\eqref{eq:thetaPlaX} in appendix~\ref{sec:phases} suggests that we can reasonably expect the low, ${\cal O}(1)$ multipoles of $\delta_\ell(\vec{k},s)$ and $\theta_\ell(\vec{k},s)$ to track the phase of the gravitational potential $\Phi(\vec{k},s)$ at the same time~$s$, in both the ``clustering limit'',
\begin{equation}
\label{eq:clustering}
\frac{k \tau_\alpha(s-s')}{m_\nu} \to  0, 
\end{equation}
and the ``free-streaming'' limit, 
\begin{equation}
\label{eq:freestreaming}
\frac{1}{a^2 \Phi}\frac{{\rm d} a^2 \Phi}{{\rm d} s} \ll  \frac{k \tau_\alpha}{m_\nu}, 
\end{equation}
of the solutions.  This leaves only the intermediate region between these two limits around the free-streaming scale~\cite{Ringwald:2004np},
\begin{equation}
\label{eq:kfs}
    k_{{\rm FS},\alpha} (s)\equiv \frac{a(s)\, m_\nu {\cal H}(s)}{\tau_\alpha} \sqrt{\frac{3}{2} \Omega_{m} (s)} \simeq 0.00023\;  \sqrt{a(s)}\, 
    \left(\frac{m_\nu}{\tau_\alpha}\right)\,
    \left(\frac{\Omega_{m,0}}{0.31}\right)^{1/2}\, 
    h/{\rm Mpc},
\end{equation}
where the question of neutrino perturbation phases is potentially grey, {\it if} $k_{{\rm FS},\alpha}$ happens to fall in a region where the gravitational clustering dynamics of the cold matter has become mildly nonlinear.
Absent further info we have only two choices: we could adopt the original phases at initialisation (such as in reference~\cite{Brandbyge:2009ce}), or the phases of $\Phi(\vec{k},s)$ at conversion time matching the expectations of the clustering and free-streaming limits. The latter seems to be the more sensible choice to us, and can be subject to convergence testing.

%%%%%%%%%%%%%%%
%%%%%%%%%%%%%%

\subsubsection{Information in MFLR multipole moments}  

Multi-fluid linear response is a perturbation theory in Fourier space.  In real space, the density perturbation $\delta_\alpha(\vec{x},\hat{\tau}_\alpha, s)$ and physical velocity $\vec{u}_\alpha(\vec{x},\hat{\tau}_\alpha,s)$ of a neutrino fluid flow characterised by the Lagrangian momentum $\vec{\tau}_\alpha = \tau_\alpha \hat{\tau}_\alpha$ are related to the MFLR quantities $\delta_{\alpha,\ell}(\vec{k},s)$ and $\theta^P_{\alpha,\ell}(\vec{k},s)$ via
\begin{equation}
\begin{aligned}
\label{eq:kick0X} 
\delta_\alpha(\vec{x},\hat{\tau}_\alpha, s) 	& = \, \sum_{\ell =0}^\infty (-{\rm i})^\ell  {\cal F}^{-1}  \left[
{\cal P}_\ell( \mu) \, \delta_{\alpha,\ell}(\vec{k},s) \right],\\
\vec{u}_\alpha(\vec{x},\hat{\tau}_\alpha,s) &	=\, \frac{\tau_\alpha}{a(s)m_\nu} \hat{\tau}_\alpha -\frac{1}{a(s) m_\nu}  \sum_{\ell =0}^\infty {\rm i}^{\ell+1} (-1)^\ell  {\cal F}^{-1}  \left[ \frac{\vec{k}}{k^2} 
 {\cal P}_\ell(\mu) \, \theta^P_{\alpha,\ell}(\vec{k},s) \right].
 \end{aligned}
\end{equation}
Here, ${\cal F}^{-1}[\cdots]$ denotes an inverse Fourier transform, and
$\delta_{\alpha,\ell}(\vec{k},s)$ and $\theta^P_{\alpha,\ell}(\vec{k},s)$ are assumed to already carry the correct phase information.
See appendix~\ref{sec:kick} for the derivation.

Observe in equation~\eqref{eq:kick0X} that the Legendre polynomials ${\cal P}_\ell(\mu\equiv \hat{k} \cdot \hat{\tau})$ appear inside the inverse Fourier transform operation. Physically, this means that once we have decided to convert to particles a neutrino flow characterised by $\tau_\alpha$, each possible direction~$\hat{\tau}_\alpha$ of the flow must experience a different gravitationally-induced velocity kick as well as density perturbation.  In reality this should of course be the case: the zeroth-order homogeneous and isotropic neutrino background is described by Lagrangian momenta direction $\hat{\tau}$ uniformly distributed over $4 \pi$ in solid angle, and the free-streaming of these neutrinos in an inhomogeneous background is precisely the origin of neutrino anisotropic stress.

As a recipe to initialise neutrino particles in a simulation, however, limited computational resource means that a literal implementation of equation~\eqref{eq:kick0X} is practically impossible, as the number of inverse Fourier transforms required to realise the MFLR-to-particle conversion in this manner is prohibitively expensive.
Even if we only place one neutrino particle with a randomly-drawn~$\hat{\tau}_\alpha$ per real-space coordinate~$\vec{x}$, the number of operations required to realised equation~\eqref{eq:kick0X} is of order the number of neutrino particles placed in the simulation box; for a typical choice of $N_\alpha=512^3$ particles, this is impossible.

One possible way to reduce the number of operations is to limit the number of available directions~$\hat{\tau}_\alpha$ in the conversion.  Alternatively, we could abandon all $\ell  > 0$ multipole moments, keeping only the monopole:  since ${\cal P}_{\ell=0}(\mu) = 1$ is independent of $\hat{\tau}_\alpha$, equation~\eqref{eq:kick0X} simplifies significantly to
\begin{equation}
\begin{aligned}
\label{eq:kick1X}
\delta_\alpha(\vec{x},\hat{\tau}_\alpha,s) 	&= \, {\cal F}^{-1}  \left[  
 \delta_{\alpha,\ell=0}(\vec{k},s) \right],\\ \vec{u}_\alpha(\vec{x},\hat{\tau}_\alpha,s) & = \, \frac{\tau_\alpha}{a(s)m_\nu} \hat{\tau}_\alpha -\frac{ {\rm i} }{a(s) m_\nu}    {\cal F}^{-1}  \left[ \frac{\vec{k}}{k^2} 
 \theta^P_{\alpha,\ell=0}(\vec{k},s) \right],
 \end{aligned}
\end{equation}
and one single inverse Fourier transform suffices for any number of choices of $\hat{\tau}_\alpha$ in the particle realisation procedure.

Clearly, throwing away the $\ell>0$ multipole moments as per equation~\eqref{eq:kick1X}
is but a resort to keeping the computational cost 
associated with MFLR-to-particle conversion
manageable. In general, missing higher multipole moments will impact on conversion fidelity even if the conversion happens deep in the linear regime, and is thus a more generic problem for particle neutrino simulations not only limited the late-time conversion/hybrid-neutrino approach advocated in this work.

Nonetheless, initialising only with the $\ell=0$ moments can be physically well motivated in some limiting situations.  On the one hand, in the low-$k$ clustering limit~\eqref{eq:clustering}, population of the $(\ell+1)$th multipole from the $\ell$th multipole occurring over a timescale of
\begin{equation}
\label{eq:repop}
 \Delta s_{\rm repop} \sim \left(\frac{k \tau_\alpha}{m_\nu}   \right)^{-1} 
\end{equation}
 effectively means that occupation of the $\ell>0$ multipoles is highly suppressed in the first place.  On the other hand, in the opposite, high $k$ free-streaming limit~\eqref{eq:freestreaming}, the same repopulation rate~\eqref{eq:repop} ensures that the $\ell>0$ multipole moments will be rapidly regenerated during the simulation run-time.  
 This again leaves the intermediate region $k \sim k_{{\rm FS},\alpha}$, defined in equation~\eqref{eq:kfs}, where improper initialisation is most likely an issue.
Indeed, as we shall see in section~\ref{sec:ConvRedshiftCriteria}, transients at $k \sim k_{{\rm FS},\alpha}$ associated with the missing $\ell>0$ multipole moments are currently what limits how late one can push back the MFLR-to-particle conversion.

%%%%%%%%%%%%%%%%%
%%%%%%%%%%%%%%%%%

\subsubsection{Eulerian perturbations to Lagrangian displacements} 

Initialisation of cold matter particles in a simulation typically uses some form of Lagrangian perturbation theory, which allows us to map the linear growth rate of the cold matter density contrast $\delta_{\rm cb}(\vec{k},s)$ to a Lagrangian displacement field and physical velocity kick.

At linear order, it is straightforward to write down an equivalent mapping between the MFLR neutrino density perturbations~$\delta_{\alpha,\ell}(\vec{k},s)$ and the linear-order Lagrangian (i.e., Zel'dovich) displacement. The issue is whether this mapping is practicable at late times when the density fluctuations of the neutrino flows  have become fairly sizeable at conversion, since displacing particles by distances more than the PM-grid spacing is generally indicative of the breakdown of perturbation theory and can cause numerical instabilities.
We can however bypass this issue by simply abandoning equal particles masses 
and switching to a variable-mass description, i.e., $m(\vec{x}) \propto [1+\delta_\alpha (\vec{x},s)]$, that reflects the neutrino flow density perturbation at the point~$\vec{x}$ where a neutrino particle is to be spawned.  Then, what remains to be decided is the relation of the real-space density $\delta_\alpha (\vec{x},s)$ to
 the MFLR outputs.

Consider $\delta_\alpha(\vec{x},\hat{\tau},s)$ in equation~\eqref{eq:kick0X}.  This quantity represents the density perturbation at~$\vec{x}$ of one neutrino fluid flow labelled by the $\vec{\tau}_\alpha= \tau_\alpha \hat{\tau}_\alpha$.  In the actual neutrino background, $\vec{\tau}_\alpha$ can point in any direction in a $4 \pi$ solid angle.  This  means that if at any one point $\vec{x}$ we are allowed to represent the density perturbation of the MFLR flow $\alpha$ using only one single particle, we must sum contributions from all directions $\hat{\tau}_\alpha$ over $4 \pi$. That is, the relevant flow density perturbation to use to assign variable particle masses is in fact
\begin{equation}
\begin{aligned}
\label{eq:density2x} 
 \delta_\alpha (\vec{x},s) & \equiv \, \frac{1}{4 \pi} \int {\rm d} \Omega_{\hat{\tau}_\alpha}\, 
\delta_\alpha(\vec{x}, \hat{\tau}_\alpha, s)  \\
& = \, \sum_{\ell =0}^\infty (-{\rm i})^\ell  {\cal F}^{-1}  \left[ \frac{1}{4 \pi} \int {\rm d} \Omega_{\hat{\tau}_\alpha}\,  {\cal P}_\ell( \hat{k}\cdot \hat{\tau}_\alpha) \, \delta_{\alpha,\ell}(\vec{k},s) \right] \\
& = \,
{\cal F}^{-1}  \left[  \delta_{\alpha,\ell=0} (\vec{k},s) \right],
\end{aligned}
\end{equation}
where the integral over solid angle $\Omega_{\hat{\tau}_\alpha}$ automatically selects out the monopole moment at the last equality. Thus, for the purpose of variable-mass assignment, using only information contained in the monopole moment $\delta_{\alpha,\ell=0}(\vec{k},s)$ is fully justifiable.

%%%%%%%%%%%%%
%%%%%%%%%%%%%%%

\subsection{Particle initialisation procedure}
\label{sec:NuInitCond}

Bearing in mind the limitations discussed above, the following is a practicable procedure for converting one neutrino fluid flow $\alpha$ to a particle representation in a simulation box already containing cold matter and possibly previously-converted neutrino particles.  The same process can be applied repeatedly for further conversions at later times.
\begin{enumerate}
\item At the conversion redshift~$z_c$, assign at a density contrast and a gravitationally-induced momentum divergence at each Fourier grid point $\vec{k}$, constructed from the MFLR monopoles of the flow $\alpha$, $\delta_{\alpha, \ell=0} (k, z_c)$ and $\theta^P_{\alpha,\ell=0}(k,z_c)$, and the phase $\breve{\phi}_{{\rm part}}(\vec{k},z_c)$ of the gravitational potential $\Phi_{\rm part}(\vec{k},z_c)$ due to {\it existing} particles in the simulation box at the same $\vec{k}$ and $z_c$.  That is, the realised neutrino density contrast and momentum divergence of flow $\alpha$ at $\vec{k}$ are
\begin{equation}
\begin{aligned}
	\breve{\delta}_{\alpha}(\vec{k}, z_c) & = \delta_{\alpha, \ell=0} (k, z_c)\,  \exp[{{\rm i} \breve{\phi}_{{\rm part}}}(\vec{k},z_c)]\, , \\
	\breve{\theta}^P_{\alpha} (\vec{k}, z_c) &= \theta^P_{\alpha, \ell = 0}  (k, z_c) \,
	\exp[{{\rm i} \breve{\phi}_{{\rm part}}}(\vec{k},z_c)]\, ,
	\end{aligned}
\end{equation}
where the estimates of the phase 
\begin{equation}
\exp[{{\rm i} \breve{\phi}_{{\rm part}}}(\vec{k},z_c)]
\equiv \frac{\breve{\Phi}_{\text{part}} ( \vec{k}, z_c)}{\sqrt{\left\langle |\breve{\Phi}_{{\rm part}}(\vec{k},z_c)|^2 \right\rangle}} \, 
	\label{eq:NuDensityRealisationPhase}
\end{equation}
excludes contributions from the neutrino particles currently under conversion.

\item Place $N_{\alpha}$ neutrino $N$-body particles uniformly on the grid points of a real-space lattice.  At the site $\vec{x}_i$ of each particle, assign the particle a mass given by 
\begin{equation}
	\breve{m}(\vec{x}_i) = \bar{m} \left(1 + {\cal F}^{-1}  \left[  \breve{\delta}_{\alpha}(\vec{k},z_c) \right]\right) \, ,
	\label{eq:NuMassRealisation}
\end{equation}
where $\bar{m} = 3 \,\Omega_{\alpha} h^2  V_{\rm box}/ (8 \pi G N_\alpha)$ denotes the average mass per particle representing the flow $\alpha$ in the simulation volume~$V_{\rm box}$.
\item Assign a physical velocity kick to each particle via
\begin{equation}
\breve{\vec{u}}_\alpha(\vec{x}_i,\breve{\hat{\tau}}_\alpha,z_c) =  \frac{\tau_\alpha}{a(z_c)m_\nu} \breve{\hat{\tau}}_\alpha -\frac{ {\rm i} }{a(z_c) m_\nu}    {\cal F}^{-1}  \left[ \frac{\vec{k}}{k^2} 
\breve{\theta}^P_{\alpha}(\vec{k},z_c) \right],
\end{equation}
where the Lagrangian momentum $\tau_\alpha$ is fixed by the neutrino fluid flow we wish to convert, and $\breve{\hat{\tau}}_\alpha$ is a random direction drawn from a solid angle of $4 \pi$.

\end{enumerate}

%%%%%%%%%%%%
\begin{figure}[t]
	\centering
	\includegraphics[trim=0mm 2mm 10mm 5mm,clip,width=120mm]{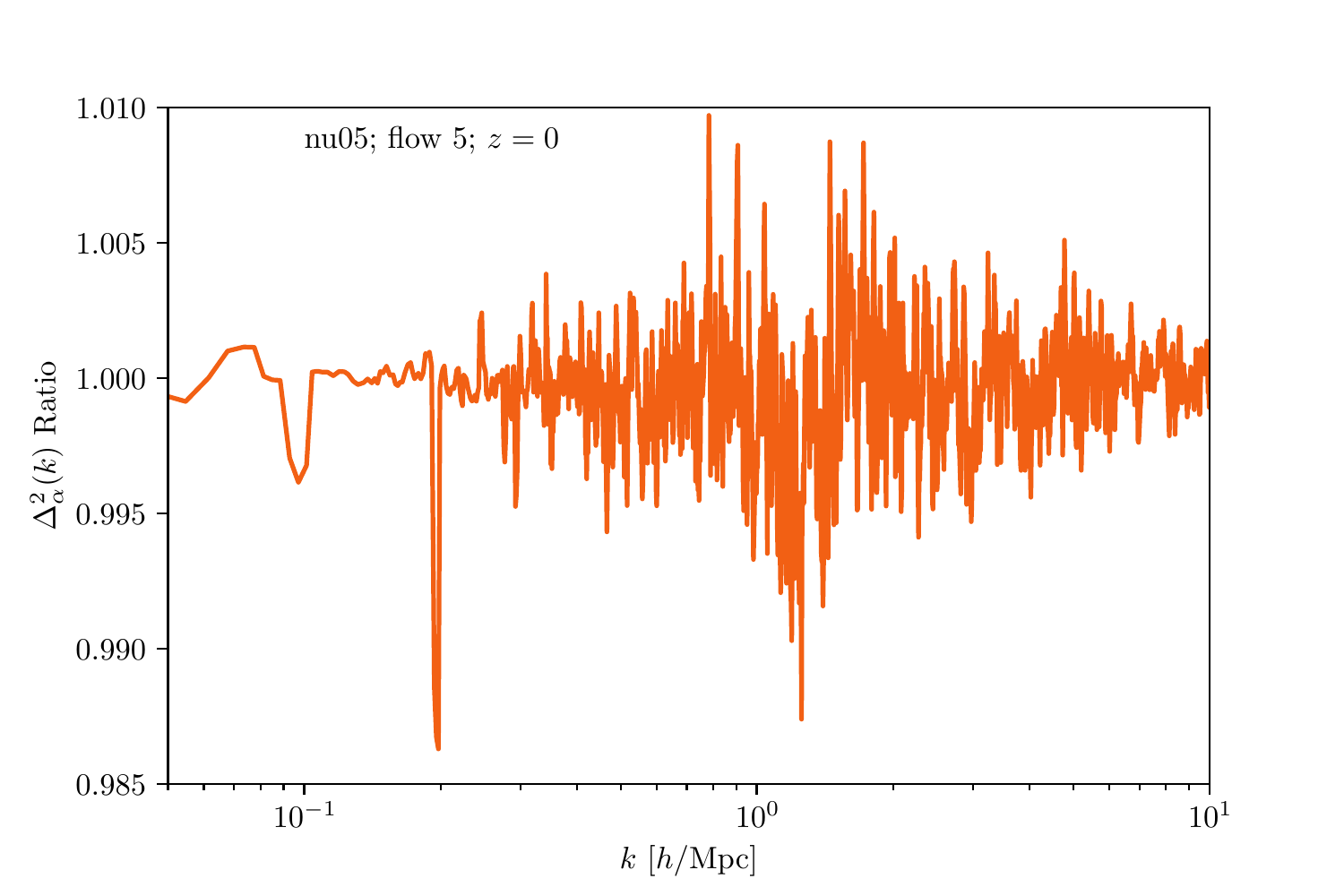}
	\caption{Power spectrum ratio of the $\alpha=5$ MFLR flow in the nu05 cosmology at $z=0$, comparing two hybrid-neutrino simulations initialised with two different neutrino particle placement procedures: putting neutrino particles on regular grid points and off grid points in randomised positions.  The comparison shows that no artefacts arise as a consequence of the regularity of the on-grid procedure. \label{fig:RandomICRatio}}
\end{figure}
%%%%%%%%%%%

At this point, the regularity of the on-grid initial neutrino particle positions may be a cause for concern. To test for the emergence of artefacts, we also consider randomly shifting the particles off the grid---before mass assignment and velocity kick---by an amount~$\in [-0.5 G, 0.5 G]$ in each Cartesian direction, where $G$ is the  grid spacing. Each off-grid neutrino then receives a mass assignment and a velocity kick according to the on-grid masses and momentum divergences interpolated via the Cloud-in-Cell method to the particle site.

We compare on-grid and off-grid initial particle placement by converting the $\alpha=5$ neutrino flow in the nu05 cosmology of table~\ref{table:SimCosmoParam}, using $N_{\alpha} = 512^3$ neutrino particles placed on a $N_{\text{PM}} = 1024^3$ PM grid at a conversion redshift of $z_c = 4$. Figure~\ref{fig:RandomICRatio} shows the flow's dimensionless power spectrum ratio formed from these two approaches at $z=0$.  Clearly, apart from some sub-percent-level fluctuation attributed to the randomly drawn initial off-grid displacements and Lagrangian momentum directions, 
the regularity of the initial neutrino particle positions do not appear to generate any systematic, unphysical effects on any scale. We shall therefore always initialise neutrino particles on grid points unless otherwise specified.

%%%%%%%%%%%%%%%
%%%%%%%%%%%%%%%%

\section{Isolated conversion criterion I: Conversion redshift}
\label{sec:criteria}

\begin{table}[t]
\begin{center}
\footnotesize
    \begin{tabular}{l|cccc|cccc}
    \hline
    \hline
          &                  &\multicolumn{2}{c}{Flows $5$--$6$}&
          &                  &\multicolumn{2}{c}{Flows $9$--$10$}& \\
    $z_c$ & $z_{\rm out}=3$ & $z_{\rm out}=2$ & $z_{\rm out}=1$ & $z_{\rm out}=0$ 
          & $z_{\rm out}=3$ & $z_{\rm out}=2$ & $z_{\rm out}=1$ & $z_{\rm out}=0$\\
    \hline
    $99$ & $577$ &$703$ &$996$ &$1565$  & $686$ &$832$ &$1168$ &$1811$\\
    $19$ & $442$ &$583$ &$816$ &$1496$  & $466$ &$622$ &$879$ &$1654$\\
    $9$  & $401$ &$542$ &$872$ &$1513$  & $401$ &$541$ &$869$ &$1502$\\
    $5$  & $315$ &$482$ &$817$ &$1545$  & $297$ &$437$ &$711$ &$1274$\\

    \hline
    \hline
    \end{tabular}
    \end{center}
    \caption{Typical total simulation run time in minutes for a selection of  simulations converting only flows 5--6 (left) and only flows 9--10 (right) of the nu05 cosmology to $N_\alpha=512^3$ neutrino particles at a range of conversion redshifts~$z_c$. Flows 9--10 represent the median of the distribution, with a Lagrangian momentum of $0.44$~meV and a velocity of 850~km/s at $z=0$. The Lagrangian momentum of flow 5--6 is 0.32~meV, corresponding to 613~km/s at $z=0$.
   All simulations have been initialised with $N_{\rm cb}=512^3$ cold particles at $z_{\rm sim}=99$, run in a $L_{\rm box}=256~{\rm Mpc}/h$ box down to $z_c$ using MFLR, and then run with neutrino particles from $z_c$ to the output redshift $z_{\rm out}$. The total run time refers to the entire simulation duration from $z_{\rm sim}$ to $z_{\rm out}$.  We perform each simulation on three nodes of UNSW's Katana cluster, with $32$ CPUs per node.  For comparison, a full MFLR treatment of neutrinos from $z_{\rm sim}$ to $z_{\rm out}\in\{3,2,1,0\}$  requires $\{236,294,429,720\}$~minutes for the same computational power.  Note also that our simulations have not all been run on exactly same three compute nodes; differences in hardware may also have contributed to small variations in the run times.  
    \label{tab:timing}
    }
\end{table}

With the MFLR-to-particle conversion procedure in place, we must now establish two sets of criteria to determine which neutrino flow(s)  would  require conversion to  particles, and the lowest redshift at which a neutrino flow should be converted. We consider in this section the question of the lowest conversion redshift.  The question of which neutrino flows require conversion is deferred to section~\ref{sec:NLCriteria}.

A low conversion redshift $z_c$ is generally desirable because of the potential saving of computation time associated with the finer time-stepping required by fast-moving neutrino particles.
Table~\ref{tab:timing} shows the typical total run times from the starting redshift $z_{\rm sim}=99$ to the output redshift~$z_{\rm out}\in \{3,2,1,0\}$, for a selection of simulations converting only flows 5--6 and only flows 9--10 of the nu05 cosmology of table~\ref{table:SimCosmoParam} to $N_\alpha=512^3$ neutrino particles at a range of conversion redshifts $z_c \in \{99,19,9,5\}$.   Two general trends are clear: (i)~the higher the conversion redshift, the longer the total run time, and (ii)~faster flows require more run time at high redshifts.  Trend~(i) is particularly well illustrated by flows 9--10, where, for output at $z_{\rm out}=0$, lowering the $z_c$ from $99$ to $5$ reduces the total computation time by $30\%$ given our simulation settings; the reduction in run time is even more significant at higher output redshifts.  Flows 5--6 also supports these trends at high $z_c >9$ and/or high  $z_{\rm out}>1$, but appear to have worse outcomes than flows 9--10 at low $z_c$ and/or low $z_{\rm out}$.  This may be due to stronger nonlinear dynamics and/or simulation transients (due, e.g., to initialising neutrino particles on slightly incorrect trajectories) that require a finer time-stepping to resolve.  The criterion for the ``right'' $z_c$ therefore needs to be determined with these effects in mind.

\begin{figure}[t]
	\centering
	\includegraphics[trim=0mm 2mm 0mm 0mm,clip,width=120mm]{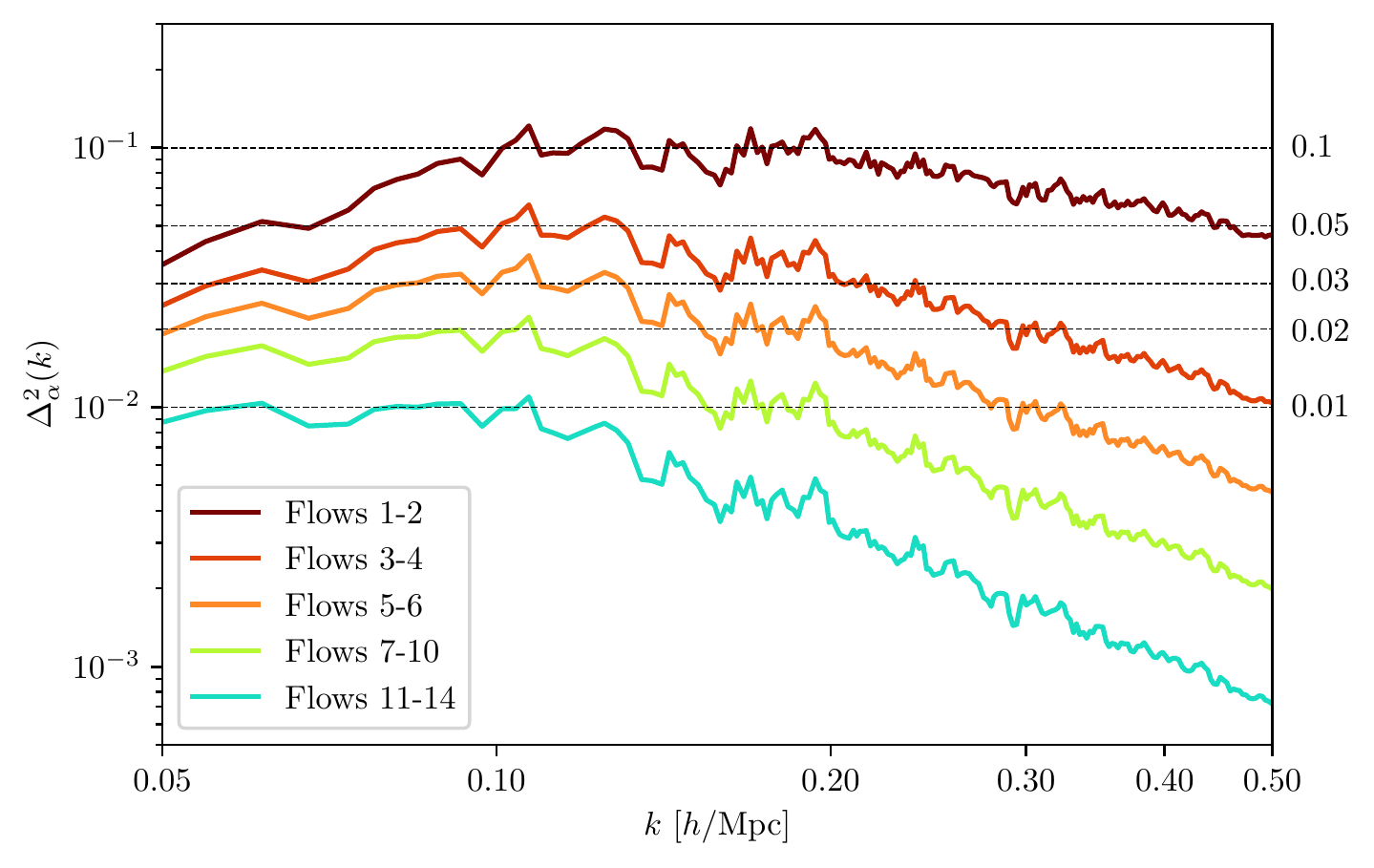}
	\caption{Dimensionless power spectra of the five representative flows of the nu05 cosmology at $z=0$, computed from an MFLR $N$-body simulation. Each representative flow is an average of its constituent flows, whose $z=0$ dimensionless power spectra are shown in figure~\ref{fig:nu05LRDnuRainbow}.
		The grouping is chosen such that the maximum amplitude of $\Delta^2_\alpha(k)$ for the representative flows span between 0.01 and 0.1, as marked by the horizontal dashed lines. \label{fig:nu05DeltaGroupedCritera}}
\end{figure}

We shall use the nu05 cosmology of table~\ref{table:SimCosmoParam} to determine the conversion redshift criterion (and the criterion for which flows to convert).  To reduce the number of simulations required for the various tests, we condense the original 20 neutrino fluid flows into five 
sets of flows, where each set is depicted by a representative flow whose physical attributes---density contrast, momentum divergence, and Lagrangian momentum---are averages of the set's constituents weighted by their number densities.  
Figure~\ref{fig:nu05DeltaGroupedCritera} shows the $z=0$ MFLR dimensionless power spectra of these five representative flows, to be compared with those of the original 20 flows shown in figure~\ref{fig:nu05LRDnuRainbow}. Their corresponding weighted Lagrangian momenta and Lagrangian velocities (i.e., velocities at $z=0$) are summarised in table~\ref{tab:repflow}.  

We shall always refer to these five representative flows when formulating the conversion redshift criterion in this section and the nonlinearity criterion in section~\ref{sec:NLCriteria}.  The criteria will be established based on a set of  ``isolated conversions'': that is,  we perform MFLR-to-particle conversion of one representative flow per simulation at a single conversion redshift.  The case of conversion of multiple flows will be discussed in section~\ref{sec:multiple_conversions}.

%%%%%%%%%%%%%%%
%%%%%%%%%%%%%%%%

%%%%%%%%%%%
\begin{table}[t]
    \begin{center}
	\footnotesize 
    \begin{tabular}{c|cc}
    \hline\hline
    Flow groups & Lagrangian momentum [meV] & Velocity at $z=0$ [km/s]\\
    \hline
    1-2    & 0.152530 & 295.015 \\
    3-4   & 0.247921 & 479.514\\
   5-6   & 0.317092 & 613.302\\
    7-10 & 0.412608 & 798.044\\
    11-14 & 0.552065 & 1067.77\\
    \hline\hline
    \end{tabular}
    \end{center}
    \caption{Weighted Lagrangian momenta and their corresponding Lagrangian velocities (i.e., velocities at $z=0$) of the five representative flows of the nu05 cosmology shown in figure~\ref{fig:nu05DeltaGroupedCritera}.
    \label{tab:repflow}}
\end{table}
%%%%%%%%%%%%

%%%%%%%%%%%%%%%%%%%
%%%%%%%%%%%%%%%%%%

\subsection{Transients from MFLR-to-particle conversion}
\label{sec:ConvRedshiftCriteria}

As discussed in section~\ref{sec:NuInitCond}, we use the MFLR solution to initialise neutrino particles at conversion.  While this solution contains some degree of nonlinearity (from the nonlinear cold matter dynamics), it should nonetheless be noted that using (purely) linear perturbations to initialise nonlinear particle dynamics at too low a redshift is known to excite transients that may not decay away in time~\cite{Crocce:2006ve}; initialising from MFLR may not be immune from this effect.
Furthermore, the neutrino particle initialisation procedure of section~\ref{sec:NuInitCond} comes with additional approximations relative to the MFLR solution, 
which may also be a source of undesirable artefacts.
Our task, therefore, is to establish the lowest conversion redshift while demanding that the simulation outcome be relatively transient-free.

%%%%%%%%%%%%%%%%%%%%%%
\begin{figure}[t]
	\centering
	\includegraphics[trim=5mm 2mm 10mm 5mm,clip,width=75mm]{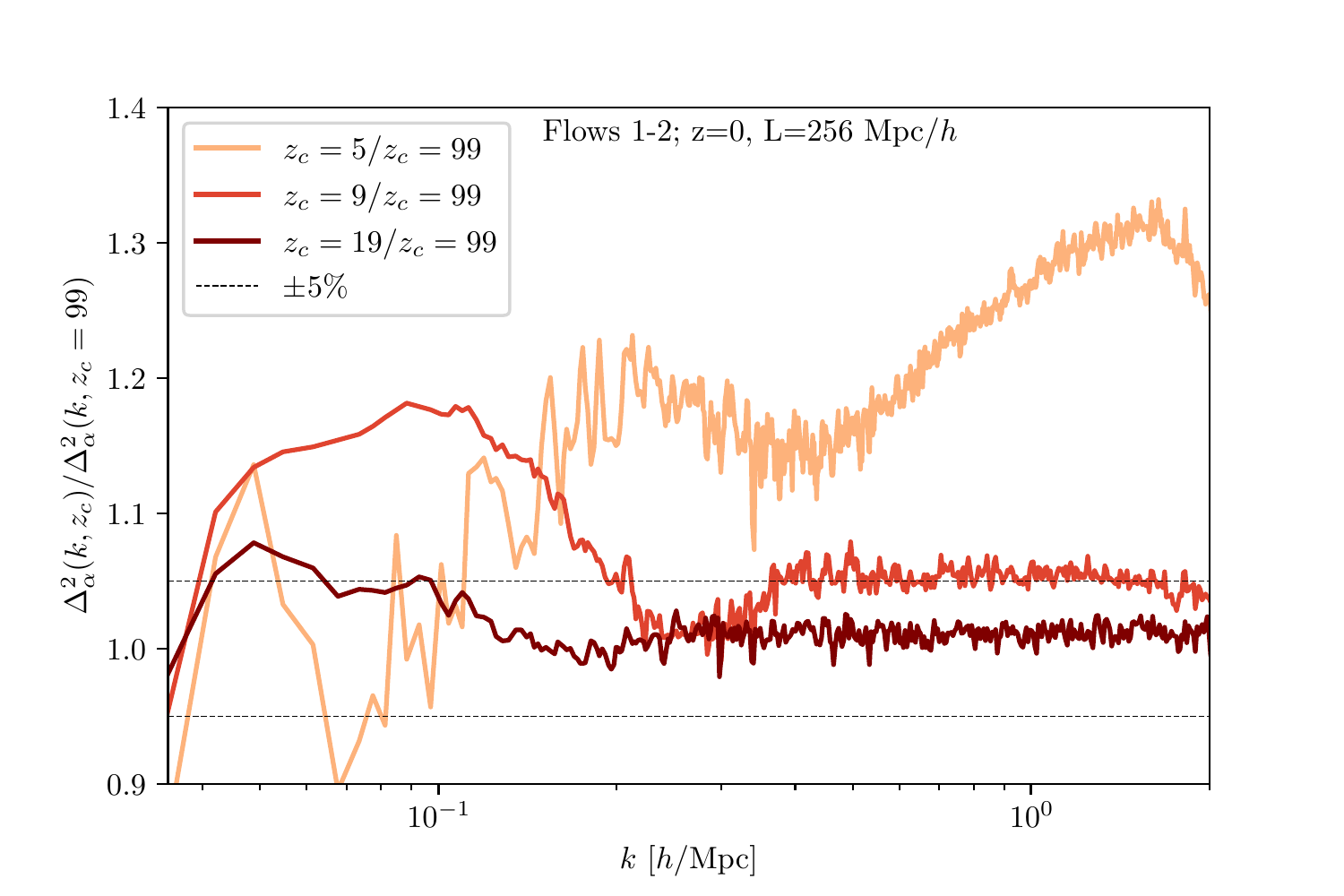}
	\includegraphics[trim=5mm 2mm 10mm 5mm,clip,width=75mm]{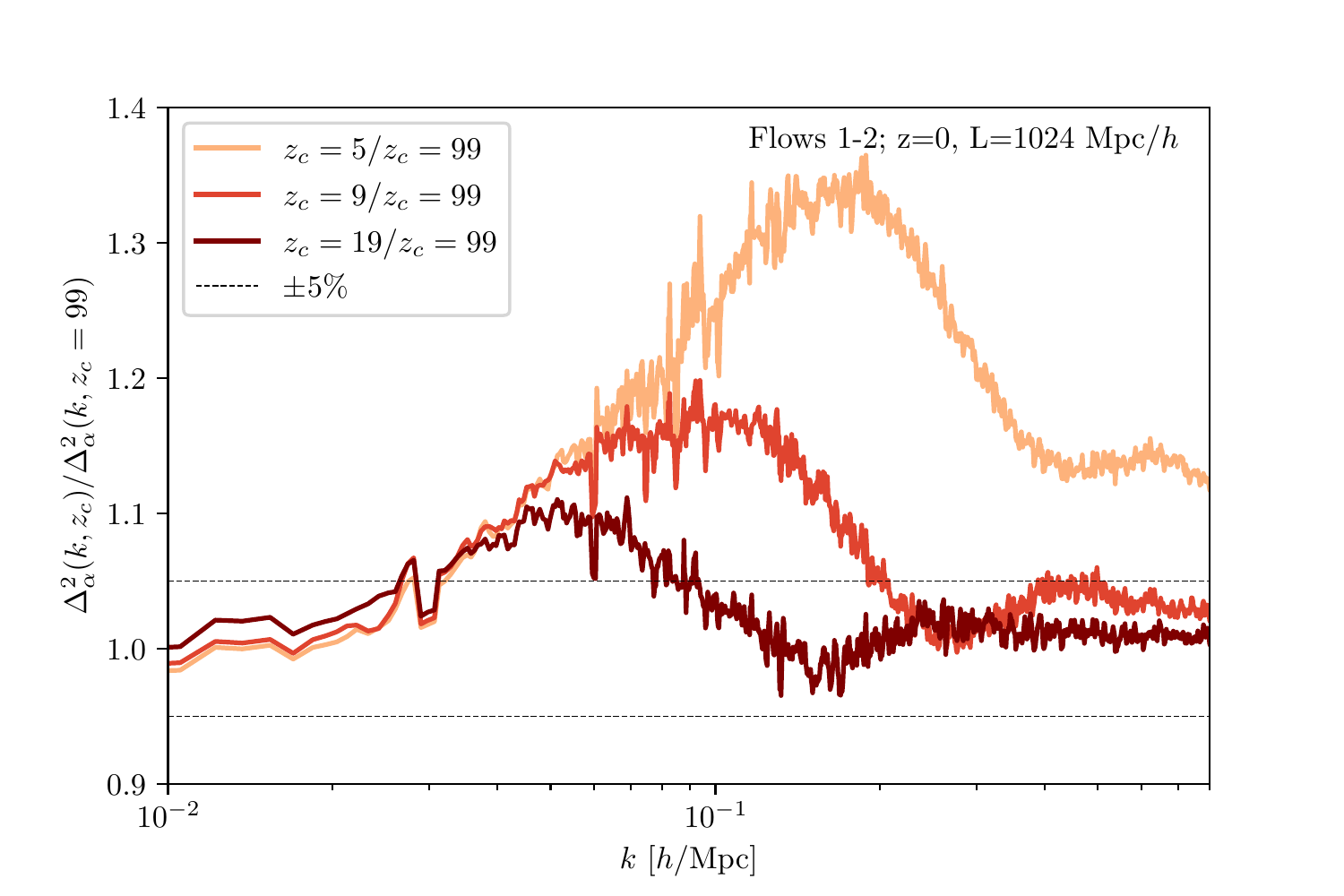}
		\includegraphics[trim=5mm 2mm 10mm 5mm,clip,width=75mm]{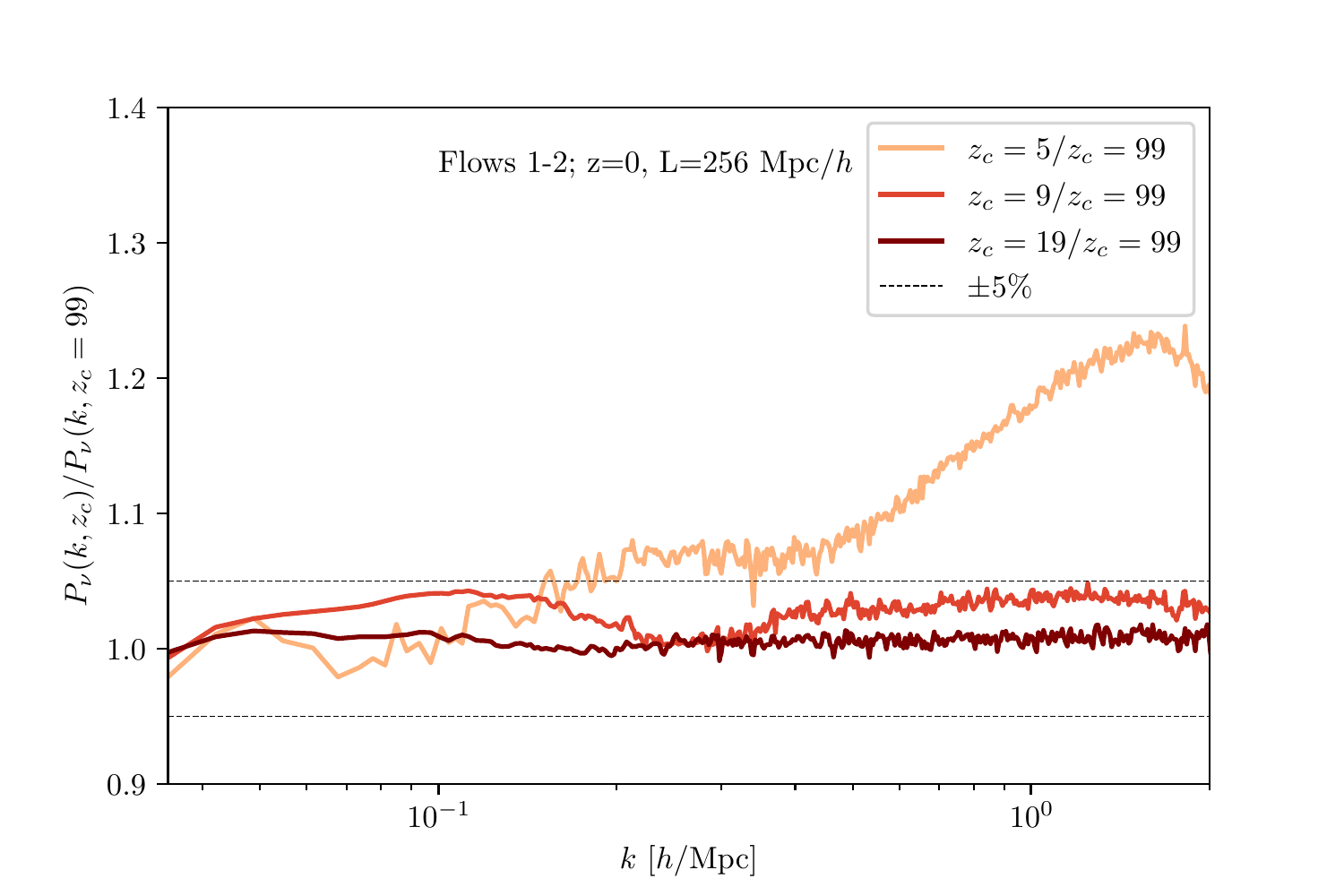}
	\includegraphics[trim=5mm 2mm 10mm  10mm,clip,width=75mm]{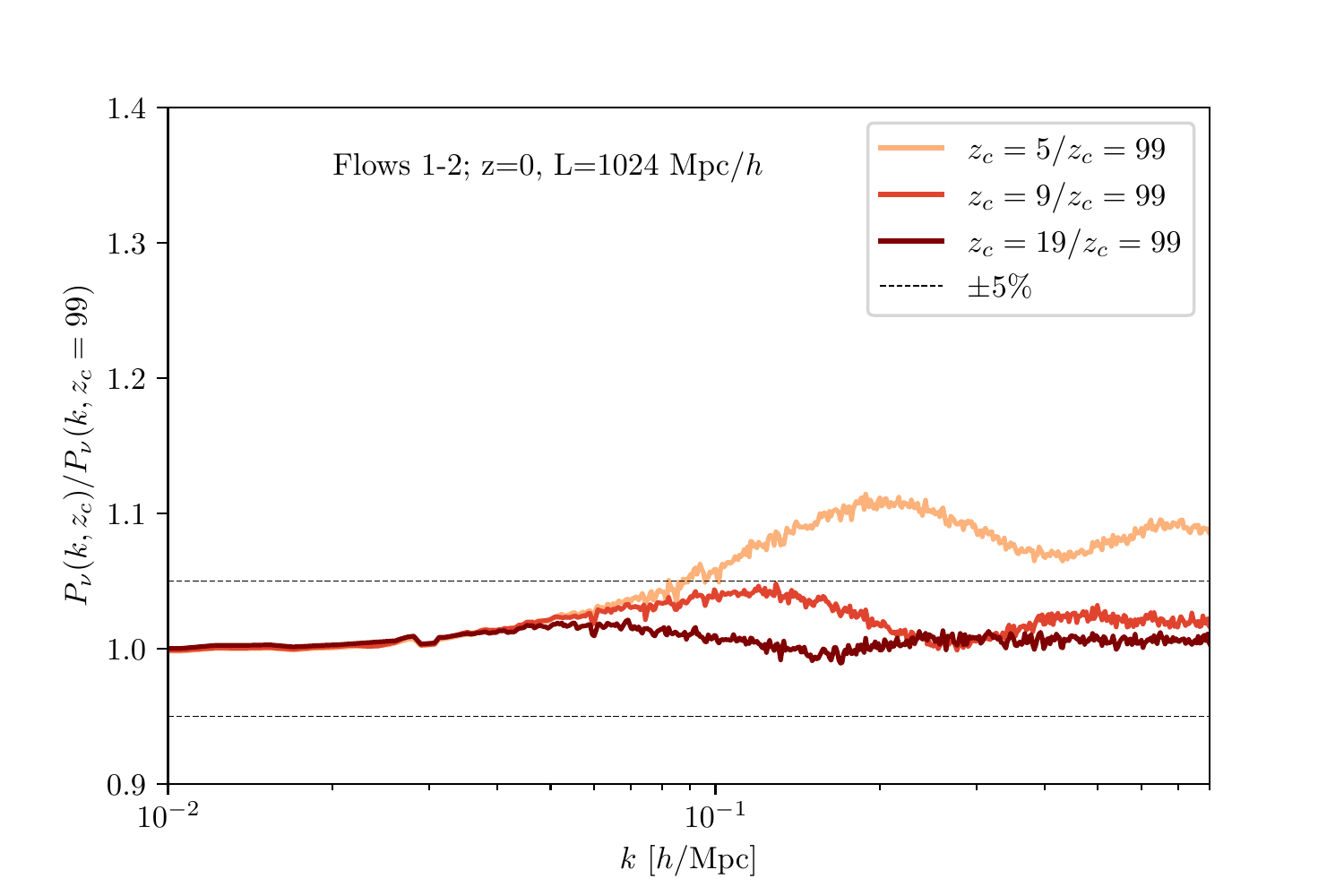}
	\caption{{\it Top row}: Power spectra of the slowest representative flow of the nu05 cosmology (see figure~\ref{fig:nu05DeltaGroupedCritera}) at $z=0$, computed from hybrid-neutrino simulations using various MFLR-to-particle conversion redshifts $z_c \in \{19,9,5\}$, relative to conversion at $z_c=99$. 
	The left panel shows simulations in boxes of side length $L_{\text{box}}=256 \, \text{Mpc}/h$; the smallest wave number sampled is $k^{256}_{\text{min}} \simeq 0.025 \, h/\text{Mpc}$. 
The right panel shows simulations in  $L_{\text{box}} = 1024 \, \text{Mpc}/h$ boxes, where $k^{1024}_{\text{min}} \simeq 0.006 \, h/\text{Mpc}$. 
Both sets of simulations contain $N_{\text{cb}} = 512^3$ cold matter and $N_{\alpha} = 512^3$ neutrino particles. 
The Poisson noise (dimensionful) power spectrum $\epsilon =  V_{\text{box}}/N_{\alpha}$  dominates over the signal at $k \gtrsim 2\, h$/Mpc and $k \gtrsim 0.7\, h$/Mpc in the small-box and large-box runs respectively.  We therefore cut off the plots at large $k$ accordingly.
{\it Bottom row}: Same as the top row, but  for the total neutrino power spectrum~$P_{\nu}(k)$, estimated as per equation~\eqref{eq:nupower}, which sums over both converted and unconverted flows.
\label{fig:Flow1to2PnuTransients}}
\end{figure}
%%%%%%%%%%%%%%%%%%

Consider first the slowest representative neutrino flow of figure~\ref{fig:nu05DeltaGroupedCritera} containing the original $\alpha=1,2$ flows.
In four separate simulations, we convert this representative flow into $N$-body particles at four different conversion redshifts, $z_{c} \in \{ 99, 19, 9, 5 \}$, using in each run $N_{\alpha} = 512^3$ particles to represent the flow.  We conduct the same exercise in boxes of two different side lengths, $L_{\rm box}=256\, {\rm Mpc}/h$ and $L_{\rm box}=1024\, {\rm Mpc}/h$.  The top row of figure~\ref{fig:Flow1to2PnuTransients} shows the resulting $z=0$ dimensionless power spectra of the flow for $z_{c} \in \{19, 9, 5\}$ normalised to the $z_c=99$ case.  The bottom row is similar, but expresses the results in terms of the {\it total} neutrino power spectrum~$P_\nu(k)$, which sums over all converted flows and unconverted MFLR flows and is 
estimated from the hybrid-neutrino simulations per
\begin{equation}
P_{\nu}^{\rm (hybrid)}(k) = \frac{1}{N_\tau} \left| \sqrt{\left\langle \left|\sum_{\rm converted}\delta_\alpha (\vec{k}) \right|^2 \right\rangle}+
\sum_{\rm unconverted}\delta_{\alpha, \ell=0} (k)\right|^2,
\label{eq:nupower}
\end{equation}
ignoring any (small) phase differences between the two contributions.

Focussing first the on the small-box ($L_{\rm box}=256\, {\rm Mpc}/h$)  runs in the top left panel, we see that the run with the lowest conversion redshift, $z_c = 5$, shows a 30\% excess in the flow's dimensionless power spectrum $\Delta_\alpha^2(k)$ at $k \sim 1 \, h/\text{Mpc}$ relative to the $z_c = 99$ run.  However, the power excess reduces dramatically to 5\% around the same scale if conversion happens at $z_c = 9$.   Improvement to percent-level agreement is possible if $z_c = 19$ is used. 
 The bottom panel shows similar behaviours in the total neutrino power spectrum $P_\nu(k)$, but at a somewhat subdued level (24\% excess for $z_c=5$ and $3\%$ for $z_c=9$), as these transients in the $\alpha=1,2$ flows are diluted by contributions to $P_\nu(k)$ from the unconverted MFLR flows.

%%%%%%%%%%%
\begin{figure}[t]
	\centering
	\includegraphics[trim=5mm 2mm 10mm 5mm,clip,width=75mm]{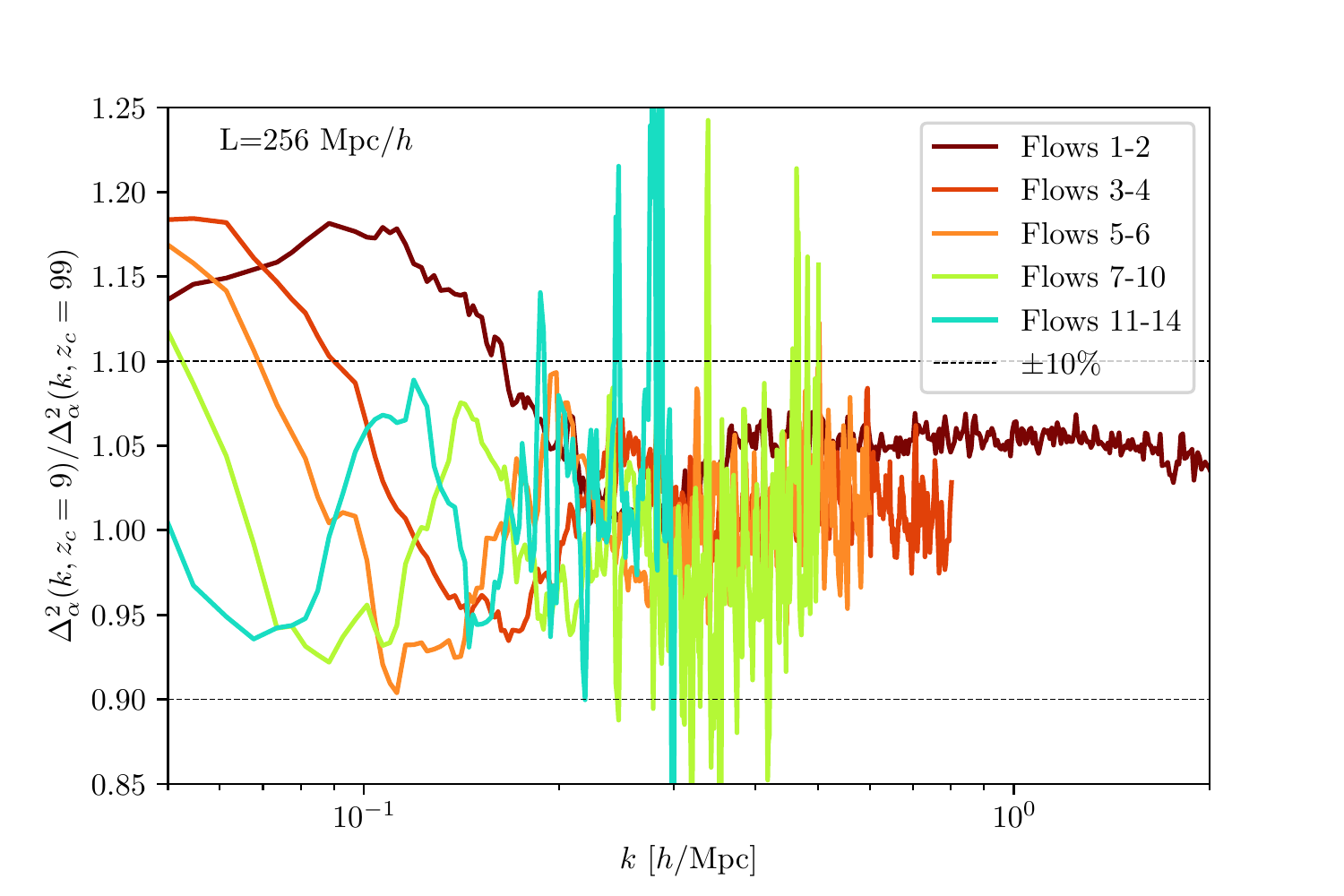}
	\includegraphics[trim=5mm 2mm 10mm 5mm,clip,width=75mm]{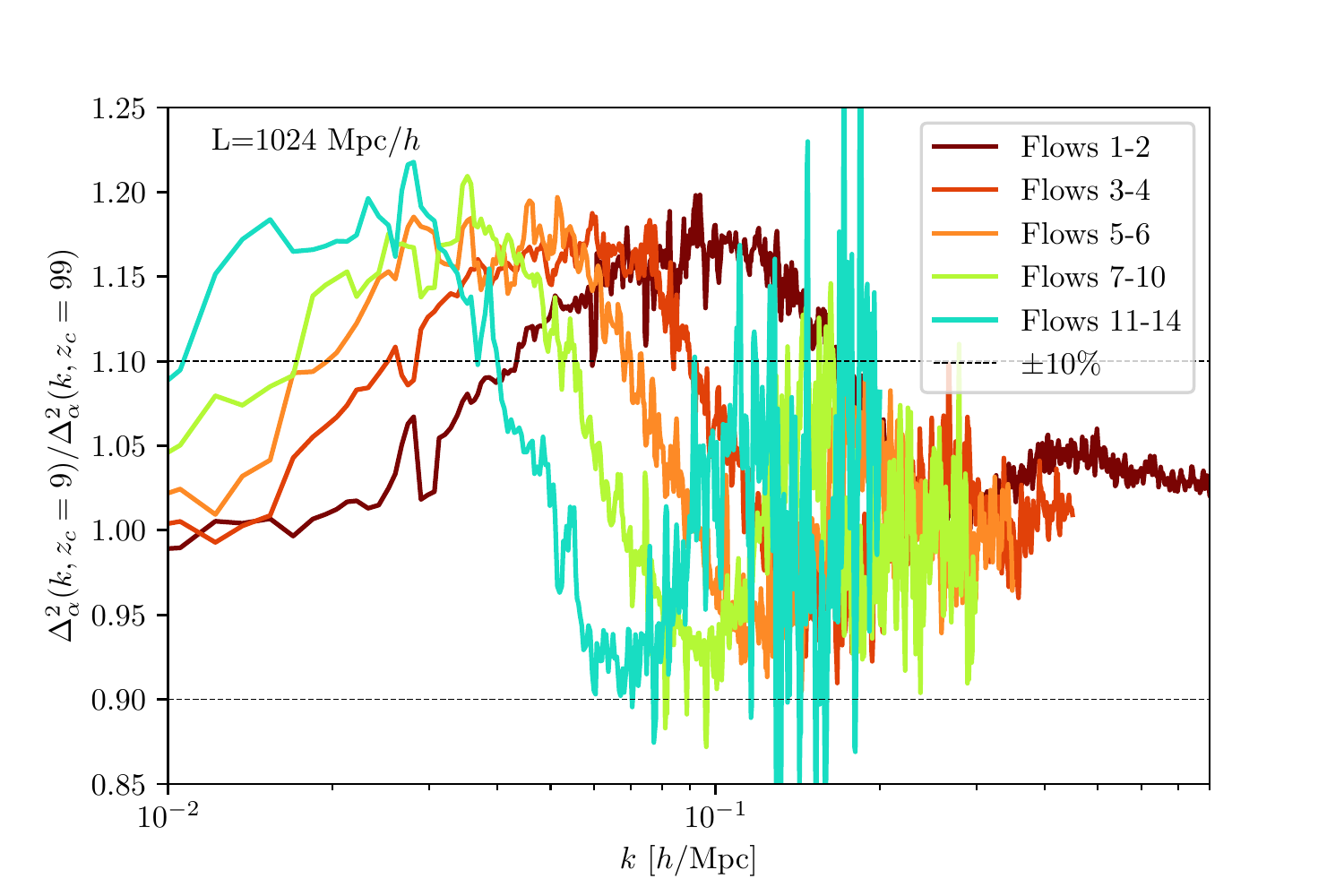}
	\includegraphics[trim=5mm 2mm 10mm 5mm,clip,width=75mm]{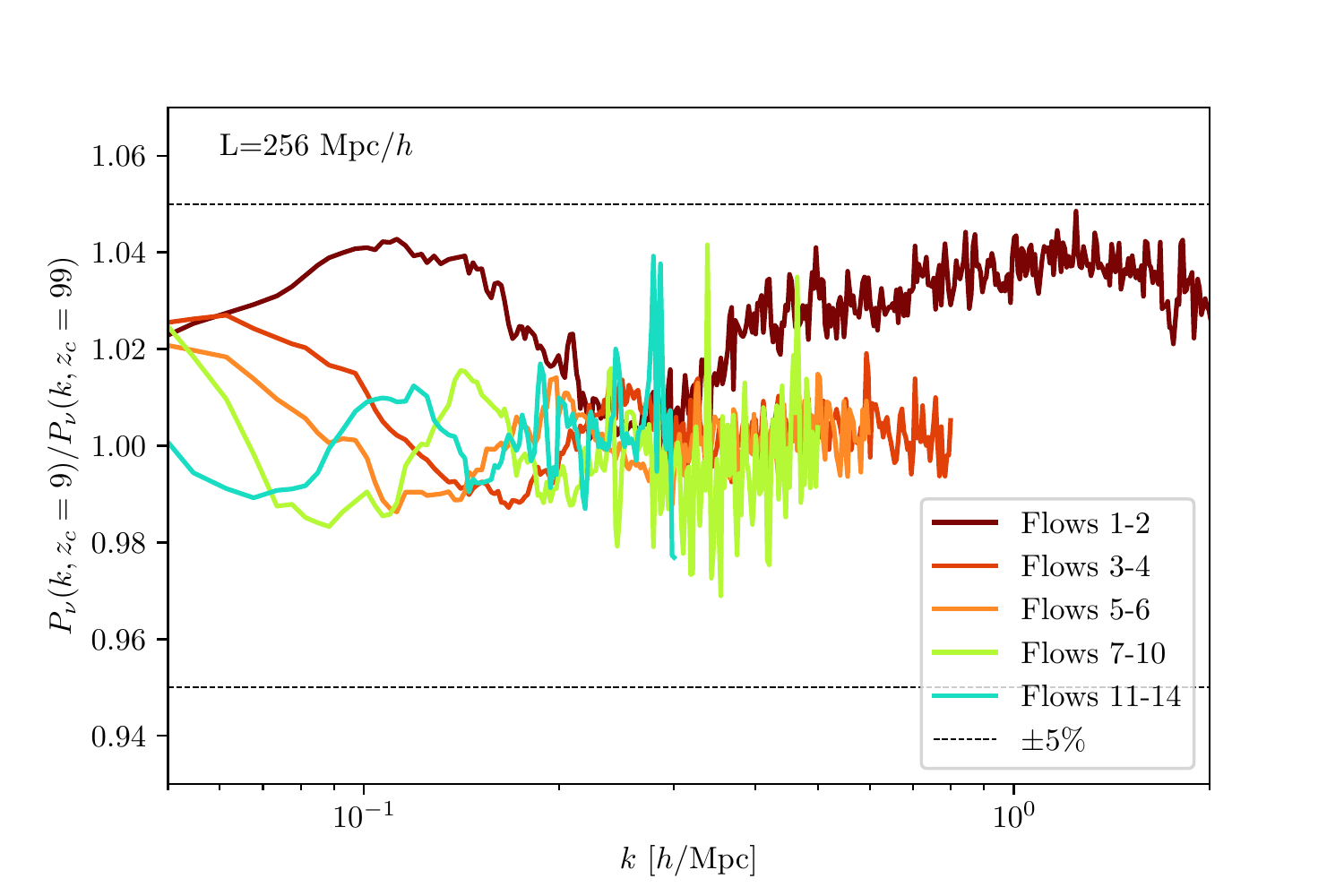}
	\includegraphics[trim=5mm 2mm 10mm 5mm,clip,width=75mm]{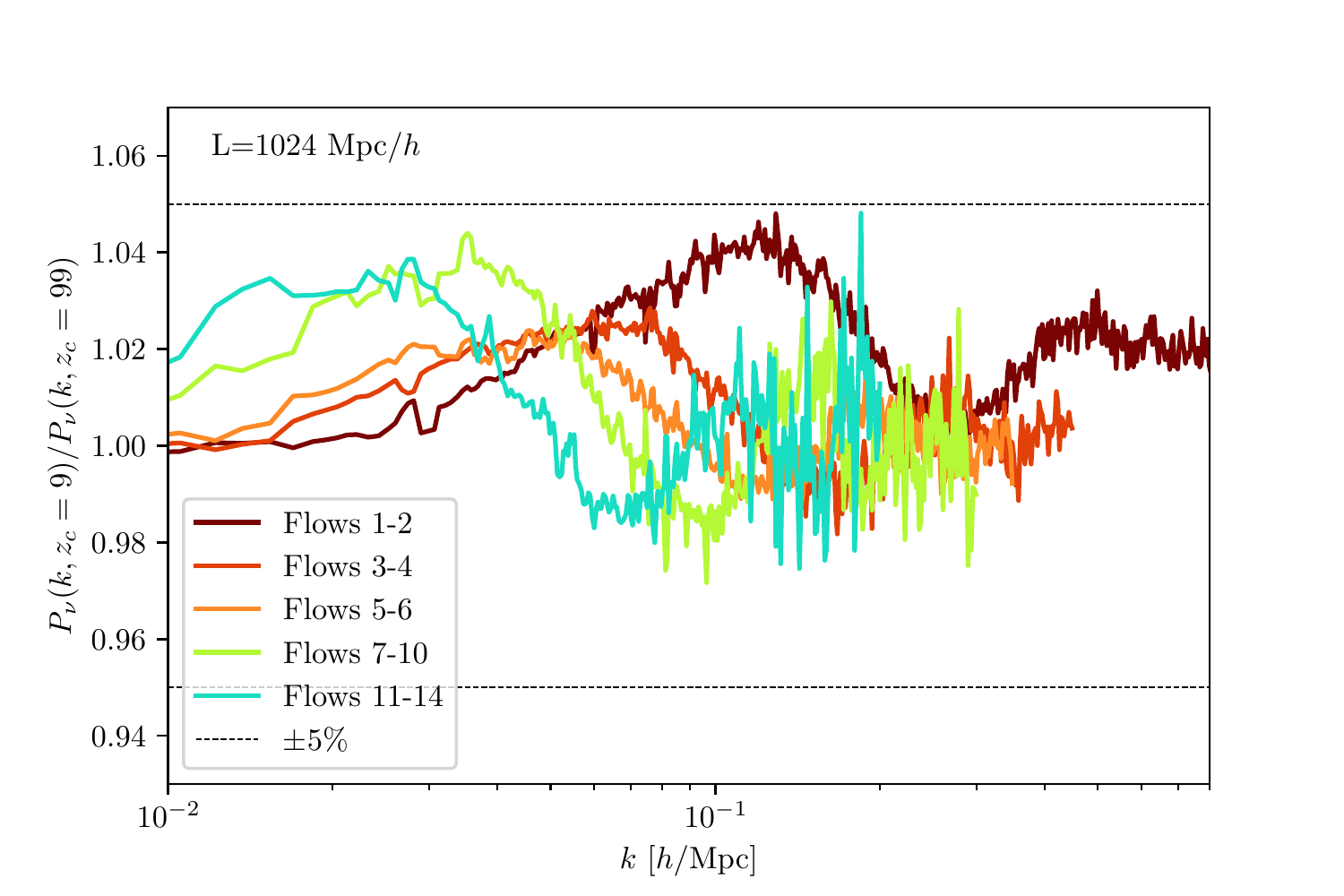}	
	\caption{{\it Top row}: Power spectra of all five representative flows of the nu05 cosmology  at $z=0$, computed from hybrid-neutrino simulations using a MFLR-to-particle conversion redshift of $z_c=9$, relative to conversion at    $z_c=99$. 
The left panel, showing simulations in a $L_{\text{box}}=256 \, \text{Mpc}/h$ box, demonstrates no discernible high-$k$ transient for the faster flows.  However, the right panel shows from the $L_{\text{box}}=1024 \, \text{Mpc}/h$ runs that the low-$k$ transient remains at the 20\% level across all flows; its location, however, scales with the flow's free-streaming scale~$k_{\text{FS},\alpha}$, such that the deviation manifests itself on larger length scales for flows with larger Lagrangian momenta~$\tau_\alpha$. All spectra have been cut off at large $k$ where Poisson noise begins to dominate over the signal.
{\it Bottom row}: Same as the top row, but for the total neutrino power spectrum $P_{\nu}(k)$.  \label{fig:AllFlowsTransientsPnuStr}}
\end{figure}
%%%%%%%%%%%

Observe also in the left panels of figure~\ref{fig:Flow1to2PnuTransients} suggestions of a second peak of power excess at smaller wave numbers,  $k \lesssim 0.2 \,h/\text{Mpc}$.  This power excess is however not well resolved in the small-box runs because of finite-volume effects. Nonetheless, we find it present again in the large-box ($L_{\rm box}=1024\, {\rm Mpc}/h$) runs displayed in the right panels, this time clearly scaling in magnitude with the scale factor at conversion, $a(z_c)$, in the region $0.03 \lesssim k/[h/{\rm Mpc}] \lesssim 0.3$: for $z_c=5$, the power excess in the flow's dimensionless power $\Delta_\alpha^2(k)$ reaches 35\%, but improves to 17\% and 10\% in the $z_c=9$ and $z_c = 19$ case respectively. In terms of the total neutrino power $P_\nu(k)$, the corresponding excesses are 12\%, 4\%, and 2\%.  Observe also that each peak is  accompanied by a trough, i.e., power deficit, immediately to its right, but at a much smaller amplitude.

The same ``low-$k$'' power excess/deficit pattern can again be clearly discerned in figure~\ref{fig:AllFlowsTransientsPnuStr}, where we show the $z=0$ power spectra from isolated conversions
of all five representative flow at $z_c=9$. Evidently, the low-$k$ excess/deficit in both $\Delta_\alpha^2(k)$ and $P_\nu(k)$ appears at ever smaller wave numbers $k$ as we increase the flow's Lagrangian momentum $\tau_\alpha$, while the ``high-$k$'' power excess is practically non-existent for the faster flows on scales unaffected by Poisson noise.  The peak excess and deficit in $\Delta_\alpha^2(k)$ are also remarkably flow-independent: in the case of $z_c=9$, the top row of figure~\ref{fig:AllFlowsTransientsPnuStr} shows a peak excess of just under 20\% for all flows,  while the maximum deficit is always just under 10\%.
The corresponding power excess/deficit in $P_\nu(k)$, shown in the bottom row,  depends however on both the flow's density contrast and reduced energy density~$\Omega_\alpha$, and hence for equal-number flows tends to decrease with the flow's Lagrangian momentum.  (The representative flows 7--10 and 11--14 appear at first glance to defy this trend.  But this is only because these groups each contribute twice as much $\Omega_\alpha$ as one of the slower groups 1--2, 3--4, and 5--6. 

Thus, we can conclude on the basis of figures~\ref{fig:Flow1to2PnuTransients} and~\ref{fig:AllFlowsTransientsPnuStr} that the low-$k$ transient impacts all flows and is, in comparison with the  high-$k$ excess which affects only the slower flows, the more strongly limiting factor on how low a redshift $z_c$ one can adopt for the conversion of MFLR neutrino flows into $N$-body particles.  This conclusion applies at least at the flow-by-flow level; it is conceivable that once we convert multiple representative flows some degree of cancellation will occur between converted flows whose low $k$ power excesses/deficits are out of phase.  We defer this discussion to section~\ref{sec:multiple_conversions}.

%%%%%%%%%%%%%%
%%%%%%%%%%%%%

\subsection{Understanding transients}
\label{sec:mimic}

Considering that the high-$k$ transient affects predominantly the slowest flows---which also have the largest MFLR dimensionless power (see figure~\ref{fig:nu05DeltaGroupedCritera})---its physical origin must almost certainly be linked to the inadequacy of linear solutions used to initialise nonlinear particle dynamics at low redshifts.  Furthermore, the question of how to assign phases to the neutrino density and velocities on nonlinear scales in the presence of free-streaming scale remains open.
As discussed in section~\ref{sec:prelim} and appendix~\ref{sec:phases}, the transition region between the clustering and free-streaming limits is highly sensitive to the evolution of the cold matter phases along a neutrino's path.  This information is unfortunately not available in our present MFLR implementation for cost-saving reasons.  But it is conceivable that some degree of phase information can be retained in a revised version of MFLR, targeting regions that require it.

In contrast, the low-$k$ transient cannot be traced to nonlinearities.  Indeed, as we have seen in figure~\ref{fig:AllFlowsTransientsPnuStr}, the transient appears at 
wave numbers that scale inversely with the flow's Lagrangian momentum~$\tau_\alpha$, peaking in the vicinity of $k_{{\rm FS},\alpha}(z_c)$, i.e., the free-streaming wave number of the flow, given in equation~\eqref{eq:kfs}, at the time of MFLR-to-particle conversion.
As  discussed in section~\ref{sec:prelim}, the cause of this low-$k$ transient is the missing $\ell>0$ moments and hence
the lack of neutrino anisotropic stress representation in the particle initialisation procedure of section~\ref{sec:NuInitCond}.  Specifically, because we have discarded all $\ell >0$ MFLR multipole moments at conversion, the now spawned $N$-body neutrino ensemble must regenerate these moments itself during simulation run-time, at a $k$-dependent rate given by $\Delta s_{\rm repop}^{-1}$ in equation~\eqref{eq:repop}. Until the regeneration is complete, the missing $\ell>0$ moments will cause the neutrino density contrasts to grow too fast, resulting in the observed power excess in $\Delta_\alpha^2(k)$.

%%%%%%%%%%%%%%%%
\begin{figure}[t]
	\centering
	\includegraphics[trim=5mm 2mm 10mm 5mm, clip,width=75mm]{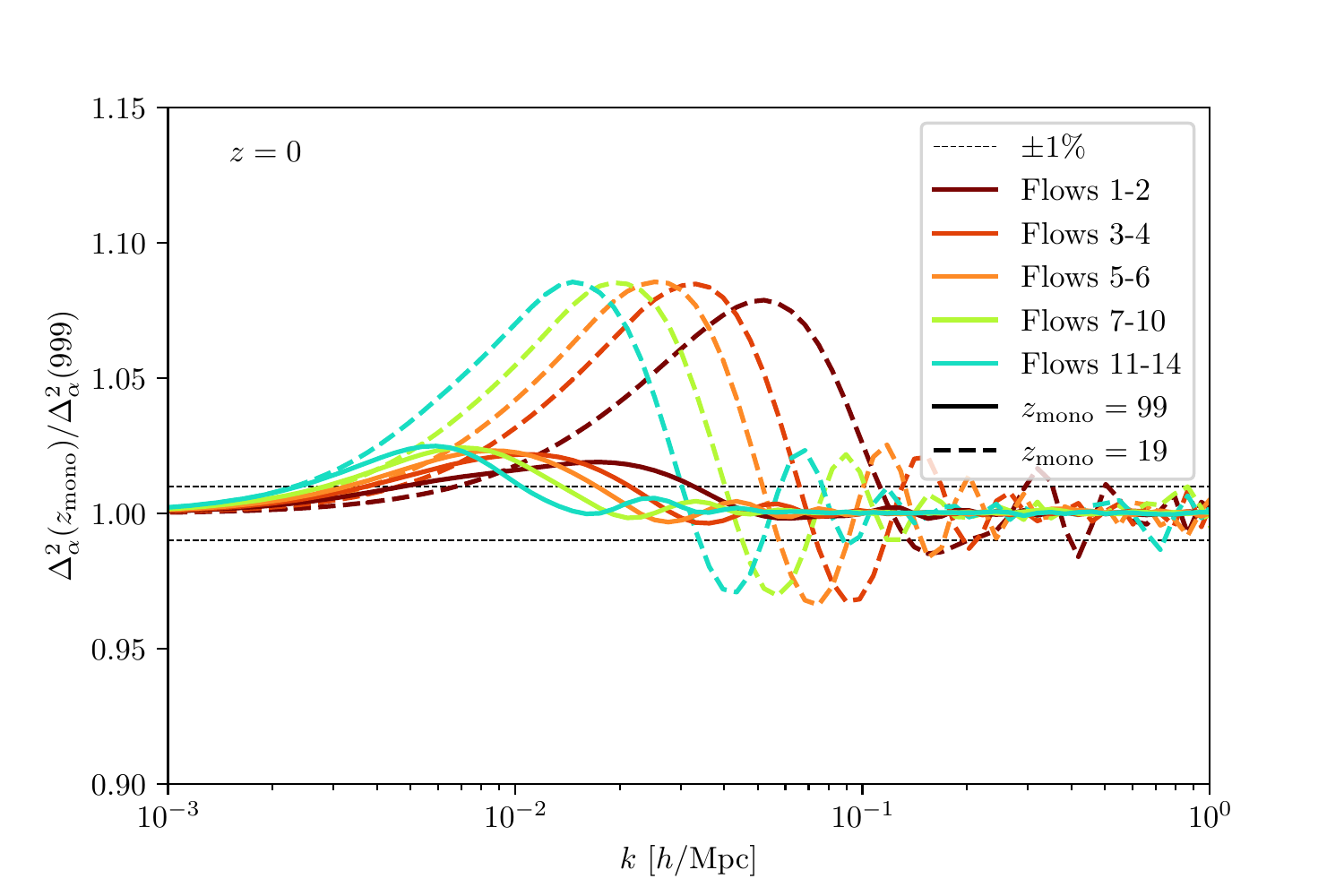}
    \includegraphics[trim=5mm 2mm 10mm 5mm,clip,width=75mm]{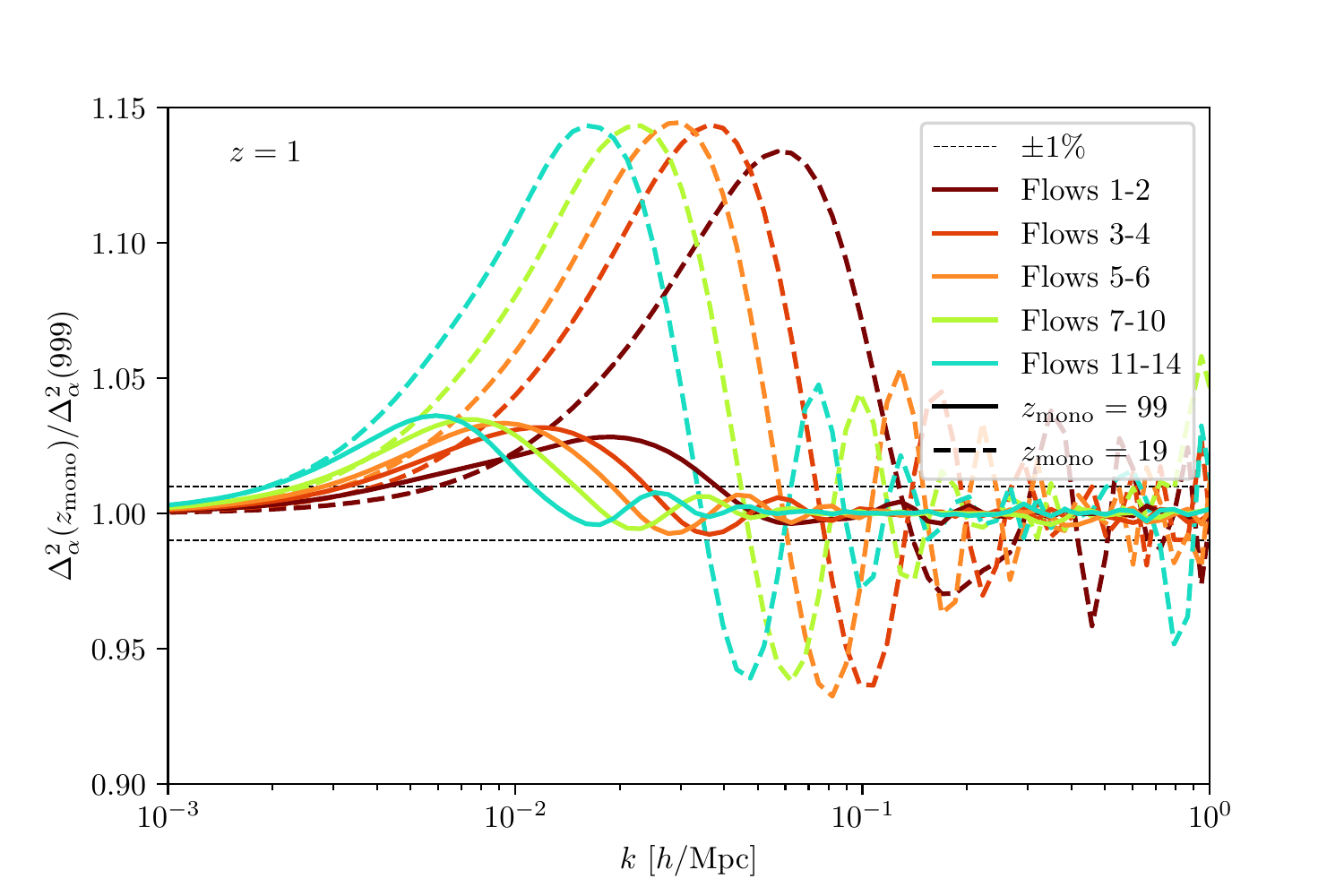}
	\includegraphics[trim=5mm 2mm 10mm 5mm,clip,width=75mm]{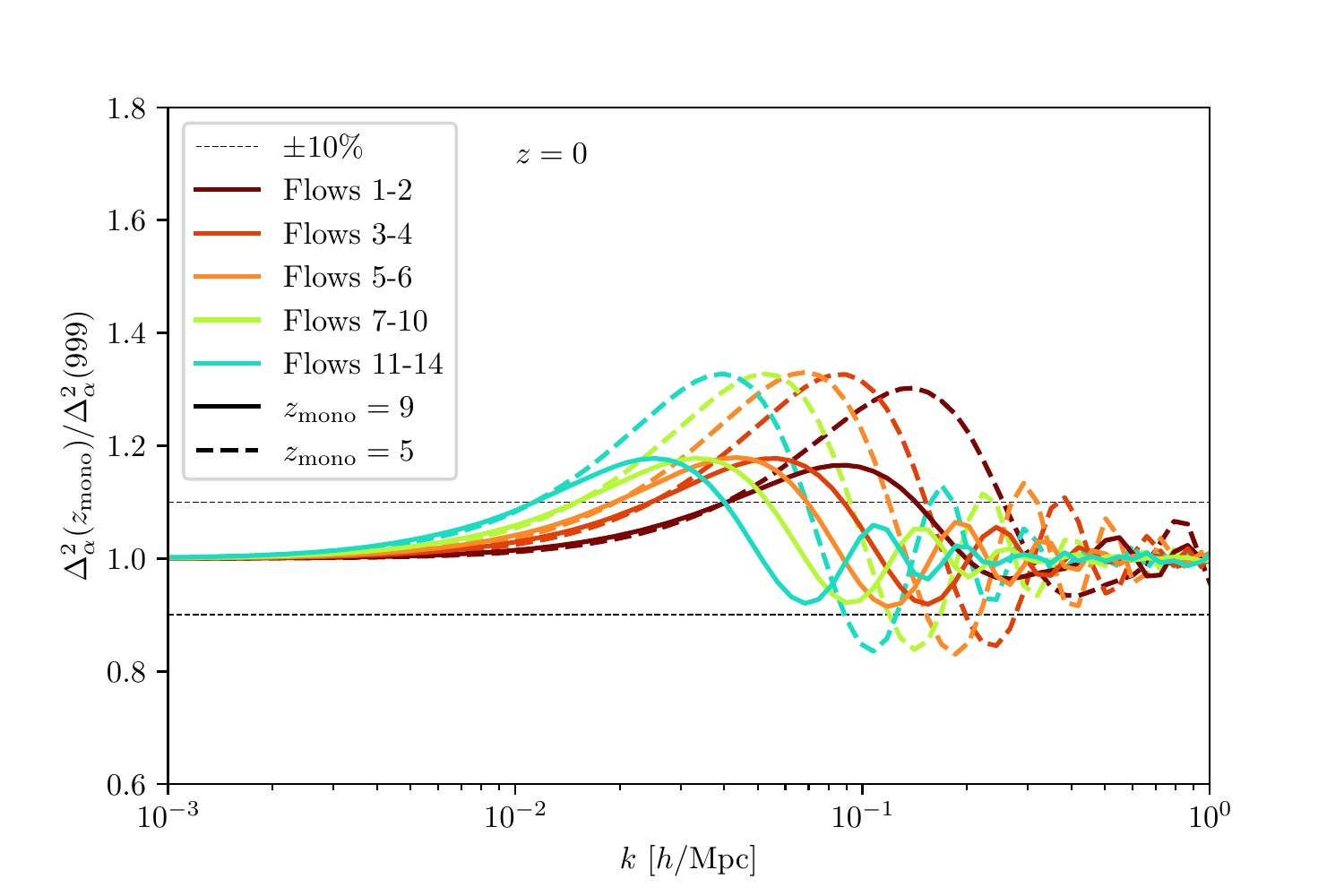}
	\includegraphics[trim=5mm 2mm 10mm 5mm,clip,width=75mm]{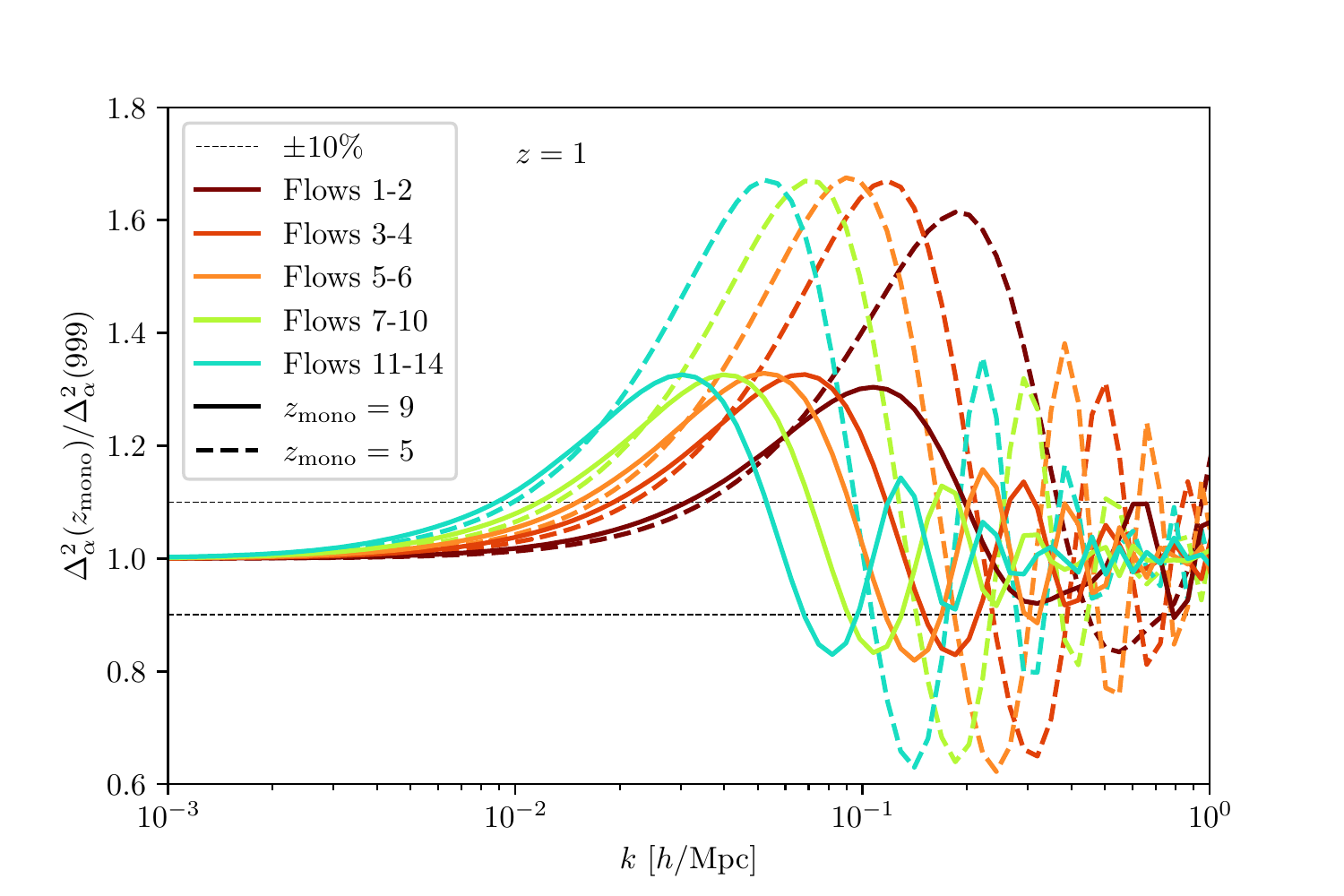}
	\caption{Dimensionless power spectra of all five representative flows of the nu05 cosmology for various conversion redshifts $z_c=99,19$ (top row) and $z_c=9,5$ (bottom row) at $z=0$ (left) and $z=1$ (right), computed from the MFLR/Time-RG mimic described in section~\ref{sec:mimic}.  All spectra have been normalised to a reference MFLR/Time-RG run initialised at $z_{\rm in} = 999$ (or, equivalently, $z_c=999$).
	\label{fig:MFLRTransients}}
\end{figure}
%%%%%%%%%%%%%%%%%%%%%%

This regeneration process and the associated power excess/deficit in $\Delta_\alpha^2(k)$ for different flows can be most clearly seen in and understood from figure~\ref{fig:MFLRTransients}. Analogous to figures~\ref{fig:Flow1to2PnuTransients} and~\ref{fig:AllFlowsTransientsPnuStr}, here in figure~\ref{fig:MFLRTransients} we
employ a ``toy model'' constructed from MFLR and the Time-Renormalisation Group (Time-RG) method~\cite{Pietroni:2008jx} for neutrinos and cold matter respectively, which mimics the MFLR-to-particle conversion by ``resetting'' all $\ell>0$ MFLR multipole moments to zero at the conversion redshifts $z_c$.  The resetting means the $\ell>0$ moments must again be regenerated between~$z_c$ and the redshift~$z$ of interest, and their initial absence leads to the same oscillatory pattern in the neutrino density as seen in the hybrid-neutrino simulations, as power from the monopole propagates to higher multipoles. Figure~\ref{fig:MFLRTransients} shows the $z=0,1$ power excesses/deficits in $\Delta_\alpha^2(k)$ of this toy model for the five representative flows of nu05 for several $z_c\in \{99,19,9,5\}$.

Given a conversion redshift $z_c$, we find the $\Delta_\alpha^2(k,z)$ power excesses in figure~\ref{fig:MFLRTransients} to be well described by
\begin{equation}
{\rm Excess} (z_c,z) \equiv \frac{\Delta_{\alpha}^2(k,z_c,z)}{\Delta_\alpha^2(k,z_c=999,z)}-1\propto \frac{a(z_c)}{a(z)}  \, j_{\ell=1} \left[\frac{k \tau_\alpha}{m_\nu}  \Delta s_{\rm elapsed}(z_c,z)\right] ,  
\end{equation}
where 
\begin{equation}
\label{eq:elapsed}
\Delta s_{\rm elapsed}(z_c,z) = \int_{a(z_c)}^{a(z)} \frac{{\rm d} a'}{a'^3 H(a')}
\end{equation}
is the superconformal time elapsed between $z_c$ and $z$. The description is also quantitatively consistent with the $z=0$ outcome of our hybrid-neutrino simulations displayed in figures~\ref{fig:Flow1to2PnuTransients} and~\ref{fig:AllFlowsTransientsPnuStr}, predicting the peak and trough locations and their amplitudes
with good accuracy.
Since $j_{\ell=1}(x)$ peaks at $x = 2.082$, it is straightforward to establish that for a flow~$\alpha$, the wave number most affected by this transient effect is
\begin{equation}
\begin{aligned}
k_{{\rm peak},\alpha}(z_c,z) & \simeq\,  2.082\,  \left(\frac{m_\nu}{\tau_\alpha}\right)\,  \Delta s_{\rm elapsed}^{-1} (z_c,z) \\
& \simeq\, 1.7\, k_{{\rm FS},\alpha}(z=0) \left(\int^{a(z)/a_\Lambda}_{a(z_c)/a_\Lambda} \frac{{\rm d} y}{\sqrt{1+ y^3}} \right)^{-1},
\end{aligned}
\end{equation}
where $a_\Lambda$ is the scale factor of matter-$\Lambda$ equality, and the integral enclosed in parentheses evaluates typically to an ${\cal O}(1)$ number that increases with $z_c$ and decreases with $z$.  Thus, again, this result highlights the transition between the clustering and the free-streaming limits of a flow as a critical region in the matter of improper initialisation.

%%%%%%%%%%%
%%%%%%%%%%%

\subsection{Can we beat down free-streaming transients?}
\label{sec:tracers}

We close this section with some comments and our outlook on how the issue of the low-$k$ transients may be dealt with in the future.

As already alluded to in section~\ref{sec:prelim}, 
one possible way to mitigate transients arising from the missing higher multipole moments in the neutrino particle initialisation procedure is to limit the number of Lagrangian momentum directions~$\hat{\tau}_\alpha$ to a reasonably small number, e.g., 50, in the particle neutrino initialisation procedure.  This would enable us to incorporate the $\ell>0$ MFLR moments in the MFLR-to-particle conversion without the prohibitive cost of an inordinately large number ($\sim N_\alpha$,  the number of neutrino particles to be initialised) of inverse Fourier transform operations required to achieve a completely randomised setting of $\hat{\tau}_\alpha$.  The gain would be that the conversion redshift can potentially be pushed down to $z_c=9$ or lower.  The price, however, is that the regularity of $\hat{\tau}_\alpha$ may come with its own set of spurious effects that need to be minimised in their own ways and warrant a separate investigation in a future work.

Another possibility might lie in adopting a variant of the tracer method used in the hybrid-neutrino scheme of~\cite{Bird:2018all}.  As described in section~\ref{sec:Bird}, the scheme of~\cite{Bird:2018all} evolves neutrino tracer particles in the simulation volume already from $z=99$---we call this the tracer placement redshift $z_t$---but with two important approximations.  Firstly, the tracers are initialised without power in either the neutrino density contrast or velocity, an approximation whose error should be negligible in the free-streaming limit, i.e., at the small scales $k \gg 0.02~h/$Mpc studied in~\cite{Bird:2018all}, but ought to be important at larger scales where neutrinos and cold matter cluster in the same way.  Since power must be regenerated entirely from scratch during simulation run-time, we would generally expect the flow power spectra to exhibit not only a deficit at $k \lesssim k_{{\rm FS},\alpha}$, but also one that is larger in amplitude than the power excess due to missing only the initial $\ell>0$ moments  in the comparable case of $z_c=z_t$.

Secondly, the trajectories of the tracers are not factored into the computation of the time steps for particle evolution under the long-range force in the $N$-body code,  in order to avoid excessively fine time steps at high redshifts.  For high-resolution simulations, this approximation could cause a neutrino tracer to stream past density clumps which ought to deflect them, leading to errors on small scales.  Testing this approximation on our hybrid-neutrino simulations is unfortunately not straightforward, as the criterion to limit a particle time step via its interparticle flyby time is implemented in \gadgetcode{}'s hierarchical time-integration scheme~\cite{2012NewA...17..711P} in a manner rather different from the PM step constraint in \gadgettwo{} on which the tracer method of~\cite{Bird:2018all} is based.
 
\begin{figure}[t]
	\centering
	\includegraphics[trim=2mm 2mm 10mm 5mm,clip,width=75mm]{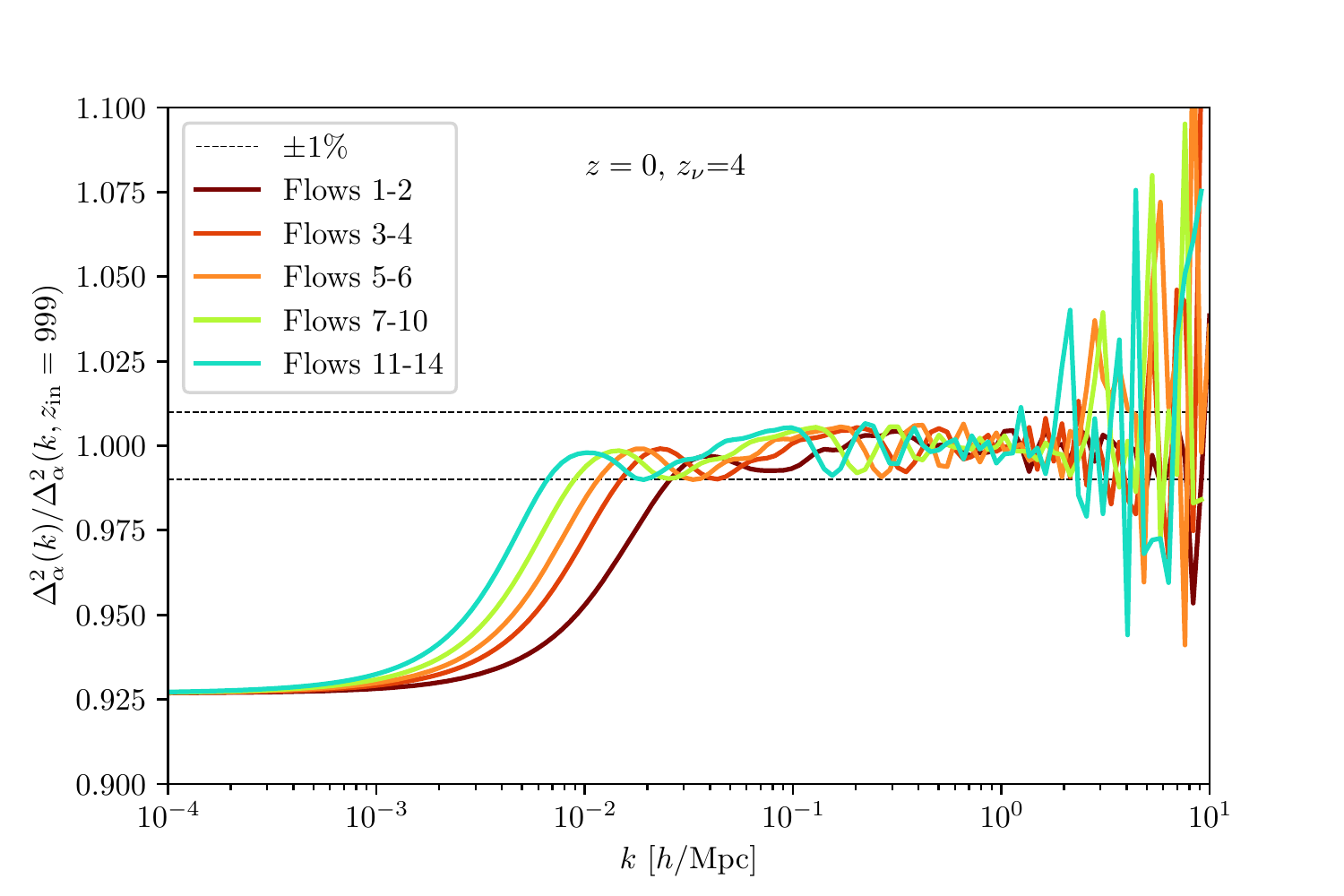}
    \includegraphics[trim=2mm 2mm 10mm 5mm,clip,width=75mm]{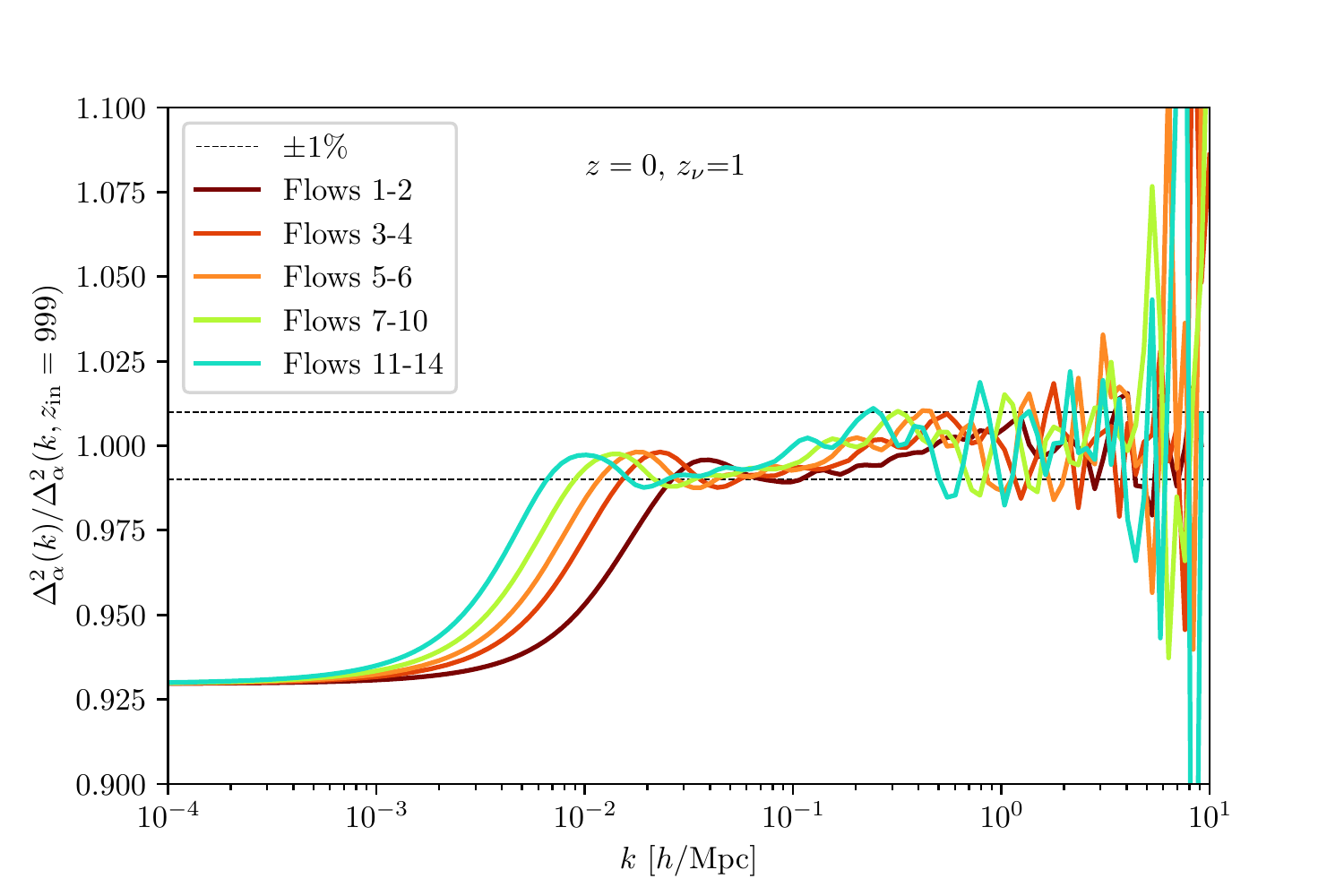}
	\caption{Dimensionless power spectra of all five representative flows of the nu05 cosmology at $z=0$, computed from the MFLR/Time-RG mimic of neutrino tracers described in section~\ref{sec:tracers}.
  The tracers are placed at $z_t=99$ and evolved to $z_\nu=4$ (left) and $z_\nu=1$ (right) in large time steps; 
	they also do not contribute to the gravitational potential until~$z_\nu$.  At $z< z_\nu$, tracers are tracked with proper time-stepping.  All spectra have been normalised to a reference (i.e., no-tracer) MFLR/Time-RG run initialised at $z_{\rm in} = 999$.\label{fig:TracerTransients}}
\end{figure}

Nonetheless, we can make use of the MFLR/Time-RG toy model again to estimate the errors of the tracer method.  Specifically, we use $2 N_\tau$ flows, with the second set $N_\tau+1 \leq \alpha \leq 2N_\tau$ having the same momenta as the first, i.e., $\tau_{\alpha-N_\tau +1} = \tau_{\alpha}$, but making no contribution to the gravitational potential, i.e., $\Omega_\alpha = 0$.  We mimic the zero-power initial condition by zeroing all tracer perturbations at $z=99$, and  the time-stepping approximation by applying the Fourier-space window function $\exp(-k v_\alpha \Delta a / (a {\mathcal H}))$ to the gravitational potential used to evolve the tracer perturbations, where $\Delta a$ is the time step size in the scale factor.   
Approximating $\Delta a/a = \{0.035, 0.018, 0.009, 0.0045, 0.0023\}$ in the redshift brackets $z=\{(99,44), (44,10), (10,3.6), (3.6, 1.8),(1.8,0)\}$ typical in a cold matter-only run,%
\footnote{For comparison, the typical time steps at $z\gtrsim10$ in our hybrid-neutrino simulations (up to flows 10--14) are at least a factor of two to four finer than in an MFLR simulation that evolves only cold matter as particles. The time steps become similar however at $z \lesssim 10$.\label{footnote:timestep}}
 figure~\ref{fig:TracerTransients} shows the  $z=0$ power excess in $\Delta_\alpha^2(k)$ in this toy model for the five representative flows of nu05, with conversion of the tracers into neutrino particles at $z_\nu \in \{4,1\}$.

%%%%%%%%%%%%%%%%
%%%%%%%%%%%%%%%%%%%%%%

Evidently, the zero-power initial condition leads to 7\% power deficit for all flows on large scales that is independent on when the tracer-to-particle conversion occurs. The deficit drops to 1\% by the flow's free-streaming scale at tracer placement $k_{{\rm FS},\alpha}(z_t)$, in a manner described approximately by 
\begin{equation}
\label{eq:deficit}
{\rm Deficit} (z_t,z) \equiv 1-\frac{\Delta_{\alpha}^2(k,z_t,z)}{\Delta_\alpha^2(k,z_t=999,z)}\propto \frac{a(z_t)}{a(z)}  \, j_{\ell=0} \left[\frac{k \tau_\alpha}{m_\nu}  \Delta s_{\rm elapsed}(z_t,z)\right] ,  
\end{equation}
where $\Delta s_{\rm elapse}(z_t,z)$ is the superconformal time elapsed between the tracer placement redshift and the redshift of interest,  given in equation~\eqref{eq:elapsed}. At $k \gtrsim k_{{\rm FS},\alpha}$, however, the excess/deficit pattern deviates from the description~\eqref{eq:deficit}, as it is now modulated by errors introduced through the smearing effect of inadequate time-stepping.  The end effect is, instead of dropping off like 
$\sim 1/k$ as implied by equation~\eqref{eq:deficit}, 
the error in each flow at $k \sim 1\, h$/Mpc remains a little below the 1\% level in the case of tracer-to-particle conversion at $z_\nu=4$ and rises to about 2\% in the case of $z_\nu=1$ for the fastest flows.

Thus, we conclude that the coarser time-stepping at high redshifts adopted in the tracer method of~\cite{Bird:2018all} does not appear to be a significant source of errors on small scales---at least not on those scales where the signal dominates over Poisson shot noise.  
However, in comparison with our hybrid-neutrino scheme where MFLR-to-particle conversion takes place at a redshift comparable to the tracer placement redshift $z_t=z_c = 99$, the low-$k$ transients associated with the initialisation procedure of~\cite{Bird:2018all} can be a factor of two to three higher in the vicinity of a flow's free-streaming scale $k_{{\rm FS},\alpha}(z_c)$.

 This conclusion suggests that tracers may be better used in a modified form, perhaps even in conjunction with our hybrid-neutrino method.  For example, one could envisage a hybrid-scheme wherein tracers are initialised with the MFLR monopole density and velocity divergence---rather than no power at all---at $z_t=99$ and then converted to particles at a redshift $z_\nu \lesssim 10$.  Based on our toy modelling in figures~\ref{fig:MFLRTransients} and~\ref{fig:TracerTransients}, such a scheme  should yield no more than about 2\% errors from transients, while saving some computational time between $z_t=99$ and $z_\nu$.  Should the problem of the missing initial $\ell>0$ moments be solved with, e.g., limiting the number of fixed $\hat{\tau}_\alpha$, tracer placement as late as $z_t=9$ might even be possible (although in this case, evolving tracers may not represent any gain in run time, given that time-stepping for cold and neutrino particles are similar at $z \lesssim 10$ as noted in footnote~\ref{footnote:timestep}).
  We leave the implementation and investigation of such extensions to future work.

%%%%%%%%%%%%%%%%%%%%%%%%
%%%%%%%%%%%%%%%%%%%%%%%%
%%%%%%%%%%%%%%%%%%%%%%%%%%%%%%%%%%
%%%%%%%%%%%%%%%%%%%%%%%%%%%%%%%%%%%%%

\section{Isolated conversion criterion II: Nonlinearity}
\label{sec:NLCriteria}

The second  criterion to be established is which fluid flows exhibit sufficient levels of nonlinear enhancement to warrant conversion to $N$-body particles.  As we have seen in figure~\ref{fig:nu05LRDnuRainbow}, some 50\% of the MFLR neutrino flows in the nu05 cosmology have dimensionless power spectra exceeding $\Delta_\alpha(k) \gtrsim 0.01$, indicative of requiring some degree of nonlinear correction.  Our goal in this section, therefore, is to formulate more concretely the conversion criteria in terms of a neutrino flow's MFLR dimensionless power spectrum at its maximum.
To do so, we again make use of five representative flows of the nu05 cosmology summarised in figure~\ref{fig:nu05DeltaGroupedCritera} and table~\ref{tab:repflow}, and perform an isolated conversion of  each flow into $N_{\alpha} = 512^3$ neutrino particles in a $L_{\text{box}}=256 \, \text{Mpc}/ h$ box at $z_c=19$.

%%%%%%%%%%%%%%%
%%%%%%%%%%%%%%%

\subsection{Enhancement to the flow}

%%%%%%%%%%%%%%%%%%%%%%%%
\begin{figure}[t]
	\centering
	\includegraphics[trim=2mm 2mm 13mm 5mm,clip,width=120mm]{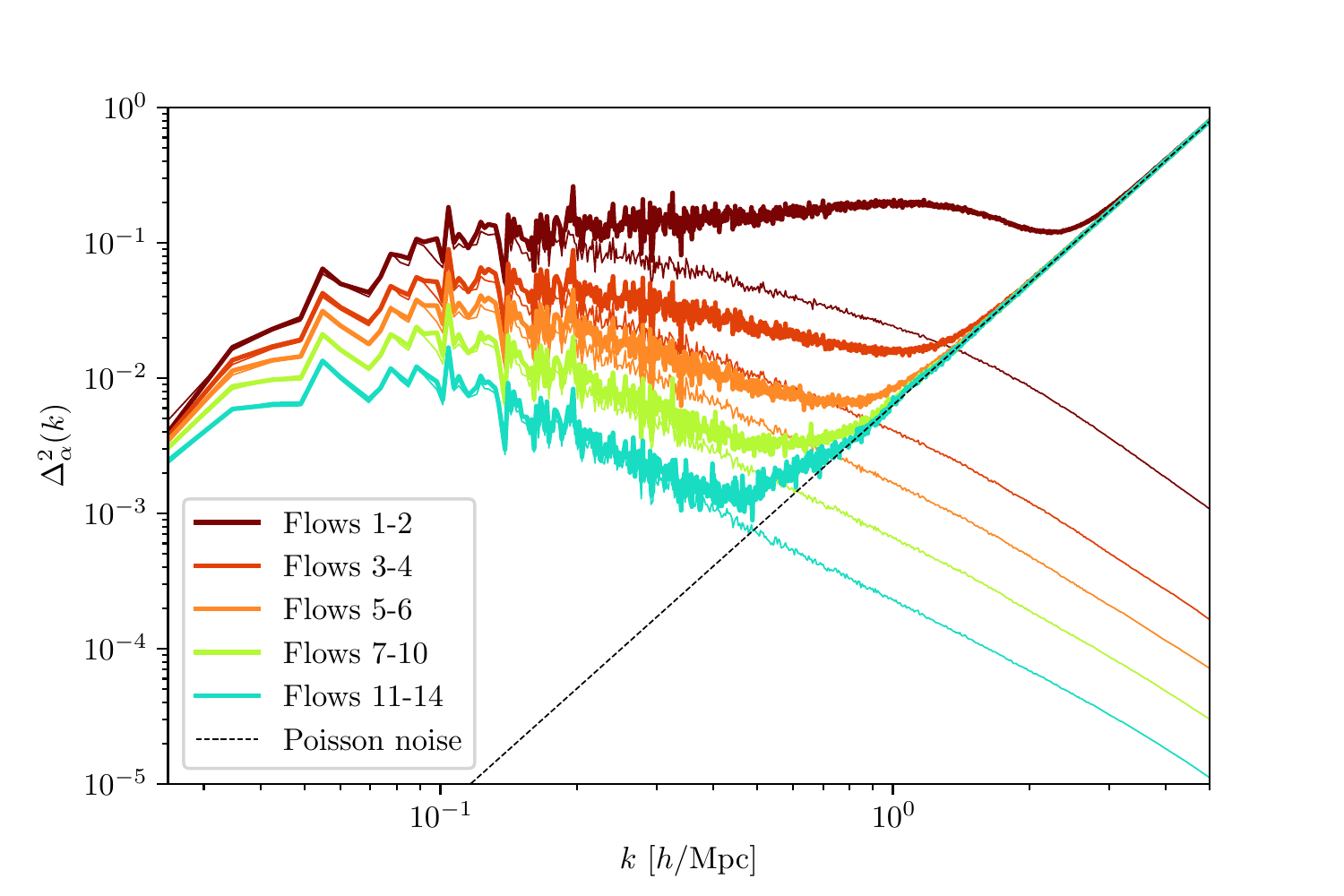}
	\includegraphics[trim=2mm 2mm 13mm 5mm,clip,width=120mm]{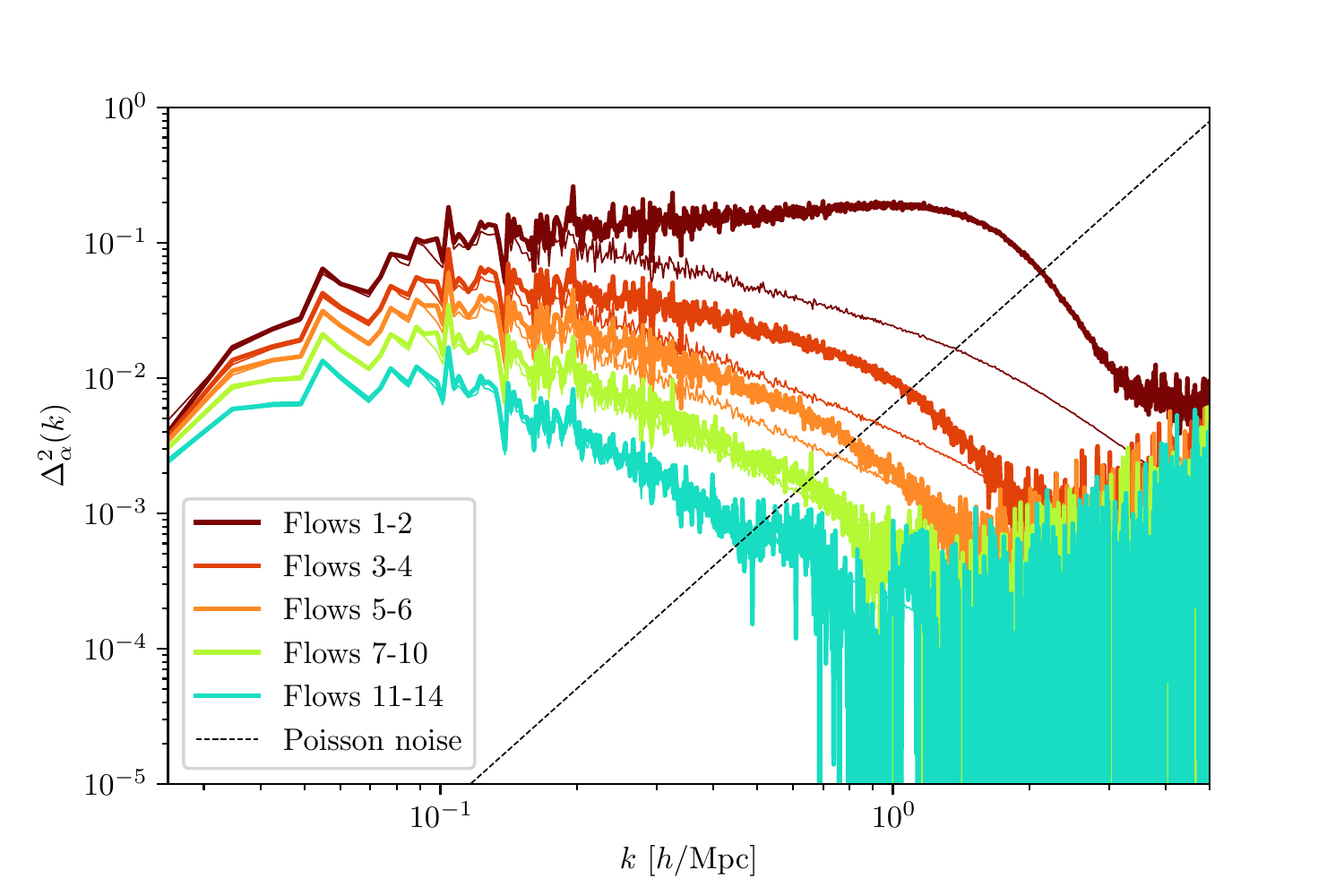}
	\caption{Dimensionless power spectra of the five representative flows of the nu05 cosmology at $z=0$.
	Solid thick lines represent the fully nonlinear power spectra computed from hybrid-neutrino simulations with a MFLR-to-particle conversion at $z_c=19$;  thin lines of the same colours denote their MFLR counterparts previously shown in figure~\ref{fig:nu05DeltaGroupedCritera}. The diagonal black dashed line denotes the Poisson noise floor for the chosen particle neutrino resolution.  The power spectra are presented ``as is'' in the top panel, while in the bottom panel the Poisson noise has been subtracted off. \label{fig:AllFlowsDelta2NLEnhancements}}
\end{figure}
%%%%%%%%%%%%%%%%%%%%%

Figure~\ref{fig:AllFlowsDelta2NLEnhancements} shows the $z=0$ dimensionless power spectra $\Delta_\alpha^2(k)$ of the five representative flows computed in this manner alongside their MFLR counterparts. As expected, the slower flows exhibit larger enhancements of power on small scales. As we move to increasingly large wave numbers, however, all five power spectra converge to the Poisson noise floor, represented by the dashed black line in the top panel of figure~\ref{fig:AllFlowsDelta2NLEnhancements}. The bottom panel shows the same dimensionless power with the Poisson noise subtracted away. 

An immediate takeaway from figure~\ref{fig:AllFlowsDelta2NLEnhancements} is that the  dimensionless power spectra of flows 7 to 14 computed from hybrid-neutrino simulations exhibit no significant enhancement over their MFLR counterparts. On this basis,
we can conclude that those neutrino fluid flows with MFLR dimensionless power spectra not exceeding $\sim 0.02$ at $z=0$ experience only minimal nonlinear evolution.  Linear response (to nonlinear cold matter) suffices to describe their evolution; a particle representation of these flows is not only unnecessary, it is also counter-productive in that the issue of free-streaming transients and the rapid domination of Poisson noise over the signal limits the usefulness of the result to a fairly narrow range in~$k$.

For flows 1 to 6, however, we find increasing  nonlinear enhancement in $\Delta_\alpha^2(k)$ as we decrease the flow's Lagrangian momentum.  For the first representative flow (encapsulating flows $\alpha=1,2$), figure~\ref{fig:AllFlowsDelta2NLEnhancements} shows a nonlinear enhancement over linear response as large as a factor of ten at $k \simeq 1\, h$/Mpc.  
Defining the hybrid-to-MFLR enhancement factor to be
\begin{equation}
\label{eq:pnlpmflr}
\gamma_{\rm NL} (k,z=0) \equiv \frac{\Delta_{\alpha}^{2\,{\rm (hybrid)}}(k, z=0)}{\Delta_ \alpha^{2\,\text{(MFLR)}}(k, z=0)} \, ,
\end{equation}
we find the power enhancement at $k \simeq 0.5 \, h$/Mpc to be well described empirically by
\begin{equation}
	\gamma_{\text{NL}}\left({k = 0.5 \, h/\text{Mpc}}, z=0\right) \simeq 0.784 + 26.575 \, \Delta^{2\,{\rm (MFLR)}}_{\alpha,\text{max}}(z=0) \, ,
	\label{eq:NLfactorfit}
\end{equation}
where  $\Delta^{2\,{\rm (MFLR)}}_{\alpha,\text{max}}(z=0)$ denotes the value of the $z=0$ MFLR dimensionless power spectrum of the flow~$\alpha$ at its maximum.  Note that the relation~\eqref{eq:NLfactorfit} 
is valid only for $\Delta^{2\,{\rm (MFLR)}}_{\alpha,\text{max}} \gtrsim 0.01$.

%%%%%%%%%%%%%
\begin{figure}[t]
	\centering
	\includegraphics[trim=2mm 2mm 2mm 2mm,clip,width=75mm]{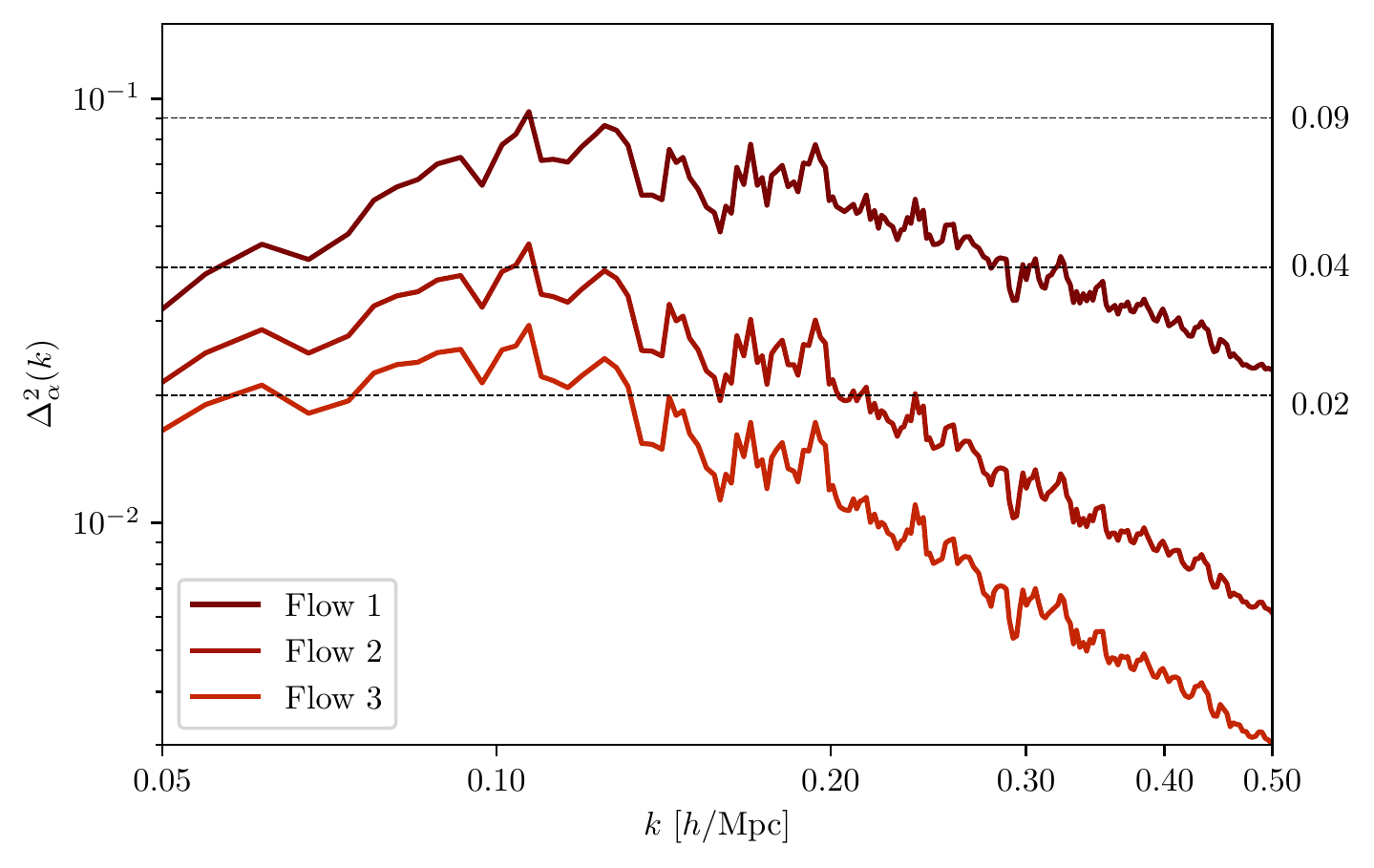}
	\includegraphics[trim=2mm 2mm 2mm 2mm,clip,height=47.5mm]{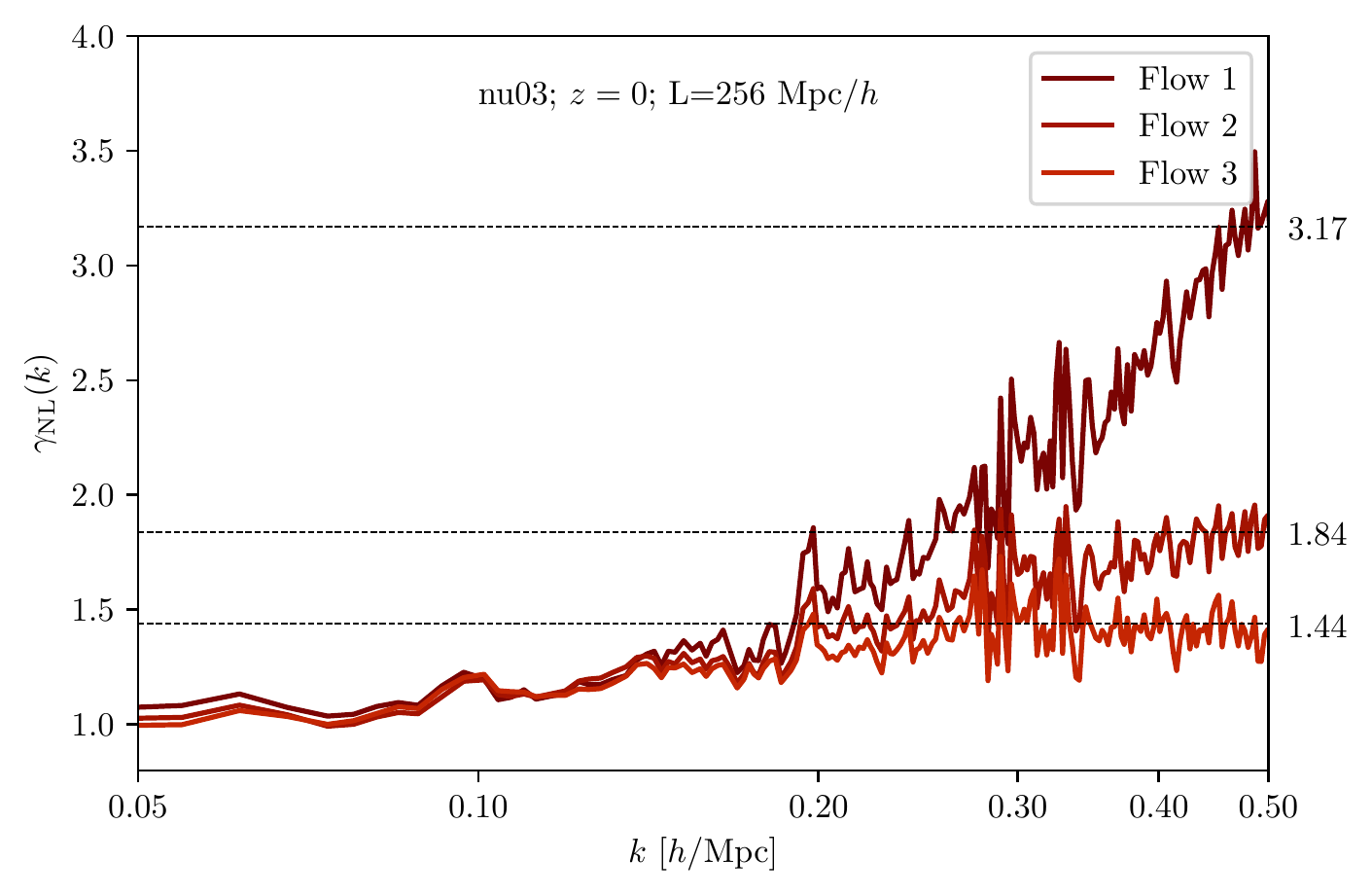}
	\caption{Dimensionless power spectra of the first three neutrino fluid flows of the nu03 cosmology at $z=0$.
	The left panel shows the dimensionless power spectra computed from an MFLR simulation and their corresponding $\Delta_{\alpha,{\rm max}}^{2\,{\rm (MFLR)}}$ values.  The right panel shows the nonlinear power enhancements found from hybrid-neutrino simulations over MFLR, with black dashed lines marking the  enhancement factor $\gamma_{\text{NL}}$ for each flow at $k = 0.5 \, h/\text{Mpc}$ as predicted by equation~\eqref{eq:NLfactorfit} given the flow's $\Delta_{\alpha,{\rm max}}^{2\,{\rm (MFLR)}}$.
\label{fig:nu03NLEnhancementsValidation}}
\end{figure}
%%%%%%%%%%%%%

While we have deduced equation~\eqref{eq:NLfactorfit} based on the nu05 cosmology, the relation between $\gamma_{\text{NL}}\left({k = 0.5 \, h/\text{Mpc}}, z=0\right)$ and $\Delta^{2,{\rm MFLR}}_{\alpha,\text{max}}(z=0)$ applies also to cosmologies with a different neutrino energy content.   Figure~\ref{fig:nu03NLEnhancementsValidation} shows the case of the ``nu03'' cosmology detailed in table~\ref{table:SimCosmoParam}, which has three equal-mass neutrinos summing to $\sum m_\nu = 0.279$~eV (cf. $\sum m_\nu = 0.465$~eV in nu05).   Here, the $z=0$ MFLR dimensionless power spectra for the first three neutrino flows are displayed in the left panel, along with their corresponding $\Delta^{2\,{\rm (MFLR)}}_{\alpha, {\rm max}}(z=0)$ values.  The right panel, on the other hand, shows the nonlinear enhancements at $z=0$ as functions of wave number $k$, computed from three hybrid simulations each with one single-flow conversion to $N_{\alpha} = 512^3$ neutrino particles at $z_c=19$; the enhancements at $k = 0.5 \, h$/Mpc match the predictions of equation~\eqref{eq:NLfactorfit}---marked by the horizontal dashed lines---exactly.

%%%%%%%%%%%%%%%%%%%%%%%%%%%%%%
%%%%%%%%%%%%%%%%%%%%%%%%%%%%%%%

\subsection{Enhancement to the total neutrino power}

The nonlinear enhancement relation~\eqref{eq:NLfactorfit} gives the fractional increase in power a particular neutrino fluid flow $\alpha$ will experience at $k=0.5 \, h/\text{Mpc}$, relative to the MFLR prediction.  This relation can also be turned into a statement about the extent to which a particular neutrino flow will contribute to enhancing the {\it total} neutrino power spectrum  constructed from summing the monopole density contrasts of all converted and unconverted flows.

Specifically, we are interested in the ratio of the total neutrino density contrast at $z=0$ constructed from a hybrid-neutrino simulation with some flows converted to particles, i.e., $\sqrt{P_\nu^{\rm (hybrid)}(k)}$ as given in equation~\eqref{eq:nupower},
to the MFLR expectation $\sqrt{P_\nu^{\rm (MFLR)}(k)}$ had we not converted these flows to particles. 
That is,
\begin{equation}
   \Gamma_{\rm NL}(k) \equiv  \sqrt{\frac{P^{({\rm hybrid})}_{\nu}}{P^{\rm (MFLR)}_\nu}}(k) 
    =   \left|1 + \frac{1}{N_{\tau}} \frac{\sum_{\rm conv.}\delta^{(\rm part)}_\alpha(k) -\sum_{\rm unconv.} \delta_{\alpha,\ell=0}(k)}{\sum_{\alpha =1}^{N_\tau}\delta_{\alpha,\ell=0}(k)} \right|\, ,
\label{eq:NuMonopoleNLRatio}
\end{equation}
where we have ignored any small phase differences between the converted and unconverted flows.
At $k = 0.5\, h$/Mpc the ratio can also be expressed as
\begin{equation}
\Gamma_{\rm NL}(k = 0.5 \, h/\text{Mpc}) 
    =   \left|1 + \frac{1}{N_{\tau}} \frac{\sum_{\rm conv.} \delta_{\alpha,\ell=0}(k)
    \left(\sqrt{\gamma_{\text{NL}}(\Delta^{2~{\rm (MFLR)}}_{\alpha,{\rm max}})} - 1 \right)
    }{\sum_{\alpha =1}^{N_\tau}\delta_{\alpha,\ell=0}(k)} \right|\, ,
\label{eq:NuMonopoleNLRatioAlpha}
\end{equation}
where $\gamma_{\rm NL}$ is the nonlinear enhancement factor for $\Delta_{\alpha}^2(k=0.5\,h/{\rm Mpc}) \propto |\delta_\alpha(k)|^2$ given in equation~\eqref{eq:NLfactorfit}, and the summation now applies to the MFLR monopole densities $\delta_{\alpha \ell=0}$ of those flows that had been converted to particles in the hybrid-neutrino simulation.

%%%%%%%%%
\begin{figure}[t]
    \centering
    \includegraphics[trim=2mm 2mm 2mm 2mm,clip,width=130mm]{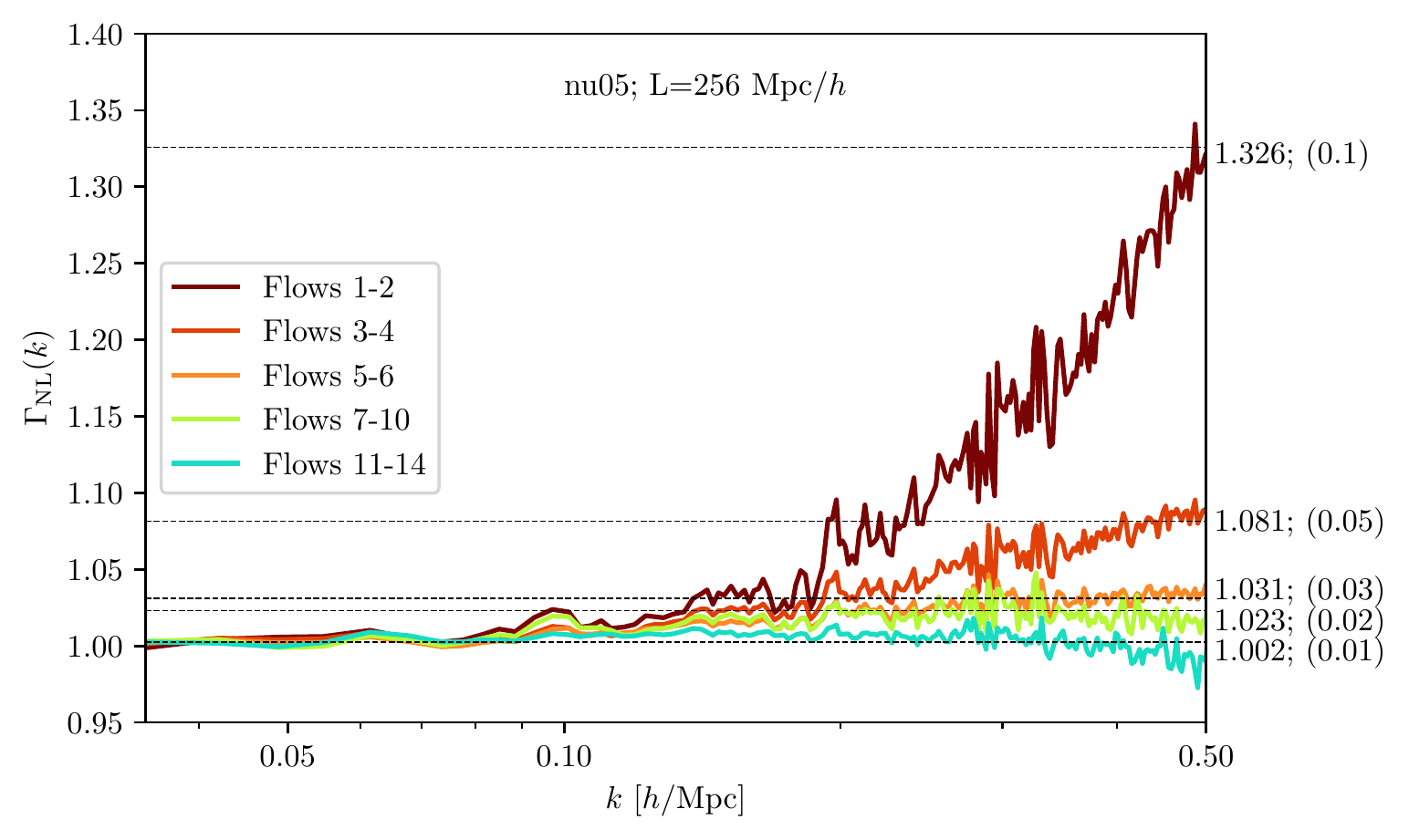}
    \caption{Nonlinear enhancement to the total neutrino density contrast at $z=0$, defined in equation~\eqref{eq:NuMonopoleNLRatio}, that would be gained from converting select flows in the nu05 cosmology to hybrid-neutrino simulations, over the MFLR expectation. 
     The black dashed lines mark the enhancement factor 
     $\Gamma_{\rm NL}(k=0.5\, h/{\rm Mpc})$ for each flow as predicted by equation~\eqref{eq:NuMonopoleNLRatioAlpha} given the flow's $\Delta^{2~{\rm (MFLR)}}_{\alpha,{\rm max}}$; their values are indicted to the right of the plot area.
    \label{fig:nu05NLCriteriaTest}}
\end{figure}
%%%%%%%%%%%

Figure~\ref{fig:nu05NLCriteriaTest} shows the nonlinear enhancement to the total neutrino density contrast~\eqref{eq:NuMonopoleNLRatio} at $z=0$ that would be gained  from the individual conversion of the five representative flows of the nu05 cosmology summarised in figure~\ref{fig:nu05DeltaGroupedCritera} and table~\ref{tab:repflow}.   
The enhancement at $k = 0.5\, h$/Mpc as predicted by  equation~\eqref{eq:NuMonopoleNLRatioAlpha} given the flow's MFLR dimensionless power spectrum at its maximum, $\Delta^{2~{\rm (MFLR)}}_{\alpha,{\rm max}}$, are marked by the horizontal dashed lines.  We note that the enhancement at $k\lesssim 0.5\, h$/Mpc ranges from $0.2$\% arising from conversion of the fastest flows, to a $33$\% gain from converting the slowest, with the middle three flows gaining between $2$\% to $8$\%.   On this basis, we can conclude that excluding neutrino flows with $\Delta^{2~{\rm (MFLR)}}_{\alpha,{\rm max}} \lesssim 0.01$ from conversion to particles will not compromise the total neutrino density contrast (and hence power spectrum) at $z=0$ beyond the sub-percent level.

On the other hand, those neutrino flows with $\Delta^{2~{\rm (MFLR)}}_{\alpha,{\rm max}} \gtrsim 0.01$ can be considered for an $N$-body particle representation depending on the accuracy threshold set by the user. For example, converting only the first representative flow---which has $\Delta^{2~{\rm (MFLR)}}_{\alpha,{\rm max}} \simeq 0.1$---will give a total neutrino density contrast accurate to about $\sim 10\%$ at $k \lesssim 0.5 \, h/\text{Mpc}$ according to equation~\eqref{eq:NuMonopoleNLRatioAlpha} and figure~\ref{fig:nu05NLCriteriaTest}, while the total neutrino power spectrum will be underestimated by about $\sim 18$\% following the same argument.

Lastly, while we have formulated our nonlinearity estimates~\eqref{eq:NuMonopoleNLRatioAlpha} in terms of a flow's $\Delta^{2~{\rm (MFLR)}}_{\alpha,{\rm max}}$ value, it is possible to translate the dependence to the flow's Lagrangian velocity, i.e., $v_\alpha$ at $a=1$, instead, via the approximate relation
\begin{equation}
     \Delta^{2~{\rm (MFLR)}}_{\alpha,{\rm max}}\simeq \left(\frac{100 \, \text{km} \,  s^{-1}}{v_{\alpha}|_{a=1}} \right)^{\frac{1}{0.52}} \, .
\end{equation}
Applying this relation to the nu05 cosmology in conjunction with equation~\eqref{eq:NuMonopoleNLRatioAlpha}, we find that converting neutrino flows with $v_{\alpha}|_{a=1} \lesssim 600 \, \text{km} \, s^{-1}$  into $N$-body particles will yield a total neutrino density contrast accurate to within 5\% at $k \lesssim 0.5 \, h/\text{Mpc}$  barring transients.

%%%%%%%%%%%%%%%%%%%%%%%%%%%%%%%%%%%%%%%%%%%%%%%%%%%%%%%%%%%%%%%%%%%%%%%%%%%%%%%%
%%%%%%%%%%%%%%%%%%%%%%%%%%%%%%%%%%%%%%%%%%%%%%%%%%%%%%%%%%%%%%%%%%%%%%%%%%%%%%%%

%%%%%%%%%%%%%%%%%%%%%%%%%%%%%%%%%%%%%%%%%%%%%%%%%%%%%%%%%%%%%%%%%%%%%%%%%%%%%%%%
\section{Multiple conversions}%%%%%%%%%%%%%%%%%%%%%%%%%%%%%%%%%%%%%%%%%%%%%%%%%%
\label{sec:multiple_conversions}%%%%%%%%%%%%%%%%%%%%%%%%%%%%%%%%%%%%%%%%%%%%%%%%

Thus far we have considered only isolated conversions to particles of small batches of neutrino flows grouped into representative flows.  Now we arrive at our primary objective, the total nonlinear neutrino power spectrum.  We begin by computing the neutrino power $P_\nu(k)$ and assessing its impact upon the power spectra of the total matter $P_{\rm m}(k)$ and the cold matter $P_{\rm cb}(k)$.  Next we quantify the effect of converting one representative flow on other representative flows.  The smallness of this interaction gives us the option of either: (i) converting multiple representative flows at once, in a single simulation with a large number of neutrino particles, or (ii) running multiple simulations of isolated conversion, each one converting a different representative flow into particles.  The latter is more time-consuming but less memory-intensive.  Finally, we consider particle conversions staggered over multiple redshifts.   The run times shown in table~\ref{tab:timing} suggest that delaying the conversions of faster, hence more linear, flows may save some computation time.  However, this saving may be at the cost of additional low-$k$ transient errors, as we have seen in section~\ref{sec:ConvRedshiftCriteria}.

\subsection{Total nonlinear power}%--------------------------------------------
\label{subsec:total_NL_P}

%%%%%%%%%%%%%%
\begin{figure}[t]
	\centering
	\includegraphics[trim=5mm 2mm 10mm 5mm,clip,width=75mm]{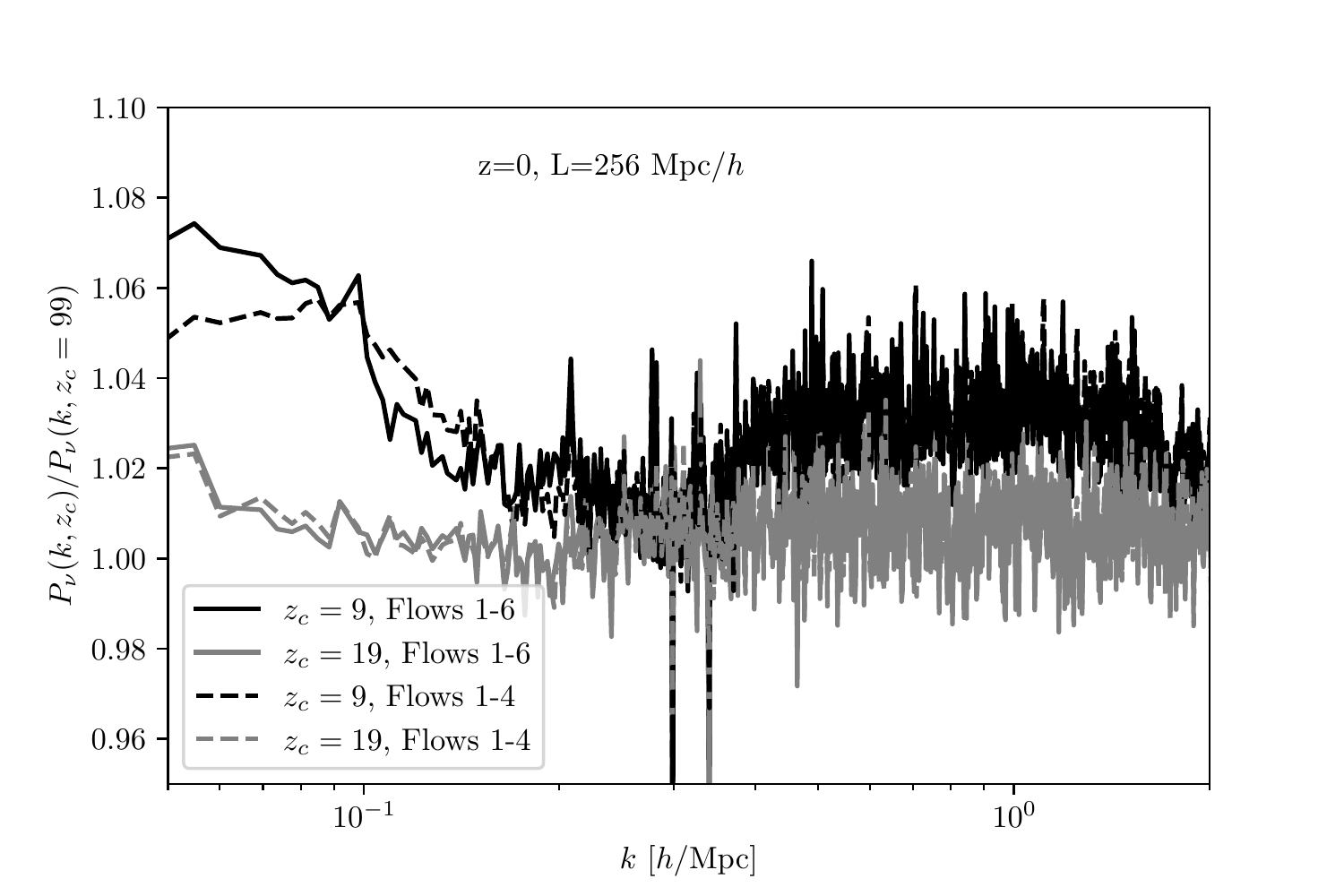}
	\includegraphics[trim=5mm 2mm 10mm 5mm,clip,width=75mm]{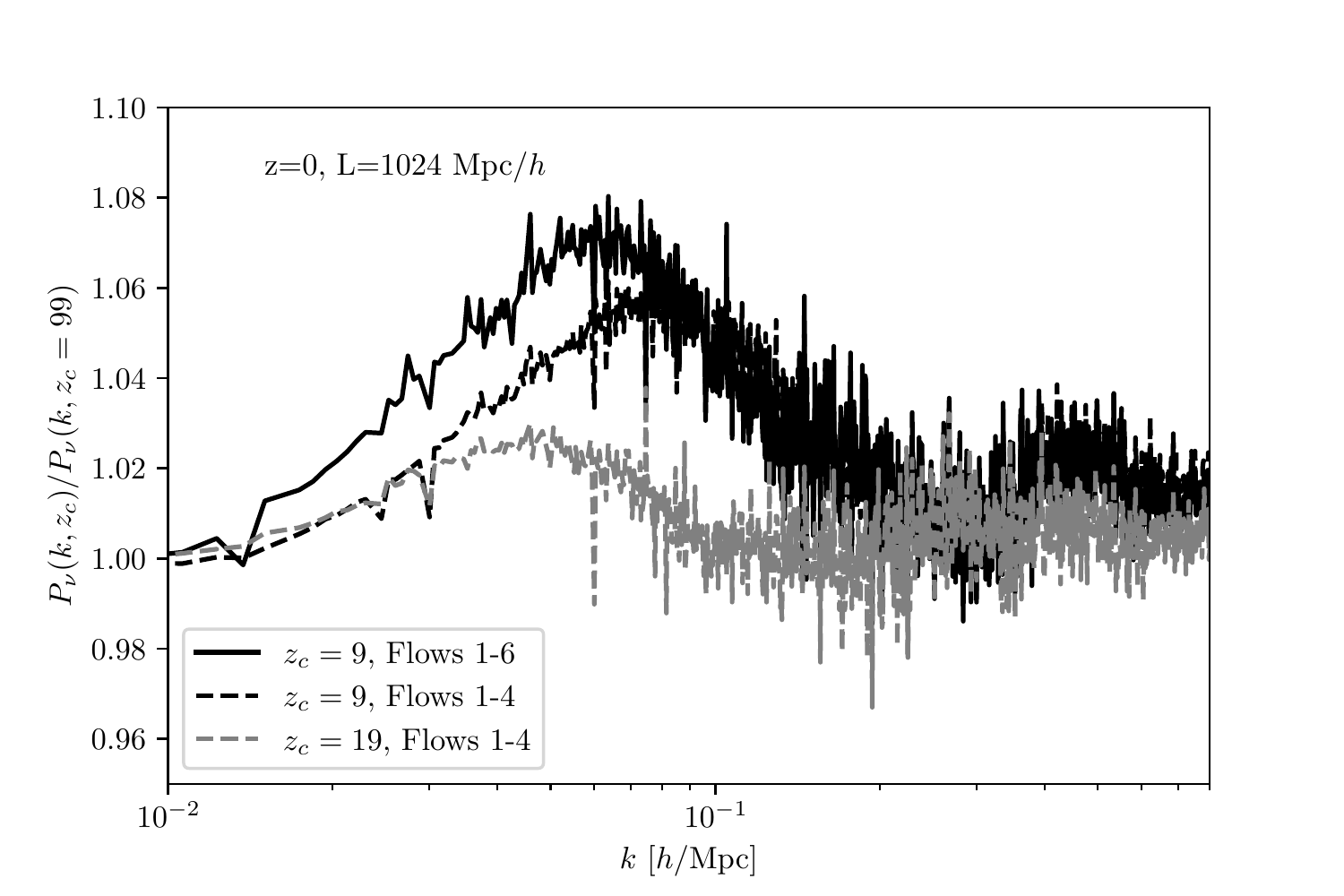}
	\caption{Total neutrino power spectrum at $z=0$ for the nu05 cosmology, computed from a hybrid-neutrino simulation converting up to the slowest three representative flows ($\alpha =1,\ldots,6$) at two different common conversion redshifts, $z_c = 9$ (grey) and $19$ (black).  The spectrum is shown normalised to the case of conversion at $z_c = 99$. The left panel employs a simulation box of side length $L_{\rm box} = 256\, {\rm Mpc}/h$ box, while the right panel shows the $L_{\rm box} = 1024 \, {\rm Mpc}/h$  run. 
 \label{fig:MultiplConversionTransients}}
\end{figure}
%%%%%%%%%%%%%%
%%%%%%%%%%%%%%%

As shown in section~\ref{sec:NLCriteria}, in the nu05 cosmology only the first six out of $N_\tau=20$ flows have significant nonlinearities.  Thus, we compute the nonlinear neutrino power by converting these six, in three representative flows, while using MFLR to track flows 7 to 20.  Using the nonlinear enhancement criterion of equation~\eqref{eq:NuMonopoleNLRatioAlpha}, we expect this procedure to be accurate at the $5\%$ level in the total nonlinear neutrino power spectrum.   As in section~\ref{sec:NLCriteria} we choose a conversion redshift of $z_c=19$. Figure~\ref{fig:MultiplConversionTransients} shows that this choice of $z_c$ limits transient effects in the $z=0$ total neutrino power spectrum $P_\nu(k)$ to the $\lesssim 3$\% level.

%%%%%%%%%%%%%%%%%%%
\begin{figure}[t]
	\centering
	\includegraphics[trim=2mm 2mm 13mm 5mm,clip,width=120mm]
	{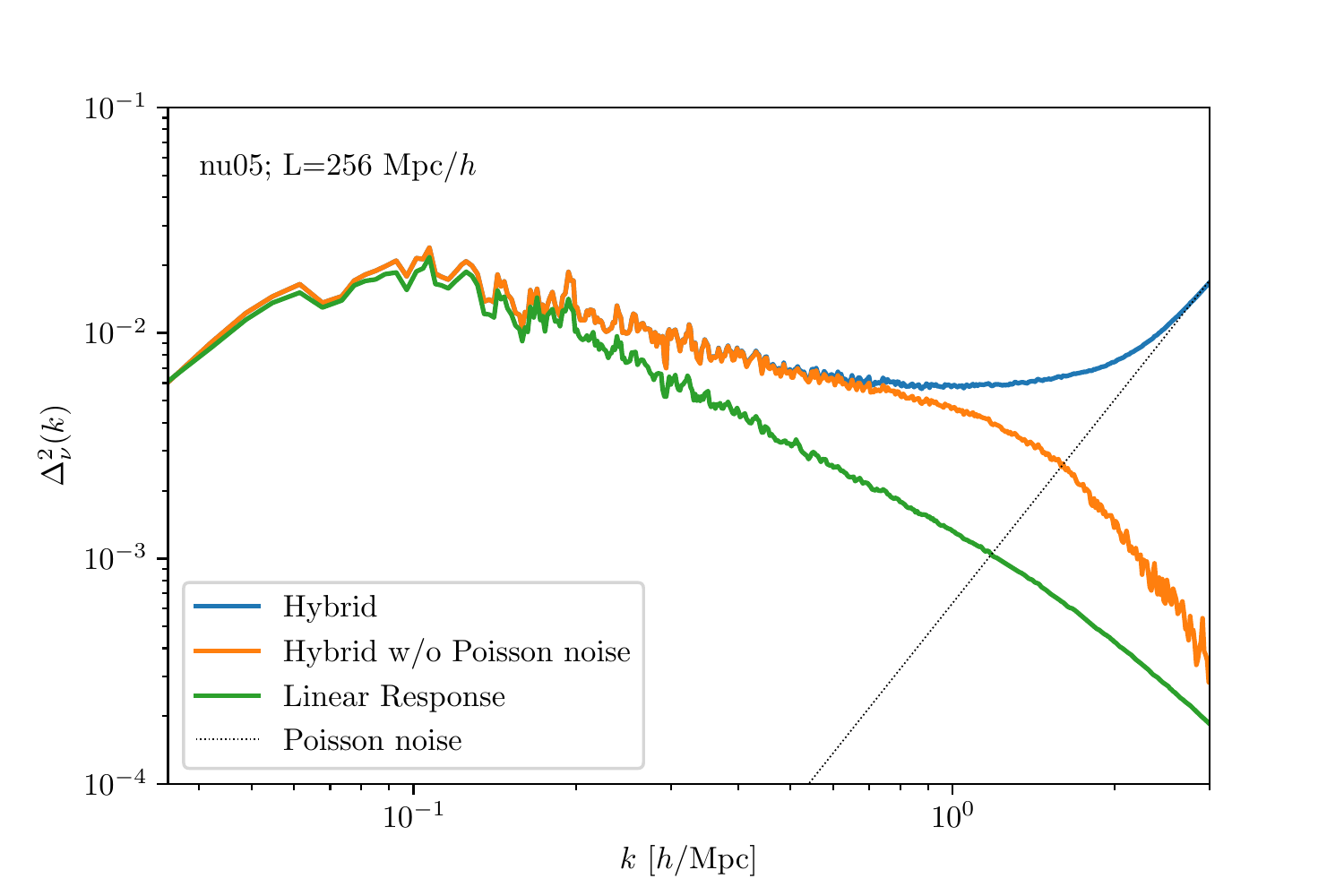}
	\includegraphics[trim=2mm 2mm 13mm 5mm,clip,width=120mm]{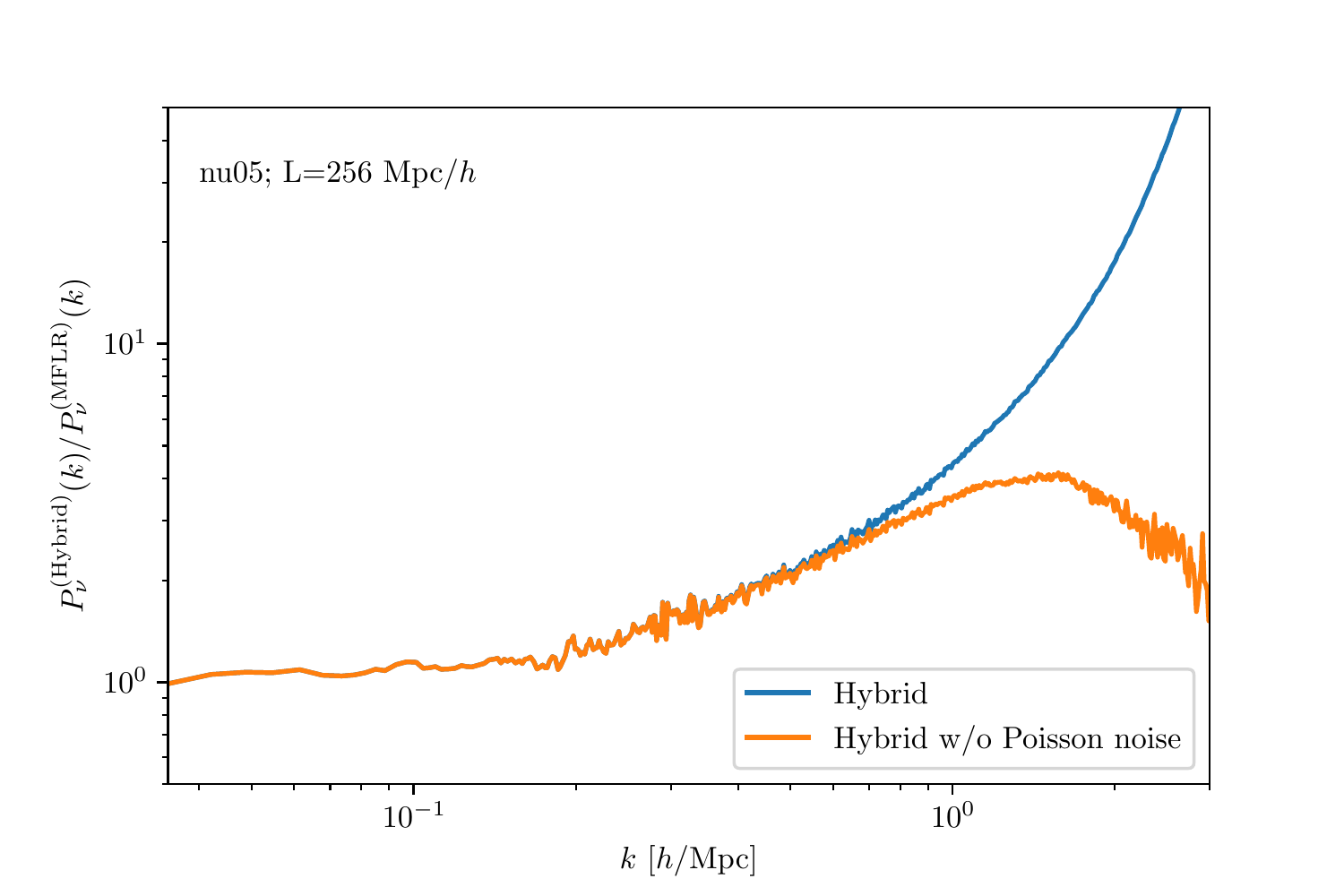}
	\caption{Dimensionless $z=0$ neutrino density power spectra, $\Delta_\nu (k) \equiv k^3 P_\nu(k)/(2 \pi^2)$, using MFLR (green),
          hybrid-neutrino simulations with (blue) and without (orange) Poisson shot noise, and
          the approximate Poisson noise (black)
          for the nu05 cosmology.
          The top panel shows all four power spectra, where Poisson shot noise can be clearly seen to dominate the hybrid-neutrino power spectrum at $k \gtrsim 1.5~h/$Mpc.  The bottom panel shows the nonlinear enhancements of the hybrid-neutrino power spectra over
          their MFLR counterpart.  Subtracting away Poisson noise reveals a peak net enhancement of $\sim 4$ at $k \sim 1.5~h/$Mpc.
          \label{fig:nu05PnuMonopoleEnhancements}
        }
\end{figure}
%%%%%%%%%%%%%%%%%

We compare in figure~\ref{fig:nu05PnuMonopoleEnhancements} our hybrid-neutrino power spectrum against that from a pure MFLR neutrino treatment and the approximate shot noise in the particle neutrino power.  Evidently, shot noise dominates the $z=0$ power at $k \gtrsim 1.5~h/$Mpc.  While we can subtract the shot noise from the result, a residual error is expected to remain on theses scales.  We therefore restrict our consideration to $k \lesssim 1~h/$Mpc.  In this range, we see that neutrino nonlinearity enhances the total shot-noise-subtracted neutrino power spectrum by factors of 3 to 4 relative to its MFLR counterpart.   This is one of the main results of this work.

%%%%%%%%%%%
\begin{figure}[t]
	\centering
	\includegraphics[trim=2mm 2mm 13mm 5mm,clip,width=120mm]{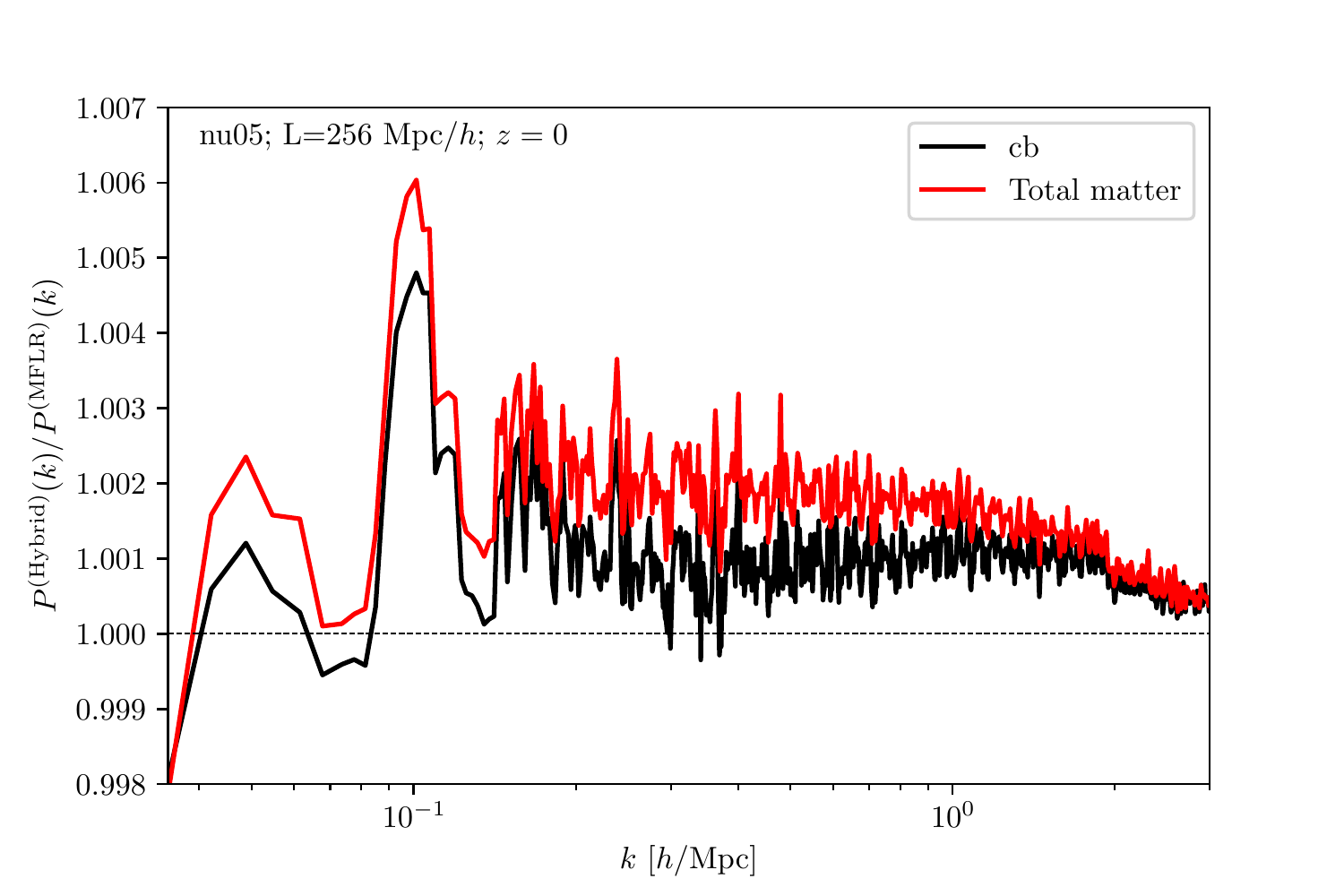}
	\caption{Ratios of the total matter (red) and the cold matter (black) power spectra at $z=0$ from the hybrid-neutrino run with flows $\alpha=1,\ldots,6$ converted to $N$-body particles at $z_c=19$, against their MFLR counterparts. The enhancements due to the presence of nonlinear neutrino perturbations are smaller than $\sim 0.2\%$ for the total matter power spectrum $P_{\rm m}(k)$, and smaller than $\sim 0.1\%$ for the cold matter power spectrum $P_{\rm cb}(k)$. 
\label{fig:CBMatterEnhancement} }
\end{figure}
%%%%%%%%%%%%%%%%%%

The enhanced neutrino clustering will enter directly into the total matter density contrast and hence increase the total matter power spectrum.  Furthermore, the cold matter power can also be expected to increase as a result of an amplified gravitational potential.  To estimate the size of these effects, we note that both the neutrino fraction $f_\nu \simeq 3.5\%$ and the clustering ratio $\delta_\nu / \delta_{\rm m} \simeq k_{\rm FS}^2/k^2$ are small in the region where the neutrino enhancement is largest, with their product being $\lesssim 0.1\%$ at $k \simeq 1~h/$Mpc; nonlinear neutrino enhancements to $P_{\rm m}(k)$ and $P_{\rm cb}(k)$ should also be of this order of magnitude.  Figure~\ref{fig:CBMatterEnhancement} confirms this expectation:  nonlinear neutrino corrections to $P_{\rm m}(k)$ and $P_{\rm cb}(k)$ spectra peak at $0.6\%$ and $0.5\%$, respectively, around the free-streaming scale, before diminishing to $0.2\%$ and $0.1\%$ at $k \gtrsim 0.2~h/$Mpc.  At the smallest scales shown, $k \gtrsim 1~h/$Mpc, neutrino enhancements to $P_{\rm m}(k)$ and $P_{\rm cb}(k)$ are further suppressed as the clustering ratio $\delta_\nu / \delta_{\rm m}$ rapidly diminishes.

%%%%%%%%%

\subsection{Interactions between converted flows}%------------------------------
\label{subsec:interactions_between_converted_flows}

%%%%%%%%%%%%%%%
\begin{figure}[tbp]
  \begin{center}
    \includegraphics[trim=2mm 2mm 13mm 5mm ,clip,width=120mm]{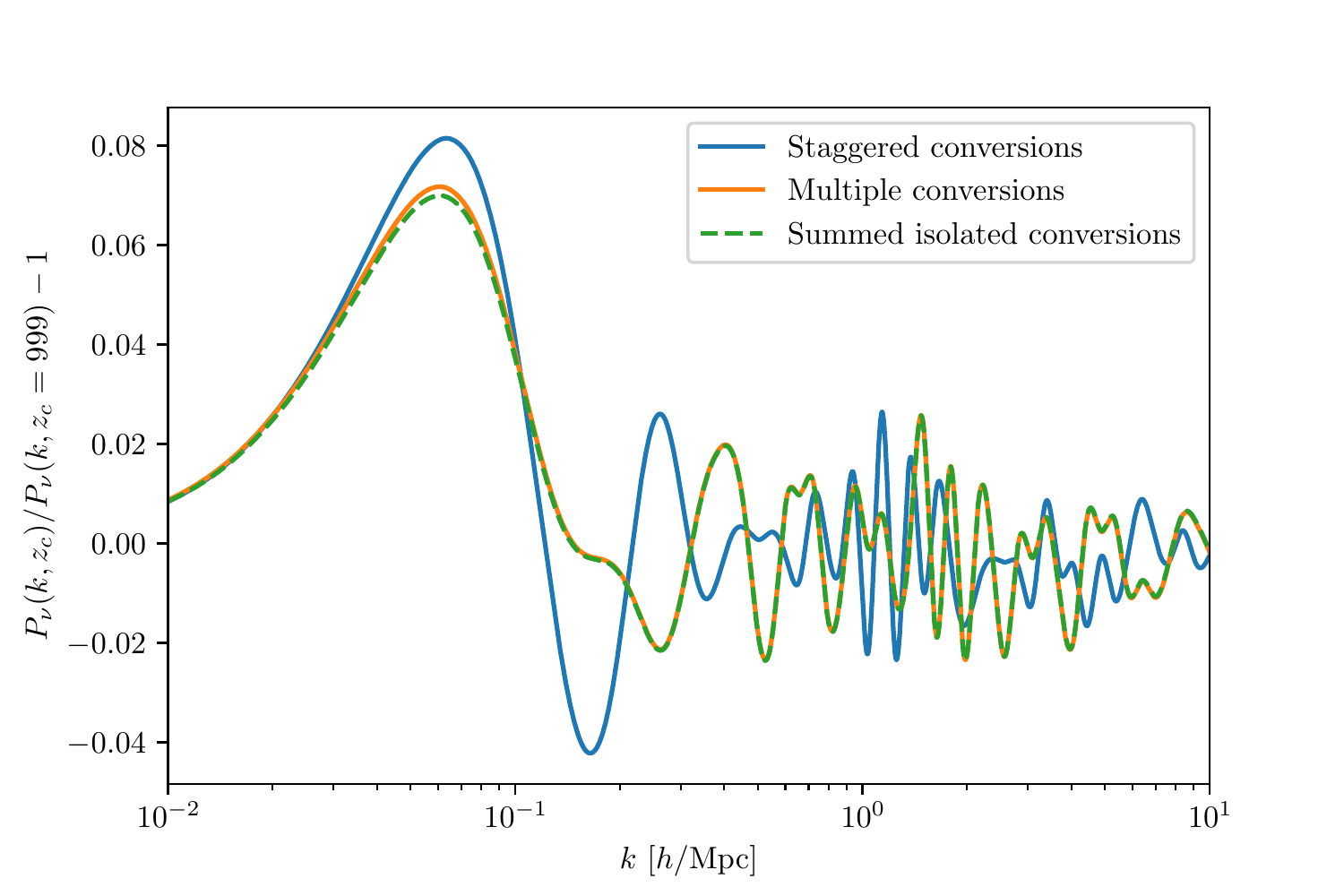}%
  \end{center}
    \caption{Fractional error in the total neutrino power spectrum at $z=0$ computed from a 
    MFLR/Time-RG mimic of neutrino simulations using the three different multiple MFLR-to-particle conversion procedures described in
    section~\ref{sec:multiple_conversions}.  Isolated (green) and multiple simultaneous (orange)
    conversions take place at $z_c=9$ for all flows $\alpha = 1, \ldots,6$.  Staggered conversions (blue) occur at $z_c=19$ for
    flows 1--2, $z_c=9$ for flows 3--4, and $z_c=5$ for flows 5--6. 
    \label{fig:transients_SIC_MC_SC}
  }
\end{figure}
%%%%%%%%%%%%%%%%%%

Different neutrino flows interact with one another only through the gravitational potential $\Phi \propto P_{\rm m}^{1/2}$, which is itself enhanced by only $\sim 0.1\%$ through the conversion of MFLR flows into particles.  Thus, the impact of one MFLR-to-particle conversion on another should be well below a percent of its power.  Negligibility of the interactions between flows opens up a powerful new technique for simulating neutrino clustering, that of summing isolated conversions, which is unique to multi-fluid treatments of neutrino clustering.  Rather than converting all nonlinearly-clustering flows at once in a single simulation, we may run multiple simulations, each of which converts just one group of flows in a representative flow, and then sum the individual density perturbations.  

This technique has parallels with the $N$-one-body approach~\cite{Ringwald:2004np}, and its power lies in its ability to use a large number of particles to realise that single group of flows, thereby reducing its shot noise.  Suppose for example that the availability of memory on a computer cluster limits our simulation to $1024^3$ neutrino particles.  Rather than converting, say, eight groups of flows all at once, each of which is represented by $512^3$ particles and has shot noise $\epsilon \sim V_{\rm box}/ 512^3$, we may run eight separate simulations, each of which uses all $1024^3$ particles for a single group of flows, and hence has a shot noise $\epsilon \sim V_{\rm box}^3 / 1024^3$ eight times smaller.

Figure~\ref{fig:transients_SIC_MC_SC} directly bounds the interaction between converted flows using a MFLR/Time-RG mimic of multiple simultaneous conversions and summed isolated conversions for the nu05 cosmology.  Once again, as in section~\ref{sec:mimic}, conversions are mimicked by setting to zero all Legendre moments $\ell > 0$ for the appropriate flows at the conversion redshift.  The resulting $z=0$ power spectrum is compared with a pure MFLR/Time-RG power spectrum with no mimic conversions (i.e., $z_c=999$).  Evidently, power spectrum errors are limited to $3\%$ at and below the free-streaming length scale, $k \geq 0.1~h/$Mpc.  Moreover, these errors are nearly identical for multiple simultaneous conversions and summed isolated conversions, agreeing everywhere to better than $0.2\%$ of the power spectrum.

%%%%%%%%%%%%%%%%%%%%%%%%%%%%

\subsection{Staggered conversions}%---------------------------------------------
\label{subsec:staggered_conversions}

\begin{figure}[t]
	\centering
	\includegraphics[trim=5mm 2mm 11mm 5mm,clip,width=75mm]{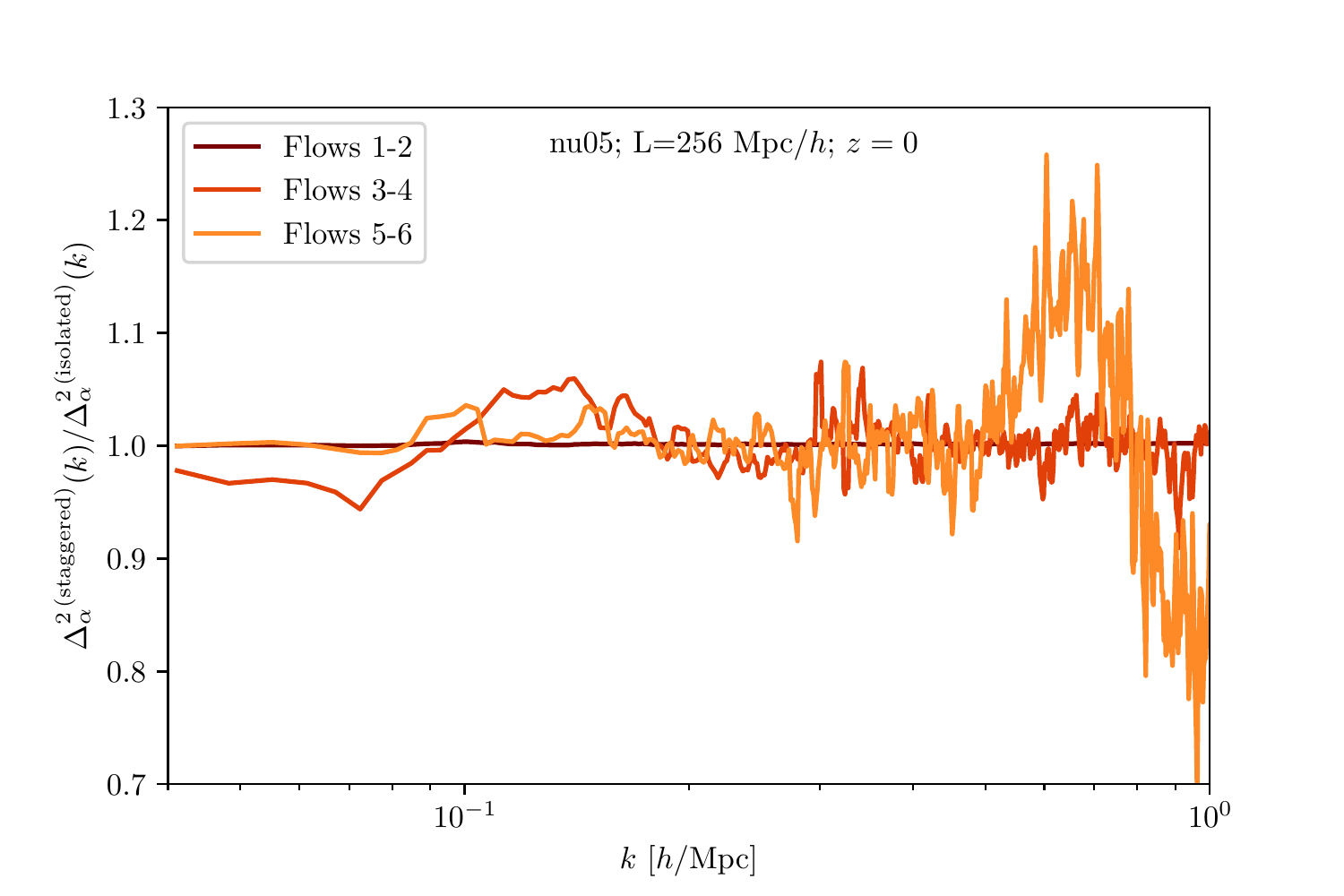}
	\includegraphics[trim=4mm 2mm 12mm 5mm,clip,width=75mm]{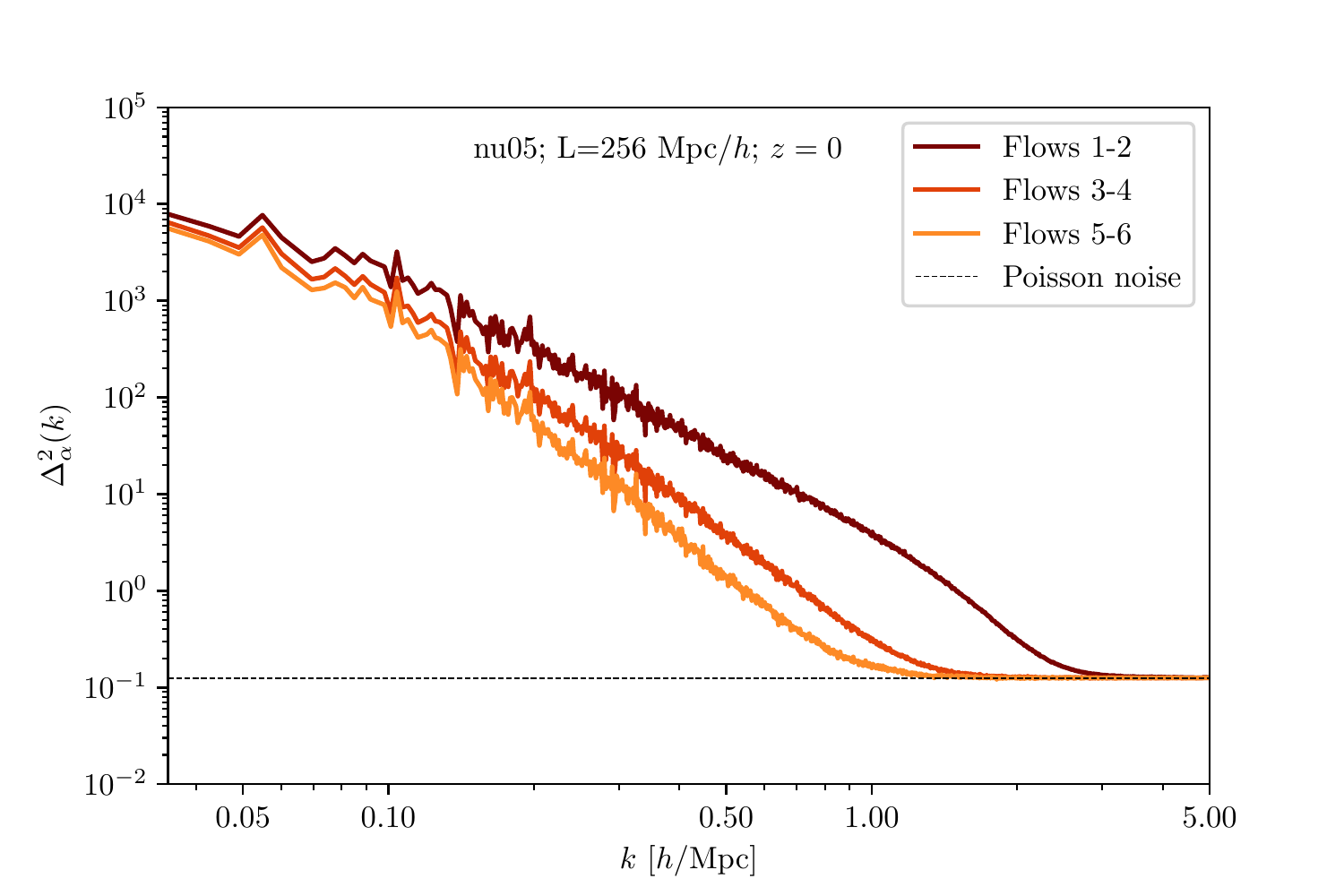}
	\caption{Dimensionless power spectra of the three slowest representative flows at $z=0$ of the nu05 cosmology resulting from staggered MFLR-to-particle conversions at $z_c=\{19,9,5\}$ from low to high $\tau_\alpha$.  The left panel shows these flow power spectra normalised to their counterparts from isolated conversions at the corresponding conversion redshift.
    The oscillatory behaviours in flows 3-4 and flows 5-6 arise from the different perturbation phases used the initialise the neutrino particles between the staggered and the isolated method: in an isolated conversion the neutrino particles see only the phases of the cold matter perturbations, whereas in a staggered conversion the neutrino particles already in the box also contribute to the initial phases of a conversion.
	 The large deviation in flows 5-6 at 
	$k \gtrsim 0.7 \, h/\text{Mpc}$ is an artefact of Poisson noise, which dominates over signal in this $k$ range, as shown in the right panel.
The conclusion is that staggered MFLR-to-particle conversion induces no further significant   nonlinear enhancement to the clustering power of the individual flows.
 \label{fig:nu05IndpVsMulti}}
\end{figure}

Multi-fluid neutrino treatments make possible yet another method for optimising $N$-body simulations, that of staggered conversions into particles at multiple redshifts in the manner of figure~\ref{fig:HybridWorkflowChart}. As shown in table~\ref{tab:timing}, faster-moving neutrinos are generally more computationally expensive at high redshifts due to the smaller time steps required to track their motion as well as the greater number of particles necessary for controlling shot noise.  However, their nonlinear clustering is also negligible until lower redshifts.  Thus, we may in principle convert higher-$\tau_\alpha$ flows into particles at lower~$z$, reducing the computational cost of the simulation with a limited impact on its accuracy.  Since MFLR perturbation theory is most accurate for precisely these fast flows, staggered MFLR-to-particle conversion at multiple redshifts efficiently combines particle and perturbative methods so as to play to the strengths of each.

At present, the principal limitations on our use of staggered conversions are transient errors associated with missing $\ell>0$ moments in our neutrino particle initialisation procedure.  The MFLR/Time-RG mimic in figure~\ref{fig:transients_SIC_MC_SC} quantifies transient errors for the three particle conversion methods studied in this section: summed isolated conversion of flows 1--2, 3--4, and 5--6 at $z_c=9$,  multiple simultaneous conversions of the same three groups of flows at $z_c=9$, staggered conversions of the three groups at $z_c =\{19,9,5\}$ respectively.

Transient errors may be divided into two regimes: large-scale power overestimates at $k \lesssim k_{\rm FS}$, and small-scale oscillatory phase errors at $k > k_{\rm FS}$.  Neglecting the $\ell>0$ power results in excessive clustering on all scales immediately after conversion to particles.  Because the $\ell$th multipole moment at wave number~$k$ can be regenerated from the $(\ell-1)$th multipole on a time scale $\Delta_{\rm repop} \sim m_\nu / (k \tau_\alpha)$, the power excess on small scales quickly dissipates, so free-streaming transient errors are typically not an issue at $z=0$.  Although our mimic does not capture the nonlinear enhancement of these initially-large errors at very high $k$, figure~\ref{fig:MultiplConversionTransients} allows us to bound the total small-scale $z=0$ power spectrum error to $<2\%$ and $<4\%$ for $z_c$ of $19$ and $9$, respectively.  

On large scales, however, regeneration of the $\ell>0$ multipoles may not be complete by $z=0$, especially if conversion happens at a low redshift.  Thus, as shown in figure~\ref{fig:transients_SIC_MC_SC},
errors in the staggered conversion run are dominated by the last conversion at $z_c=5$: at $k \simeq 0.06~h/$Mpc, we find an $8\%$ error for staggered conversions, versus $7\%$ for summed isolated and  multiple simultaneous conversions. Moreover, staggered-conversion errors fall more slowly with wave number.

Nevertheless, staggered-conversion transient errors do eventually fall below 3\% in the range $k \gtrsim 0.2~h/$Mpc, comparable to those in the other two multiple-conversion runs. This is due partly to the fact that these small-scale errors are oscillatory in $k/k_{{\rm FS},\alpha}$, so that summation over multiple flows as in figure~\ref{fig:transients_SIC_MC_SC} reduces their overall magnitude.  Thus, our MFLR/Time-RG mimic demonstrates staggered conversion to be an efficient simulation method that accurately reproduces $P_\nu(k)$ at small scales.

Lastly, figure~\ref{fig:nu05IndpVsMulti} examines potential nonlinear interactions between flows in the case of staggered conversion at the {\it flow-by-flow} level.  Here, we compare the $z=0$ flow dimensionless power spectra $\Delta_\alpha^2(k)$ of the slowest three representative flows from a staggered-conversion simulation converting flows 1--2, 3--4, and 5--6 at $z_c=\{19,9,5\}$ respectively, to isolated conversions of the same three flows, again at $z_c=19$ for flows 1--2, etc.  Where the simulation outcome does not suffer from excessive Poisson noise, we observe a small amount ($<5$\%) of noise associated with the small phase differences between initialising neutrino particles with only cold particles in the simulation box and initialising with both cold matter and previously-converted neutrino particles present.  There is however no significant nonlinear power enhancement in the later-converted flows in the staggered-conversion run, indicating no significant interactions between the flows.  In view of the issue of transients associated with low-redshift conversions, this result in fact lends support to the method of isolated conversion at a reasonably ``high'' redshift, e.g., $z_c=19$, if accuracy in the power spectra of the individual flows (rather than their summation) is desired.

In summary, staggered MFLR-to-particle conversions at low redshifts are useful for computing the total neutrino power spectrum $P_\nu(k)$, particularly at small scales $k \gtrsim k_{\rm FS}$, where nonlinear corrections are most significant.  This is the primary goal of $N$-body neutrino simulations.  However, an application demanding accurate momentum resolution in the neutrino distribution will not benefit from the cancellation of small-scale errors in the summed power spectrum.  Given that interactions between flows are insignificant, for such applications that demand accuracy in the power spectra of the individual flows, we suggest isolated conversions at $z_c \gtrsim 19$ for all flows with significant nonlinearities.

%%%%%%%%%%%%%%%%%%%%%%%%%%%%%%%%%%%%%%%%%%%%%%%%%%%%%%%%%%%%%%%%%%%%%%%%%%%%%%%%
%%%%%%%%%%%%%%%%%%%%%%%%%%%%%%%%%%%%%%%%%%%%%%%%%%%%%%%%%%%%%%%%%%%%%%%%%%%%%%%%
%%%%%%%%%%%%%%%%%%%%%%%%%%%%%%%%%%%%%%%%%%%%%%%%%%%%%%%%%%%%%%%%%%%%%%%%%%%%%%%%
%%%%%%%%%%%%%%%%%%%%%%%%%%%%%%%%%%%%%%%%%%%%%%%%%%%%%%%%%%%%%%%%%%%%%%%%%%%%%%%%
%%%%%%%%%%%%%%%%%%%%%%%%%%%%%%%%%%%%%%%%%%%%%%%%%%%%%%%%%%%%%%%%%%%%%%%%%%%%%%%%
%%%%%%%%%%%%%%%%%%%%%%%%%%%%%%%%%%%%%%%%%%%%%%%%%%%%%%%%%%%%%%%%%%%%%%%%%%%%%%%%
%%%%%%%%%%%%%%%%%%%%%%%%%%%%%%%%%%%%%%%%%%%%%%%%%%%%%%%%%%%%%%%%%%%%%%%%%%%%%%%%
%%%%%%%%%%%%%%%%%%%%%%%%%%%%%%%%%%%%%%%%%%%%%%%%%%%%%%%%%%%%%%%%%%%%%%%%%%%%%%%%
%%%%%%%%%%%%%%%%%%%%%%%%%%%%%%%%%%%%%%%%%%%%%%%%%%%%%%%%%%%%%%%%%%%%%%%%%%%%%%%%
%%%%%%%%%%%%%%%%%%%%%%%%%%%%%%%%%%%%%%%%%%%%%%%%%%%%%%%%%%%%%%%%%%%%%%%%%%%%%%%%

%%%%%%%%%%%%%%%%%%%%%%%%%%%%%%%%
%%%%%%%%%%%%%%%%%%%%%%%%%%%%%%%%%

%%%%%%%%%%%%%%
\section{Comparison with \concept{}}
\label{sec:concept}

In this final section, we compare the results of our hybrid-neutrino approach with those of the $N$-body code~\concept{}~\cite{Dakin:2015uka,Dakin:2017idt,Dakin:2021ivb}.  \concept{} is a  P$^3$M code for cold matter.  However, unlike hybrid-neutrino or traditional particle-based simulations, \concept{} uses a fluid description for the relic neutrino population, where fluid quantities such as 
the density, velocity, pressure, and anisotropic stress are tracked on a grid.  A similar method was also presented earlier in~\cite{Banerjee:2016zaa}. Grid-based approaches circumvent the issue of Poisson noise due to sampling with discrete particles.  The price, however, is that the infinite number of degrees of freedom is now reexpressed as an infinite fluid moment hierarchy, and specifying a closure and/or truncation condition requires some care to keep the system tractable without losing accuracy.

In the \concept{} implementation~\cite{Dakin:2015uka,Dakin:2017idt,Dakin:2021ivb}, only the neutrino density~$\rho$  and velocity~$u^i$ are tracked by nonlinear fluid equations; the pressure $P$ and anisotropic stress $\sigma^i_j$ are determined by linear closure conditions.  For numerical stability \concept{} uses  ``conserved'' versions of these quantities, defined as $\varrho \equiv a^{3(1 + \mathscr{w})}\rho$, $J^i \equiv a^4(\rho + P)u^i$, $\mathcal{P} \equiv a^{3(1+\mathscr{w})}P$, and $\varsigma^i_j \equiv (\varrho + \mathcal{P})\sigma^i_j$, where
\begin{equation}
	\mathscr{w}(a) \equiv \frac{1}{\ln a}\int_1^a \frac{w(a')}{a'}\, \mathrm{d}a' \,
\end{equation}
is effective equation of state parameter of the neutrino fluid.  Then, the  nonlinear fluid equations for $\varrho$ and $J^i$ are given by
\begin{equation}
\begin{aligned}
	\dot{\varrho} & =\,  -a^{3\mathscr{w}-1}\partial_i J^i 
	+ 3aH(w\varrho - \mathcal{P}), \\
	\dot{J}^i & =\, -\partial^j\biggl[a^{3\mathscr{w} - 1}\frac{J^iJ_j}{\varrho + \mathcal{P}} + a^{-3\mathscr{w}+1}\varsigma^i_j\biggr] 
    - a^{-3\mathscr{w} + 1}\partial^i\mathcal{P} 
    - a^{-3\mathscr{w} + 1}\bigl(\varrho + \mathcal{P}\bigr)\partial^i\psi\, ,
\label{eq:euler_concept}
\end{aligned}
\end{equation}
under the assumption that $\psi=\phi$, and neglecting terms of order $\partial_i \phi u^i$ and higher.  The closure conditions are implemented in the form
\begin{equation}
\begin{aligned}
\label{eq:closure}
	\delta P (\vec{k}) & \simeq \, \delta \rho(\vec{k}) \left(\frac{\delta P(k)}{\delta\rho(k)}\right)_{\rm L}\,, \\
	\sigma^i_j(\vec{k}) &  \simeq \,\delta \rho(\vec{k}) \left(\frac{\sigma^i_j(k)}{\delta\rho(k)}\right)_{\rm L}\,,
	\end{aligned}
\end{equation}
where $\delta \rho \equiv \rho - \bar{\rho}$ and $\delta P \equiv P - \bar{P}$ are the density and pressure perturbation respectively away from their corresponding mean, and the subscript ``L'' denotes linear quantities computed with, e.g., a linear Boltzmann code such as \classcode{}.  See reference~\cite{Dakin:2017idt} for details on the \concept{} code and the implementation of nonlinear fluid neutrinos therein. The method of~\cite{Banerjee:2016zaa} also solves a similar set of nonlinear fluid equations, but, in contrast to \concept{}, closes the hierarchy using the nonlinear pressure and anisotropic stress estimated from tracer neutrino particles evolved alongside the fluid equations. The tracers do not however contribute to the gravitational potential.

We perform simulations using the public version of \concept{} for  the nu00 and nu05 cosmologies.  Our simulations are carried out in boxes of side length $L_{\text{box}}=256 \, \text{Mpc}/h$, with $N_{\text{cb}} = 512^3$ cold  matter particles. The neutrino grid size, $N_{\nu{\rm grid}} = 512^3$, matches the PM grid used to solve for the particle-neutrino and neutrino-neutrino gravitational forces, while a separate P${}^3$M grid of size $N_{\mathrm{P}^3\mathrm{M}} = 1024^3$ is employed for resolving  the PM component of the particle-particle gravity. Initialisation of the simulation takes place at  $z_{\rm sim}=49$, using (i)~the linear fluid quantities outputted  by \classcode{} directly for the grid-based neutrino fluid and (ii) the corresponding \classcode{} linear growth function to displace and kick the cold matter particles in the Zel'dovich approximation.

%%%%%%%%%%%%%%%%%%%%
\begin{figure}[t]
	\centering
	\includegraphics[trim=5mm 2mm 10mm 5mm,clip,width=75mm]{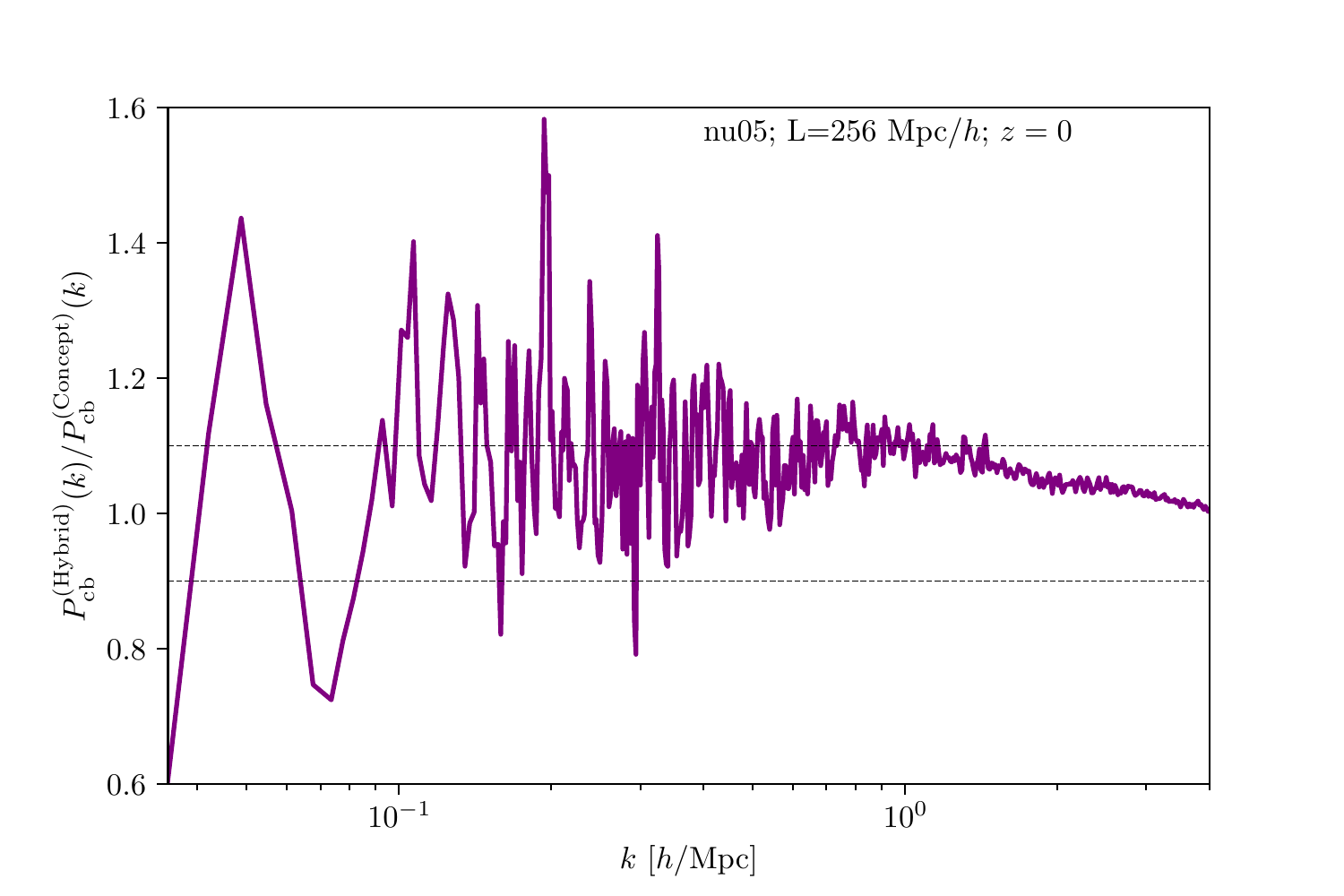}
	\includegraphics[trim=1mm 2mm 10mm 5mm,clip,width=75mm]{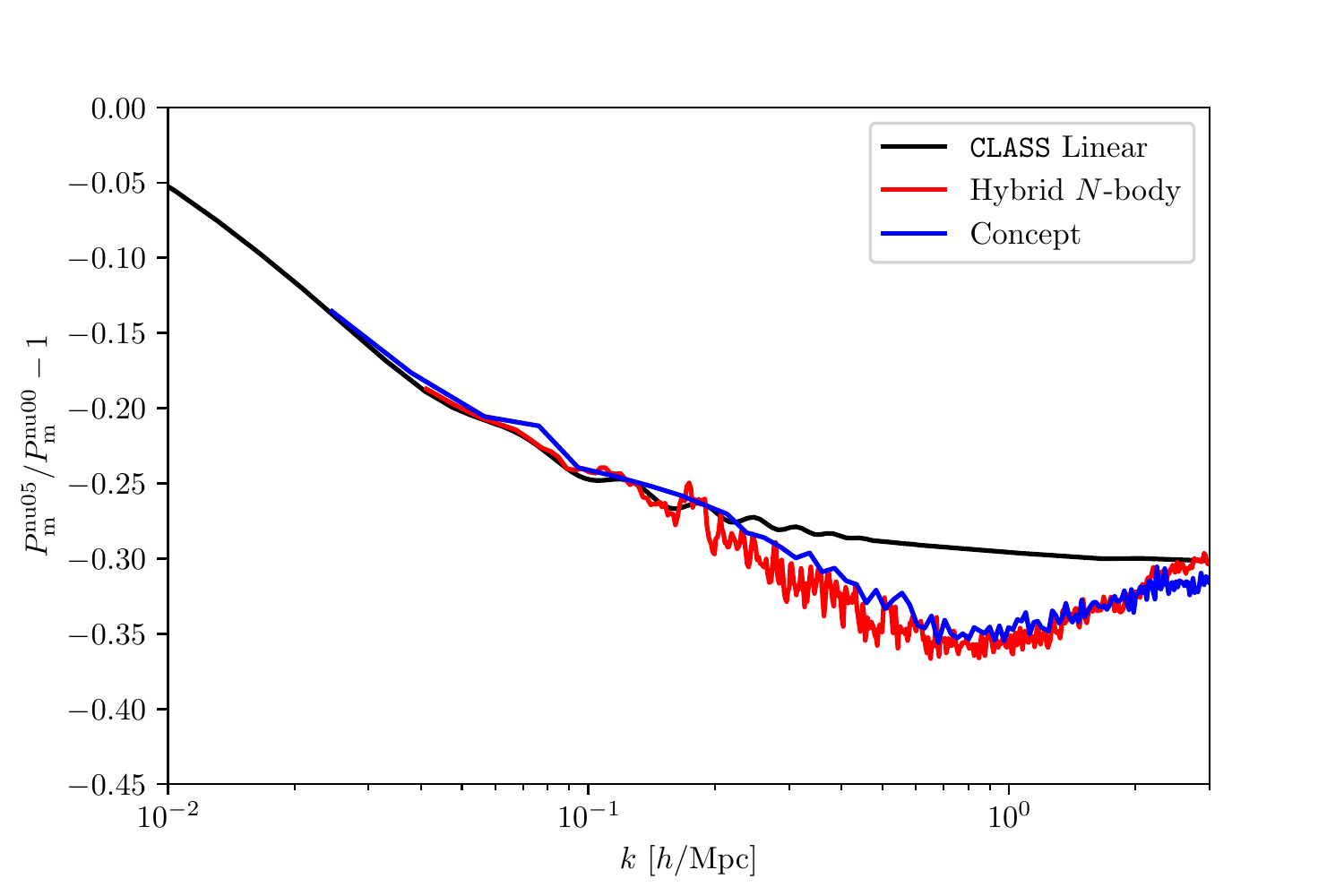}
	\caption{Hybrid-neutrino versus \concept{} comparison.
	{\it Left}: Ratio of the $z=0$ cold matter power spectra of the nu05 cosmology from the two approaches, using identical numbers of cold particles~$N_{\mathrm{cb}}$, box side length~$L_{\text{box}}$, and PM-grid size~$N_{\mathrm{PM}}$.
	{\it Right}: The $z=0$ total matter power spectrum of the nu05 massive neutrino cosmology, relative to the massless case nu00, computed from our hybrid-neutrino simulation (red), \concept{} (blue), and linear perturbation theory with \classcode{} (black).  The hybrid-neutrino result is very similar to the MFLR result (cf.~figure~\ref{fig:nu05LRSpoons}), as expected given the small nonlinear  enhancement in $P_{\rm m}^{\rm nu05}(k)$ due to nonlinear neutrino clustering seen in figure \ref{fig:CBMatterEnhancement}. \label{fig:GadgetvsConcept}}
\end{figure}
%%%%%%%%%%%%%%%%%%

The left panel of figure~\ref{fig:GadgetvsConcept} shows the ratio of the $z=0$  cold matter power spectra  of the nu05 cosmology, $P^{({\rm hybrid})}_{\rm cb}/P^{({\rm concept})}_{\rm cb}$,
formed from our hybrid-neutrino and the \concept{} simulations.  Fair
agreement between two codes can be seen, with differences of $\lesssim 10\%$ typically seen in $N$-body code comparisons.  In the right panel we plot the  ratios of the total matter power spectrum between the nu05 massive neutrino and nu00 massless neutrino cosmologies, $P_{\rm m}^{\rm nu05}(k)/P_{\rm m}^{\rm nu00}(k)-1$, computed from our hybrid-neutrino approach and from \concept{}.  Again, we seen a 1--2\%-level agreement between the two nonlinear approaches in predicting the spoon-shaped power suppression.

%%%%%%%%%%%%%%%%%%%%
\begin{figure}[t]
	\centering
	\includegraphics[trim=2mm 2mm 2mm 2mm,clip,width=130mm]{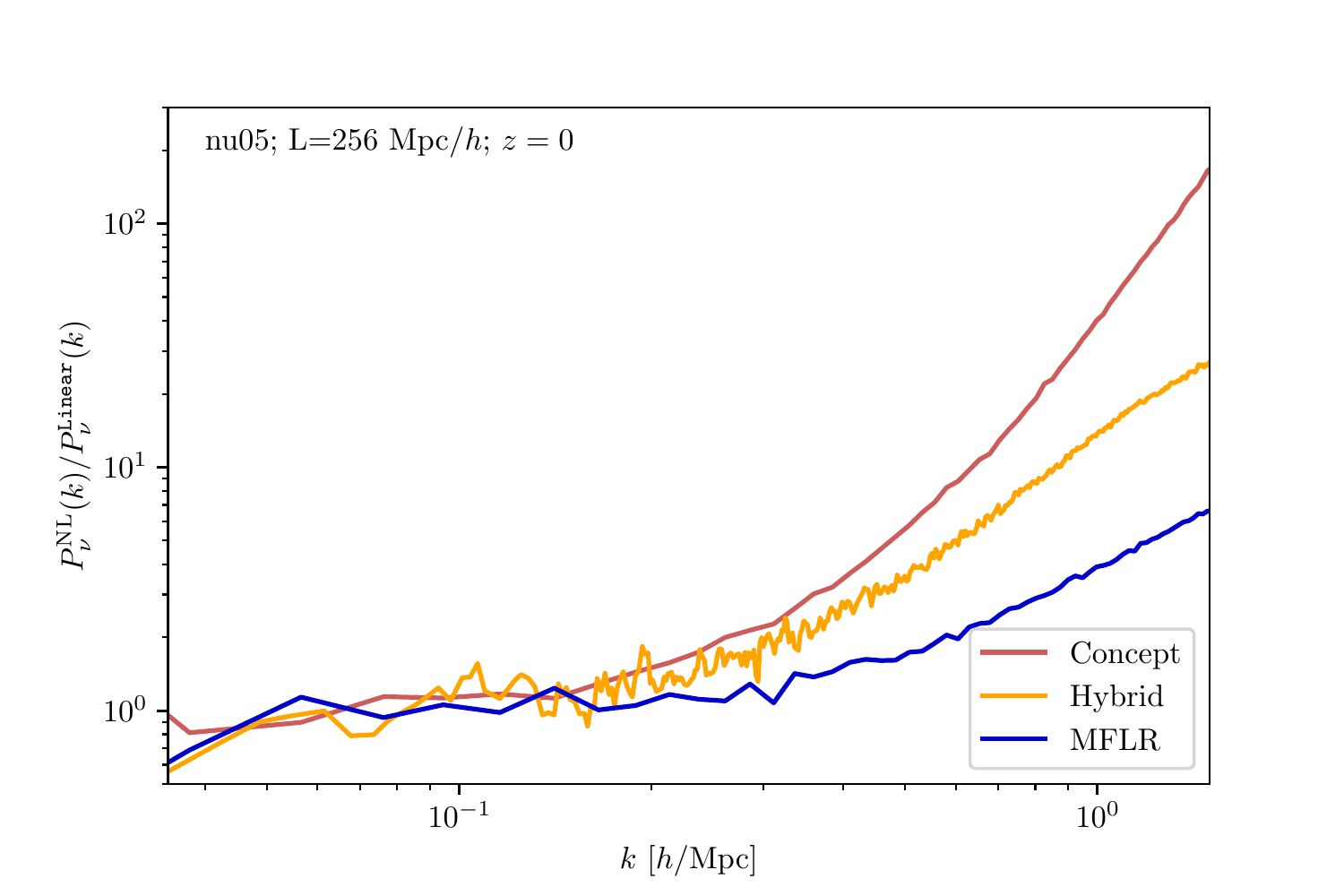}
	\caption{Nonlinear enhancement in the $z=0$ total neutrino power spectrum over the linear-theory prediction of~\classcode{} for the nu05 cosmology.  The red line represents the \concept{} result, orange the hybrid-neutrino approach of this work as described in section~\ref{sec:multiple_conversions}, and blue the MFLR prediction.\label{fig:GadgetvsConceptNu}}
\end{figure}
%%%%%%%%%%%%%%%%%%

The neutrino power spectra, however, show much larger differences. As shown in figure~\ref{fig:GadgetvsConceptNu}, hybrid-neutrino, \concept{}, and MFLR simulations all produce comparable total neutrino density power spectra at 
$k\lesssim 0.2~h/$~Mpc.  Immediately beyond this range, however, \concept{} begins to overestimate $P_\nu (k)$ relative to our hybrid-neutrino code, by up to a factor of two at $k \lesssim 0.8~h/$Mpc.   Interestingly though, despite the overestimation, the agreement between hybrid-neutrino and \concept{} is still better than that between hybrid-neutrino and MFLR up to this point.
At even larger wave numbers, the discrepancy between hybrid-neutrino and \concept{} grows, such that at the largest wave number shown in figure~\ref{fig:GadgetvsConceptNu}, $k \simeq 1.5~h/$Mpc, we see \concept{} overestimating $P_\nu(k)$ by a factor five relative to the hybrid-neutrino approach.

Reference~\cite{Dakin:2017idt} also found that \concept{} overestimated the total neutrino power in comparison with the original hybrid treatment of~\cite{Brandbyge:2009ce}, from $k \sim 0.3~h/$Mpc up to the $k$-scale where the latter became noise-dominated.  This overestimation was attributed to an unphysical effect whose cause was not identified.  We conjecture that the neutrino power excess may be due to the linear closure condition~\eqref{eq:closure} being inadequate to redistribute power in the lowest kinetic moments (density, velocity, pressure and anisotropic stress) to the higher kinetic modes, whose further investigation will have to be deferred to a future work.  However, with continuing progress in neutrino simulation methods, we are confident that better agreement between nonlinear methods can be achieved  in the neutrino sector in the near future.

%%%%%%%%%%%%%

\section{Conclusions}
\label{sec:conclusions}

We have implemented and thoroughly studied a new technique for simulating the cosmological clustering of massive neutrinos, the ``hybrid-neutrino'' $N$-body simulation, 
which combines a grid-based multi-fluid linear response theory together with a partial particle representation of the relic neutrino population.   Designed to model the velocity dispersion of massive neutrinos, the multi-fluid approach partitions the neutrino background into multiple ``flows'', each characterised by its initial momentum and obeys its own continuity and Euler equations~\cite{Dupuy:2013jaa,Dupuy:2014vea,Dupuy:2015ega}; the linearised version was previously implemented by some of us~\cite{Chen:2020bdf} into the Particle-Mesh component of \gadgetcode{}~\cite{Springel:2020plp} as a linear response to nonlinear cold matter perturbations.

The hybrid-neutrino method of the present work extends and improves upon the multi-fluid linear response simulations of~\cite{Chen:2020bdf} by selectively converting only the slowest-moving of the neutrino flows into a particle representation and only for a fraction of the simulation run-time when true nonlinear neutrino dynamics defy the linear response description.  When and where perturbations remain small and linear response suffices to describe their evolution, the flows are intentionally left as a linearised fluid solved on the mesh.
Thus, not only does the hybrid-neutrino approach offer fine-grained control over the distribution of computational resources where it is genuinely needed, it also circumvents the issue of noise-dominated outcomes in instances where the flow clustering power is too small to overcome the Poisson noise floor inherent in all particle-based simulations.

Using as a working example a massive neutrino cosmology with $\Omega_\nu h^2 = 0.005$, or equivalently $\sum m_\nu = 0.465$~eV, we find that only $\sim 30\%$ of the neutrino population has significant nonlinear clustering.  Moreover, out of this $30\%$, the fastest flows only begin to cluster nonlinearly at low redshifts (i.e., $z \lesssim 10$). Thus, one can construct a computationally least-intensive hybrid-neutrino simulation timeline wherein neutrino flows are evolved using multi-fluid linear response together with $N$-body cold matter (cold dark matter and baryons) particles from, e.g., $z=99$, and then converted from a fluid description to a particle representation in a staggered fashion: the slowest $10\%$ at a high redshift (e.g., $z = 19$), the next $10\%$ at an intermediate redshift (e.g., $z = 9$), and the last $10\%$ at a lower redshift (e.g., $z = 5$).  Such a staggered set of conversions at multiple redshifts agrees well with other particle conversion procedures (e.g., all 30\% converted at $z=19$) at the level of a few percent.  The final total neutrino density power spectrum from a hybrid-neutrino simulation, the corresponding nonlinear enhancement over its purely linear and multi-fluid linear response counterparts, and its contrast with the \concept{}~\cite{Dakin:2015uka,Dakin:2017idt,Dakin:2021ivb} outcome can be found in figures~\ref{fig:nu05PnuMonopoleEnhancements} and~\ref{fig:GadgetvsConceptNu}.

Last but not least, the power of the multi-fluid approach lies in its retention of fine-grained information about the neutrinos' momentum distribution at all times.  This information is preserved to a good extent also after conversion of the fluid flows to particles, but is a feature missing in other nonlinear neutrino methods such as \concept{}~\cite{Dakin:2015uka,Dakin:2017idt,Dakin:2021ivb}.
For our working example cosmology, figure~\ref{fig:AllFlowsDelta2NLEnhancements} shows the 
dimensionless power spectra of five representative groups of neutrino flows computed from hybrid-neutrino simulations, with peak powers spanning an order of magnitude at the free-streaming scale $k \sim 0.1~h/$Mpc and over two orders of magnitude at $k\sim 1~h/$Mpc.

At present, the main limitation on the momentum resolution of our hybrid-neutrino method is a free-streaming transient error arising from our neutrino particle initialisation procedure at particle conversion: when mapping the multi-fluid linear response outcomes to particle initial conditions, some information concerning the anisotropic free-streaming of neutrinos is invariably lost, exciting transients that take time to dissipate.  This problem is by no means unique to our particular method of neutrino particle simulations, but is especially prominent (up to $\sim 20$\%) in the power spectra of the individual neutrino flows around the associated free-streaming scale if conversion takes place at low redshifts (e.g., $z=5$).
It may, of course, be mitigated by converting all nonlinearly-clustering flows to particles at a suitably high redshift (e.g., $z \gtrsim 19$).  
However, we have also proposed an alternative neutrino particle initialisation procedure that would allow us to incorporate higher moments in the description of the neutrino anisotropies at the fluid-particle interface, which may eliminate these free-streaming transients and enable us to take advantage of low-redshift conversion.
We leave the exploration of this alternative initialisation scheme as well as potential nonlinear corrections to it to a future work.

%%%%%%%%

\acknowledgments

JZC acknowledges support from an Australian Government Research Training Program Scholarship. MM acknowledges support from C\'{e}line B{\oe}hm. 
AU is supported by the European Research Council (ERC) under the European Unions Horizon 2020 research and innovation programme (grant agreement No 769130).
Y$^3$W is supported in part by the Australian Research Council's Future Fellowship (project FT180100031).  This research is enabled by the Australian Research Council’s Discovery Project (project DP170102382) funding scheme, and includes computations using the computational cluster Katana supported by Research Technology Services at UNSW Sydney.

\appendix

\section{Interfacing MFLR and particles}
\label{sec:appendix}

We give in this appendix technical details to support the discussions in section~\ref{sec:prelim} that motivate our MFLR-to-particle conversion procedure.

\subsection{Perturbation phases}
\label{sec:phases}

Treating the gravitational potential as an external variable, the fluid equations~\eqref{eq:mflrFluidEqn} have the formal solution
\begin{equation}
\begin{aligned}
\label{eq:formalsoln}
\theta^P_\alpha (\vec{k},\mu,s) & =\,  m_\nu \, k^2 \int_{s_{\rm i}}^s {\rm d} s' \, a^2 (s') \,  \Phi (\vec{k},s') \, e^{-{\rm i}  k \mu \tau_\alpha (s-s')/m_\nu},\\
\delta_\alpha (\vec{k},\mu,s) & = \, - \frac{1}{m_\nu} \int_{s_{\rm i}}^s {\rm d} s' \,  \theta_{\alpha}^P (\vec{k},\mu,s') \, e^{-{\rm i}  k \mu \tau_\alpha (s-s')/m_\nu}\\
& =\,   - k^2 \int_{s_{\rm i}}^s {\rm d} s' \, a^2 (s') \, (s-s')\, \Phi (\vec{k},s') \, e^{-{\rm i}  k \mu \tau_\alpha (s-s')/m_\nu},
\end{aligned}
\end{equation}
where we have reinstated the $\vec{k}$ dependence of $\Phi(\vec{k})$ and 
assumed $\delta_{\alpha}(\vec{k},\mu,s_{\rm i})=\theta_{\alpha}(\vec{k},\mu,s_{\rm i})=0$ at the initial time $s_{\rm i}$.  The plane wave can be expanded as
\begin{equation}
\label{eq:planewave1}
\exp({\rm i}  k \mu x)	= \sum_{\ell =0}^\infty {\rm i}^\ell\,  (2 \ell +1) \, j_\ell (k x) \, {\cal P}_\ell (\mu),
\end{equation}	
where ${\cal P}_\ell$ is a Legendre polynomial, and $j_\ell$ a spherical Bessel function, leading to
\begin{equation}
\delta_\alpha(\vec{k},\mu,s) 
= - \sum_{\ell =0}^\infty (-{\rm i})^\ell\,  (2 \ell +1) \,  {\cal P}_\ell (\mu)\, \frac{1}{m_\nu}  \int_{s_{\rm i}}^s {\rm d} s' \, a^2 (s') \, \theta^P_\alpha (\vec{k},\mu,s') \,  j_\ell  \left[k \tau (s-s')/m\right],
\end{equation}
and similarly for $\theta_\alpha^P(\vec{k},\mu,s)$.  Then, decomposing
the perturbations in terms of a Legendre expansion~\eqref{eq:legendre} and using the orthogonality condition
\begin{equation}
\int_{-1}^1 {\rm d}\mu\, {\cal P}_\ell (\mu) \, {\cal P}_{\ell'}(\mu) = \frac{2}{2 \ell + 1} \delta_{\ell \ell'}^{\rm (K)},    
\end{equation}
it is straightforward to establish 
\begin{equation}
\begin{aligned}
\label{eq:thetaPla}
 \theta^P_{\alpha,\ell} (\vec{k},s)  & = \,  (2 \ell +1) \, m_\nu \, k^2  \int_{s_{\rm i}}^s {\rm d} s' \, a^2 (s') \,  \Phi(\vec{k},s') \,  j_\ell  \left[k \tau_\alpha (s-s')/m_\nu\right], \\
 \delta_{\alpha,\ell} (\vec{k},s)  & = \, -(2 \ell +1) \, \frac{1}{m_\nu}  \int_{s_{\rm i}}^s {\rm d} s' \, \theta^P_{\alpha, \ell} (\vec{k},s') \,  j_\ell  \left[k \tau_\alpha (s-s')/m_\nu\right] \\
 & = \, - (2 \ell +1) \, k^2  \int_{s_{\rm i}}^s {\rm d} s' \, a^2 (s') \, (s-s')\, \Phi(\vec{k},s') \,  j_\ell  \left[k \tau_\alpha (s-s')/m_\nu\right]
 \end{aligned}
\end{equation}
as the formal solutions for $\delta_{\alpha,\ell}(\vec{k},s)$ and $\theta^P_{\alpha,\ell}(\vec{k},s)$.

Following reference~\cite{Chen:2020kxi}, the solutions~\eqref{eq:thetaPla} can be analysed in the clustering and the free-streaming limits.  In the following, we shall focus on the momentum divergence solution~$\theta^P_{\alpha,\ell}(\vec{k},s)$; similar expressions can be straightforwardly deduced for the density perturbation $\delta_{\alpha,\ell}(\vec{k},s)$.

\paragraph{Clustering limit}  Identifying the clustering limit with  $ x  \equiv k \tau_\alpha (s-s')/m_\nu\to 0$, and noting that 
\begin{equation}
j_\ell(x) \to  \frac{2^{-(\ell+1)} \sqrt{\pi}}{\Gamma \left( \ell + \frac{3}{2} \right)} \,  x^\ell
\end{equation}
in the same limit, we find the clustering limit of the $\theta^P_{\alpha,\ell} (\vec{k},s)$ solution~\eqref{eq:thetaPla} to be
\begin{equation}
\label{eq:thetaPlc}
\theta^P_{\alpha,\ell} (\vec{k},s) \simeq 
\frac{2^{-\ell} \sqrt{\pi}}{\Gamma \left( \ell + \frac{1}{2} \right)}   \left(\frac{k \tau_\alpha}{m_\nu} \right)^\ell m_\nu  \, k^{2}   \int_{s_{\rm i}}^s {\rm d} s' \, a^2 (s') \,  \Phi(\vec{k},s') \,  (s-s')^\ell,
\end{equation}
which can be recast into the form
\begin{equation}
\label{eq:thetaPlcdiff}
\frac{\partial^{\ell+1}}{\partial s^{\ell+1}}\theta^P_{\alpha,\ell} (\vec{k},s)  \simeq  2^{-\ell} \sqrt{\pi} \, \frac{\Gamma\left(\ell+1 \right)}{\Gamma \left( \ell + \frac{1}{2} \right)} \, \left( \frac{k \tau_\alpha}{m_\nu}\right)^\ell  m_\nu  \, k^{2} \, a^2 (s) \,  \Phi (\vec{k},s).
\end{equation}
For $\ell=0$, equation~\eqref{eq:thetaPlcdiff} reduces to
\begin{equation}
\label{eq:thetaPlcdiff0}
\frac{\partial}{\partial s}\theta^P_{\alpha,\ell=0} (\vec{k},s)   \simeq   m_\nu  \, k^{2}   \, a^2 (s) \,  \Phi(\vec{k},s),
\end{equation}
which takes the same form as the linearised equation of motion for the cold matter velocity  divergence~$\theta_{\rm cb}$ up to the particle mass; a similar correspondence also holds between the flow density perturbation $\delta_{\alpha,\ell=0}(\vec{k},s)$ and cold matter density perturbation~$\delta_{\rm cb}(\vec{k},s)$.

Thus, a reasonable assumption is that $\delta_{\alpha,\ell=0}(\vec{k},s)$ and  $\theta^P_{\alpha,\ell=0}(\vec{k},s)$ will track and take on the same phase as 
$\delta_{\rm cb}(\vec{k},s)$ and $\theta_{\rm cb}(\vec{k},s)$ respectively.  In practice however, while the phase of $\Phi(\vec{k},s)$ is dominated by $\delta_{\rm cb}(\vec{k},s)$ and is hence easy to extract from a simulation snapshot, extracting the phase of $\theta_{\rm cb}(\vec{k},s)$ from a snapshot may not be straightforward.
Nonetheless, given that the clustering limit likely falls in or close to the linear regime, it is perhaps not unreasonable to expect the phase of $\theta_{\rm cb}(\vec{k},s)$ to approximate that of $\Phi_{\mathbf{k}}(s)$ as well.  This is the approximation we shall make in our MFLR-to-particle conversion procedure.

Lastly, observe that equation~\eqref{eq:thetaPlcdiff} and its $\delta_{\alpha,\ell}$ counterpart can be recast into the form
\begin{equation}
\begin{aligned}
\label{eq:thetaPlcdiffsimp}
\frac{\partial}{\partial s}\theta^P_{\alpha,\ell+1} (\vec{k},s)&  \simeq   \, \frac{\ell+1}{2\ell+1} \left(\frac{k \tau_\alpha}{m_\nu}\right) \theta^P_{\alpha,\ell} (\vec{k},s) ,\\
\frac{\partial}{\partial s}\delta_{\alpha,\ell+1} (\vec{k},s)&  \simeq   \, \frac{\ell+1}{2\ell+1} \left(\frac{k \tau_\alpha}{m_\nu}\right) \delta_{\alpha,\ell} (\vec{k},s) ,
\end{aligned}
\end{equation}
assuming $\theta^P_{\alpha,\ell} (\vec{k},s_{\rm i})$, $\delta_{\alpha,\ell}(\vec{k},s)$, and all of their time derivatives to be vanishing.  
Equation~\eqref{eq:thetaPlcdiffsimp} tells us that it takes one free-streaming timescale,  $\Delta s \sim m_\nu/(k \tau_\alpha)$, to populate the $(\ell+1)th$ multipole from the $\ell$th multipole.  Since the clustering limit is defined by $ x  \equiv k \tau_\alpha (s-s')/m_\nu\to 0$, we deduce that the $\ell>0$ multipoles are effectively never populated in this limit.  See also the left panel of figure~1 in reference~\cite{Chen:2020bdf}.

\paragraph{Free-streaming limit}
The free-streaming solution is constructed from applying integration by parts to equation~\eqref{eq:thetaPla}, i.e., 
\begin{equation}
\label{eq:thetaPlbyparts}
\theta^P_{\alpha,\ell} (\vec{k},s) \simeq (2 \ell +1) \, \frac{m^2_\nu k}{\tau_\alpha}  \Bigg\{\Big[ a^2 (x) \,  \Phi(\vec{k},x) \,  F_\ell  (x) \Big]^{x=x_{\rm i}}_{x=0} - \int_0^{x_{\rm i}} {\rm d} x\; \frac{{\rm d} (a^2 \Phi)}{{\rm d} x}\, F_\ell(x)
\Bigg\},
\end{equation}
where
\begin{equation}
F_\ell (x) \equiv \int {\rm d} x \, j_\ell (x)
\end{equation}
is an indefinite integral of the spherical Bessel function of order~$\ell$, defined up to a constant offset: we choose the offset such that $F_\ell(x \to \infty)  \to 0$.

To evaluate equation~\eqref{eq:thetaPlbyparts}, we note that ${\rm d} (a^2 \Phi)/{\rm d}x$ is a monotonically decreasing function of $x$ (or monotonically increasing function of $s'$).  On the other hand, the function $F_\ell(x)$ is fairly flat and tends to $-(\sqrt{\pi}/2)\Gamma(1/2 + \ell/2)/\Gamma(1+\ell/2) \sim {\cal O}(-1)$ at $x \lesssim \ell$, while at $x  \gtrsim \ell$ it drops quickly to zero.  (For $\ell = 0$, the switch occurs at $x\simeq 1$.)  Thus, the dominant contribution to the second integral in equation~\eqref{eq:thetaPlbyparts} comes from the interval $x=[0,\ell]$, i.e.,
\begin{equation}
 \int_0^{x_{\rm i}} {\rm d} x\; \frac{{\rm d} (a^2 \Phi)}{{\rm d} x}\, F_\ell(x) \simeq  \int_0^{\ell} {\rm d} x\;  \frac{{\rm d} (a^2 \Phi)}{{\rm d} x}\, F_\ell(x) \simeq \ell \, \left[ \frac{{\rm d} (a^2 \Phi)}{{\rm d} x}\, F_\ell(x) \right]_{x=0}. 
\end{equation}
Then, in comparison with the first term of equation~\eqref{eq:thetaPlbyparts}, we see immediately that the second term can be neglected if
\begin{equation}
\frac{\ell}{a^2 \Phi} \frac{{\rm d} (a^2 \Phi)}{{\rm d} s} \frac{m_\nu}{k \tau_\alpha} \ll 1,
\end{equation}
leading to the free-streaming solution
\begin{equation}
\begin{aligned}
\label{eq:thetaPlfs}
\theta^P_{\alpha,\ell} (\vec{k}, s)  &\simeq \,  - (2 \ell +1) \, \frac{m_\nu^2 k}{\tau_\alpha}  a^2 (s) \,  \Phi (\vec{k},s) \,  F_\ell  (x=0) \\
& =\, \sqrt{\pi} \left(\ell +1/2\right)   \frac{\Gamma\left(\frac{1}{2} + \frac{\ell}{2} \right)}{\Gamma \left(1 + \frac{\ell}{2} \right)} \, \frac{m_\nu^2 k}{\tau_\alpha}  a^2 (s) \,  \Phi (\vec{k},s)
\end{aligned}
\end{equation}
upon letting $x_{\rm i} \to \infty$.  

A solution similar to equation~\eqref{eq:thetaPlfs} can be constructed for $\delta_{\alpha, \ell}(\vec{k},s)$ in the same free-streaming limit,
\begin{equation}
\begin{aligned}
\label{eq:deltafs}
\delta_{\alpha,\ell} (\vec{k}, s)  &\simeq \,  - \sqrt{\pi} \left(\ell +1/2\right)   \frac{\Gamma\left(\frac{1}{2} + \frac{\ell}{2} \right)}{\Gamma \left(1 + \frac{\ell}{2} \right)} \, \frac{1}{k \tau_\alpha} \theta^P_{\alpha,\ell} (\vec{k},s)\\
&\simeq\, - \pi \left[(\ell+1/2) \frac{\Gamma(\frac{1}{2} + \frac{\ell}{2})}{\Gamma(1+\frac{\ell}{2})} \right]^2 \left(  \frac{m_\nu}{\tau_\alpha}\right)^2 a^2(s) \Phi(\vec{k},s),
\end{aligned}
\end{equation}
where we have used the approximate solution~\eqref{eq:thetaPlfs} at the second equality. Thus, as for $\delta_{\alpha, \ell=0}(\vec{k},s)$ the clustering limit, we again come to the conclusion that, in the free-streaming limit, $\delta_{\alpha, \ell}(\vec{k},s)$ and $\theta^P_{\alpha, \ell}(\vec{k},s)$ share the same phase as the gravitational potential $\Phi(\vec{k},s)$ at the same time~$s$.  What strongly distinguishes between the two limits, however, is that the $\ell>0$ multipoles in the free-streaming case are highly populated---more so than the monopole---as suggested by the solutions~\eqref{eq:thetaPlfs} and~\eqref{eq:deltafs} (see also the right panel of figure~1 in reference~\cite{Chen:2020bdf}), in comparison with their suppressed population in the clustering limit.

%%%%%%%%%%%
%%%%%%%%%%%%

\subsection{Initial physical velocity kick}
\label{sec:kick}

Suppose $\theta_\alpha^P(\vec{k},\mu,s)$ is the actual momentum divergence of a particular realisation (i.e., not just the $\hat{k}$-averaged quantity solved by our MFLR module) and we wish to use it to give neutrino particles initial peculiar velocities at MFLR-to-particle conversion.  To do so, let us consider what the momentum divergence $\theta^P(\vec{k},\mu,s)$ actually means. 

Performing an inverse Fourier transform back to real space, the real-space counterpart of $\theta^P_\alpha(\vec{k},\mu,s)$, $\theta^P_\alpha(\vec{x}, \mu,s) = {\cal F}^{-1}\left[\theta^P_\alpha (\vec{k},\hat{\tau}_\alpha,s)\right]$, is the divergence of the ``peculiar comoving momentum'' of the neutrino fluid flow with Lagrangian momentum $\vec{\tau}_\alpha = \tau_\alpha \hat{\tau}_\alpha$ at the spatial coordinate point $\vec{x}$.  That is, if we were to take a neutrino fluid with these properties at time~$s$, this fluid would have a comoving momentum given by 
\begin{equation}
\label{eq:pp}
\vec{q}_\alpha(\vec{x},\hat{\tau}_\alpha,s) = \tau_\alpha \hat{\tau}_\alpha + \Delta \vec{q}_\alpha(\vec{x},\hat{\tau}_\alpha,s),
\end{equation}
where, in our scheme, $\Delta \vec{q}_\alpha(\vec{x},\hat{\tau}_\alpha,s)$ is curl-free and related to the momentum divergence via  $\theta^P_\alpha(\vec{x},\hat{\tau}_\alpha,s) \equiv \nabla \cdot \Delta \vec{q}_\alpha(\vec{x},\hat{\tau}_\alpha,s)$. 

In terms of Fourier components and their Legendre moments, equation~\eqref{eq:pp} is equivalently
\begin{equation}
\begin{aligned}
\label{eq:ppsf}
\vec{q}_\alpha(\vec{x},\hat{\tau}_\alpha,s) &=\, \tau_\alpha \hat{\tau}_\alpha 
 - {\rm i} {\cal F}^{-1} \left[\frac{\vec{k}}{k^2} \theta^P_\alpha(\vec{k},\mu,s)\right] \\
& = \, \tau_\alpha \hat{\tau}_\alpha - \sum_{\ell =0}^\infty {\rm i}^{\ell+1} (-1)^\ell  {\cal F}^{-1}  \left[ \frac{\vec{k}}{k^2} 
 {\cal P}_\ell( \mu) \, \theta^P_{\alpha,\ell}(\vec{k},s) \right],
 \end{aligned}
\end{equation}
which we can also rewrite as 
\begin{equation}
\label{eq:kick0} 
\vec{u}_\alpha(\vec{x},\hat{\tau}_\alpha,s) 	 = \frac{\tau_\alpha}{a(s)m_\nu} \hat{\tau}_\alpha -\frac{1}{a(s) m_\nu}  \sum_{\ell =0}^\infty {\rm i}^{\ell+1} (-1)^\ell  {\cal F}^{-1}  \left[ \frac{\vec{k}}{k^2} 
 {\cal P}_\ell(\mu) \, \theta^P_{\alpha,\ell}(\vec{k},s) \right]
\end{equation}
for the physical velocity, $\vec{u}_\alpha \equiv \vec{q}_\alpha/(a m_\nu)$, of the neutrino fluid of Lagrangian momentum~$\vec{\tau}_\alpha$.

%%%%%%%%%%%%%%%%%%%%

\bibliographystyle{utcaps}
\bibliography{HybridBib}

\end{document}